\documentclass[a4paper,11pt]{article}
 \pdfoutput=1
 \usepackage{jheppub} 
 \usepackage{graphicx}
 \usepackage{amssymb}
\usepackage{dcolumn}
\usepackage{bm}
\usepackage{color}
\usepackage{multirow}

\allowdisplaybreaks

 \newcommand{\lsim}{{\;\raise0.3ex\hbox{$<$\kern-0.75em\raise-1.1ex\hbox{$\sim$}}\;}}
\newcommand{\gsim}{{\;\raise0.3ex\hbox{$>$\kern-0.75em\raise-1.1ex\hbox{$\sim$}}\;}}
\newcommand{\beq}{\begin{equation}}
\newcommand{\eeq}{\end{equation}}
\newcommand{\bea}{\begin{eqnarray}}
\newcommand{\eea}{\end{eqnarray}}
\def\l{\left}
\def\r{\right}

\def\baa{\begin{array}}
\def\eaa{\end{array}}

\def\hs{\hspace*{-0.05cm}}
\mathchardef\minus="002D

\title{\boldmath Using sorted invariant mass variables to evade combinatorial ambiguities in cascade decays
}
 
\author[a]{Doojin Kim,}
\author[a]{Konstantin T.~Matchev,} 
\author[b]{Myeonghun Park} 

\affiliation[a]{Physics Department, University of Florida, Gainesville, FL 32611, USA}
\affiliation[b]{Center for Theoretical Physics of the Universe, Institute for Basic Science (IBS), Daejeon, 34051, Korea} 

\emailAdd{imworry@ufl.edu}
\emailAdd{matchev@ufl.edu}
\emailAdd{parc.ctpu@gmail.com}

\abstract{The classic method for mass determination in a SUSY-like cascade decay chain
relies on measurements of the kinematic endpoints in the invariant mass distributions of
suitable collections of visible decay products. However, the procedure is complicated by
combinatorial ambiguities: e.g., the visible final state particles may be indistinguishable
(as in the case of QCD jets), or one may not know the exact order in which they
are emitted along the decay chain. In order to avoid such combinatorial ambiguities, 
we propose to treat the final state particles fully democratically and consider the
sorted set of the invariant masses of all possible partitions of the visible particles in the decay chain.
In particular, for a decay to $N$ visible particles, one considers the sorted sets 
of all possible $n$-body invariant mass combinations ($2\le n\le N$) and determines the
kinematic endpoint $m_{(n,r)}^{max}$ of the distribution of the $r$-th largest $n$-body 
invariant mass $m_{(n,r)}$ for each possible value of $n$ and $r$. 
For the classic example of a squark decay in supersymmetry, 
we provide analytical formulas for the interpretation of these
endpoints in terms of the underlying physical masses. 
We point out that these measurements can be used 
to determine the structure of the decay topology, e.g., the number 
and position of intermediate on-shell resonances. }

\preprint{CTPU-15-20} 
\date{December 3, 2015}

\begin{document} 
\maketitle
\flushbottom

\section{Introduction}
\label{sec:introduction}

\begin{figure}[t]
\centering
\includegraphics[width=10cm]{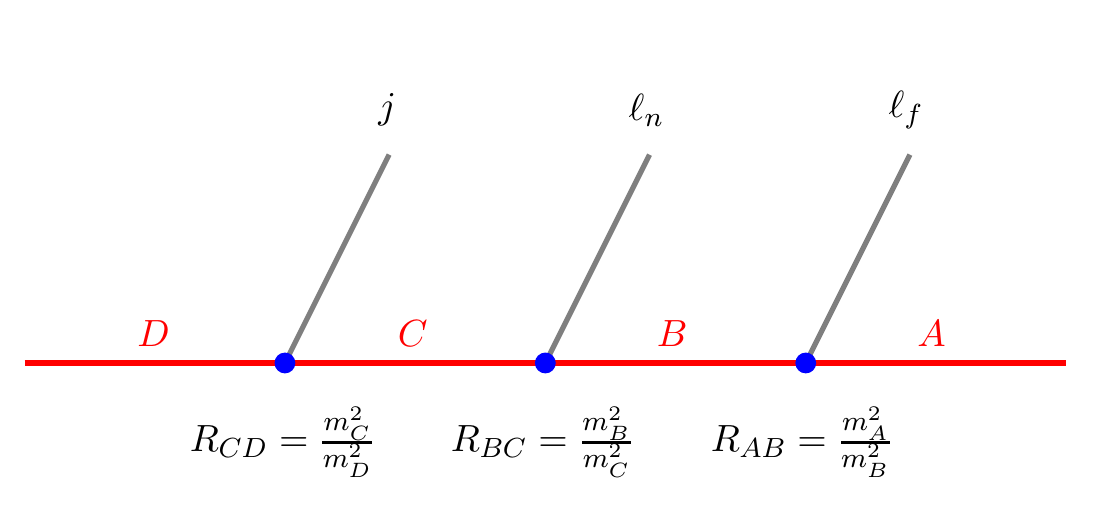}
\caption{\label{fig:chain} The typical cascade decay chain used in mass measurement studies.
Here $D$, $C$, $B$ and $A$ are new BSM particles, while the
corresponding SM decay products are: a QCD jet $j$, 
a ``near'' lepton $\ell_n^\pm$ and a ``far'' lepton $\ell_f^\mp$.
This chain is quite common in SUSY, with the identification
$D=\tilde q$, $C=\tilde\chi^0_2$, $B=\tilde \ell$ and $A=\tilde\chi^0_1$,
where $\tilde q$ is a squark, $\tilde \ell$ is a slepton, and
$\tilde\chi^0_1$ ($\tilde\chi^0_2$) is the first (second) 
lightest neutralino. Results for the masses of the new particles are
often quoted in terms of the $D$ mass $m_D$ and the three
dimensionless squared mass ratios 
$R_{CD}$, $R_{BC}$ and $R_{AB}$.}
\end{figure}

Now that the long awaited Higgs boson of the Standard Model (SM) appears to have been discovered
\cite{Chatrchyan:2012jja,Aad:2013xqa,Chatrchyan:2013mxa}, the best evidence for new physics Beyond the Standard Model (BSM) 
is provided by the dark matter problem \cite{Arrenberg:2013rzp}.
Dark matter particles can be produced directly at high energy colliders like the Large Hadron Collider (LHC) at CERN 
\cite{Bartl:1986hp,Birkedal:2004xn,Feng:2005gj}. However, the expected rates are relatively low, since dark matter 
is (super)weakly interacting. In generic models, therefore, the {\em indirect} dark matter production 
at colliders is much more copious, with dark matter particles appearing in the decay chains of 
heavier (perhaps colored or charged) new particles.
One such possible decay chain is shown in Fig.~\ref{fig:chain}, where a heavy new particle $D$ decays successively
to lighter particles $C$, $B$, and $A$, the latter perhaps being a dark matter candidate.
This particular decay chain is rather ubiquitous in models of low energy supersymmetry\footnote{An analogous 
decay chain can be present in many other BSM models, e.g.~in Universal Extra Dimensions (UED) \cite{Appelquist:2000nn}, where
$D$ is a Kaluza-Klein (KK) quark, $C$ is a KK $Z$-boson, $B$ is a KK lepton, and $A$ is a KK photon 
\cite{Cheng:2002ab}. } (SUSY), 
where particle $D$ is a squark, $C$ is a heavy neutralino, $B$ is a slepton, and $A$ is the lightest neutralino
(the typical dark matter candidate in SUSY).

The main difficulty in the analysis of the decay chain of Fig.~\ref{fig:chain} stems from the fact that
particle $A$, being a dark matter candidate, is invisible in the detector, hence its energy and momentum are not 
directly measured. This makes the problem of determining the masses and spins of the new particles $A$ through $D$
rather challenging. Over the last 20 years, a fairly large body of literature\footnote{See, e.g., the recent reviews 
\cite{Barr:2010zj,Wang:2008sw} and references therein.} has been devoted to this problem. Among the different 
approaches which have been proposed, the classic method of kinematic endpoints is arguably the most popular
and robust technique for mass determination. With this method, one studies the invariant mass distributions 
of different combinations of the visible decay products and attempts to locate their kinematic endpoints
(generally in the presence of some background continuum). In the example of the decay chain in Fig.~\ref{fig:chain}, 
one can form three 2-body invariant mass variables ($m_{j\ell_n}$, $m_{j\ell_f}$, and $m_{\ell\ell}$) 
and one 3-body invariant mass variable, $m_{j\ell\ell}$, so that the basic set of invariant mass variables is
\beq
\left\{ m_{j\ell_n}, m_{j\ell_f}, m_{\ell\ell}, m_{j\ell\ell}  \right\}.
\label{IMbasic}
\eeq
Ideally, one would like to study each of the variables (\ref{IMbasic}) {\em individually} and obtain the
corresponding kinematic endpoints $m_{j\ell_n}^{max}$, $m_{j\ell_f}^{max}$, $m_{\ell\ell}^{max}$, and $m_{j\ell\ell}^{max}$,
which can be simply related to the unknown masses $\{m_A, m_B, m_C, m_D\}$
by analytic expressions available in the literature (see, e.g., \cite{Allanach:2000kt,Gjelsten:2004ki,Burns:2009zi}).
Unfortunately, the situation is not that simple, as it becomes muddled by various combinatorial ambiguities:
\begin{itemize}
\item {\em Partitioning ambiguity.} In general, in addition to the three visible objects from the decay chain in
Fig.~\ref{fig:chain}, there will be a number of additional objects in the event --- the decay products from the 
other\footnote{The lifetime of the dark matter candidate is typically protected by a $Z_2$ parity, which implies that
new particles are necessarily pair-produced. Therefore, each event contains a second decay chain, similar to the 
one depicted in Fig.~\ref{fig:chain}.} decay chain in the event, jets from initial state radiation, pile-up, etc.
The question then becomes, how to select the correct objects $j$, $\ell_n$ and $\ell_f$ for the analysis variables
(\ref{IMbasic}). This is a really challenging problem, for which no universal solution exists, although several
ideas have been tried, including the ``hemisphere method" \cite{MP,Matsumoto:2006ws}, (a combination of)
invariant mass and $M_{T2}$ cuts \cite{Rajaraman:2010hy,Bai:2010hd,Baringer:2011nh,Choi:2011ys}, and 
neural networks \cite{Shim:2014aua}. In this paper, we will ignore the partitioning ambiguity and instead focus 
on the ordering ambiguity described next. Our assumption is justified in the case when particle $D$ is produced singly,
or when $D$ is produced in association\footnote{A well-known such example in SUSY is provided by the associated squark-neutralino 
production \cite{Baer:1990rq,Agrawal:2013uka}.} with the stable particle $A$, so that a second decay chain simply does not exist.
\item {\em Ordering ambiguity.} Having selected the correct visible objects arising in a given decay chain, 
we still have to decide on the {\em order} in which they are emitted along the chain. For example, motivated by SUSY,
in Fig.~\ref{fig:chain} one makes the specific assumption that the jet comes first, followed by the two leptons.
However, even with this extra assumption, the ambiguity is not completely resolved, as we still do not know the
exact ordering of the two leptons. In other words, we are not justified in using the labels ``near" and ``far" to denote 
the two leptons, which makes it impossible to study separately the distributions
of $m_{j\ell_n}$ and $m_{j\ell_f}$ in the real experiment.

Two possible ways out of this conundrum have been suggested. The standard approach \cite{Allanach:2000kt} 
is to trade the variables $m_{j\ell_n}$ and $m_{j\ell_f}$ for their ordered cousins
\begin{eqnarray}
m_{j\ell(lo)} &\equiv& \min \left\{m_{j\ell_n}, m_{j\ell_f} \right\}  , \label{mjllodef} \\ [2mm]
m_{j\ell(hi)} &\equiv& \max \left\{m_{j\ell_n}, m_{j\ell_f} \right\}  . \label{mjlhidef}
\end{eqnarray}
Then, instead of the set (\ref{IMbasic}), one can consider the alternative set of variables
\beq
\left\{ m_{j\ell(lo)}, m_{j\ell(hi)}, m_{\ell\ell}, m_{j\ell\ell}  \right\},
\label{IMhilo}
\eeq
measure their respective endpoints, and from those extract the physical mass spectrum \cite{Allanach:2000kt,Gjelsten:2004ki,Burns:2009zi}. 
A more recent alternative approach \cite{Matchev:2009iw} introduces new invariant mass variables which are symmetric functions of
$m_{j\ell_n}$ and $m_{j\ell_f}$, thus avoiding the need to distinguish $\ell_n$ from $\ell_f$ on an event per event basis.
\end{itemize}

However, while both of these approaches are designed to address the ordering ambiguity problem, 
it is our view that they do not go far enough --- in the sense that the assumption of the jet 
being the first emitted particle is still hardwired in the analysis from the very beginning. From the point of view of an experimenter
whose duty is to perform unbiased measurements without theoretical prejudice, there is no compelling reason to make
that assumption. One can easily construct theory models\footnote{Granted, such models will contain intermediate 
particles with unusual, ``leptoquark", quantum numbers.} in which the jet is the second (or the third) visible particle in the diagram of 
Fig.~\ref{fig:chain}. In order to account for such scenarios, one needs to further generalize the reordering in 
Eqs.~(\ref{mjllodef},\ref{mjlhidef}) to include swapping the jet with one of the leptons. To be concrete, 
for the example of Fig.~\ref{fig:chain}, we propose to
further replace the two-body invariant mass variables $m_{j\ell(lo)}$, $m_{j\ell(hi)}$ and $m_{\ell\ell}$ from (\ref{IMhilo}) with
the {\em three} ordered variables 
\begin{eqnarray}
m_{(2,1)} &\equiv& {\max}_1 \left\{m_{j\ell_n}, m_{j\ell_f}, m_{\ell\ell} \right\}  , \label{mord1} \\ [2mm]
m_{(2,2)} &\equiv& {\max}_2 \left\{m_{j\ell_n}, m_{j\ell_f}, m_{\ell\ell}  \right\}  , \label{mord2} \\ [2mm]
m_{(2,3)} &\equiv& {\max}_3 \left\{m_{j\ell_n}, m_{j\ell_f}, m_{\ell\ell}  \right\}  , \label{mord3}
\end{eqnarray}
where we have used the function ${\max}_r$ to denote the $r$-th largest among a given set of elements\footnote{Obviously, 
the $r$-th largest among $r$ elements is the {\em smallest} of those elements:
$$
{\max}_r \left\{ x_1, x_2, \ldots, x_r \right\} \equiv \min \{ x_1, x_2, \ldots, x_r \},
$$
so that Eq.~(\ref{mord3}) is the generalization of Eq.~(\ref{mjllodef}).}.
The variable set (\ref{IMhilo}) is then replaced by 
\beq
\left\{ m_{(2,1)}, m_{(2,2)}, m_{(2,3)}, m_{j\ell\ell}  \right\},
\label{IM123}
\eeq
whose kinematic endpoints can then be used to extract the physical spectrum.
To this end, however, one would first need to derive the analytic formulas for these new 
endpoints in terms of the physical masses, and this will be one of the main goals of this paper.

\begin{figure}[t]
\centering
\includegraphics[width=7cm]{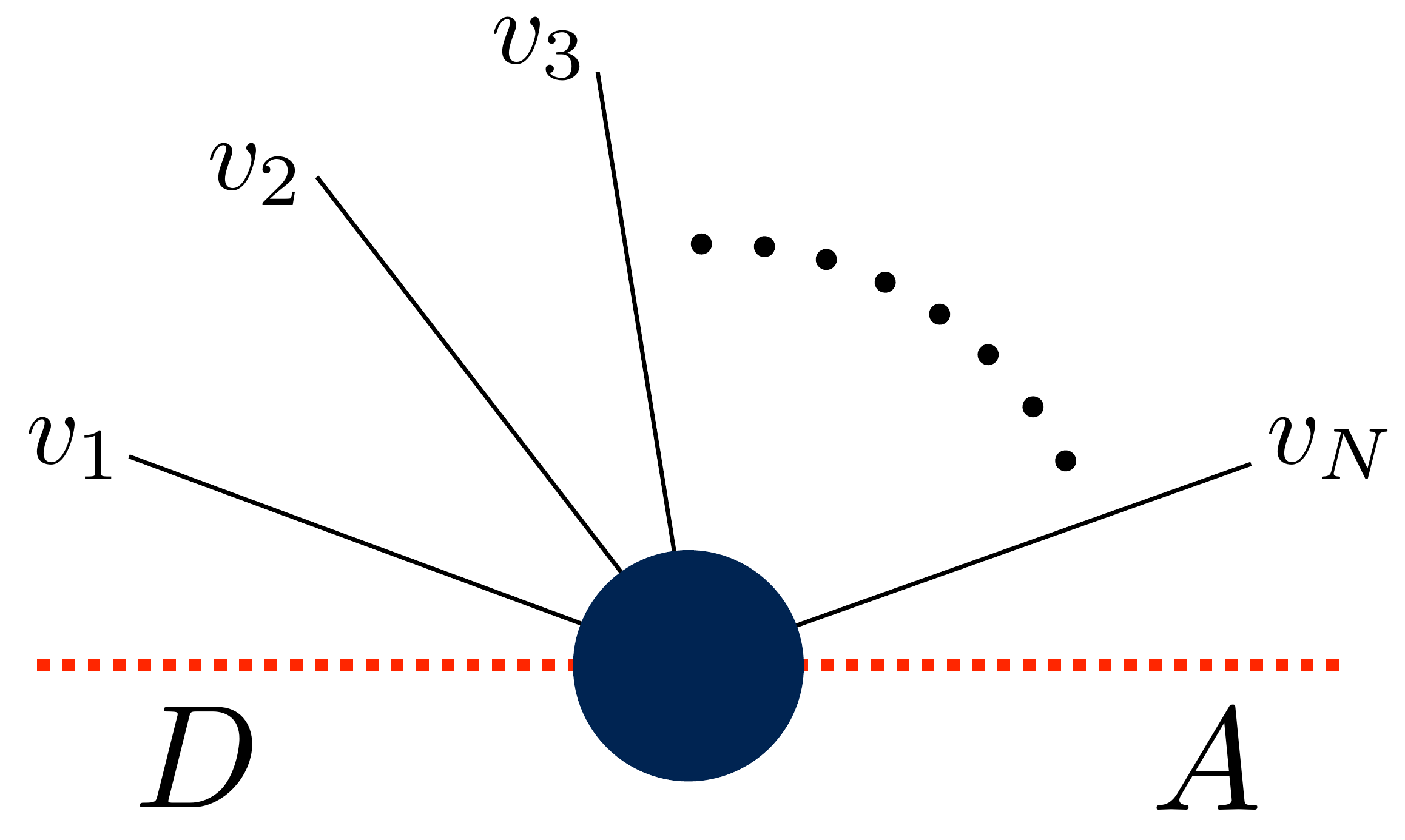}
\caption{\label{fig:gendecay} The generic decay of a heavy resonance $D$ to $N$ indistinguishable 
visible particles (solid black lines) and one invisible particle $A$ (red dashed line).
}
\end{figure}	

In summary, in this paper we propose to evade the ordering ambiguity problem by 
considering all possible partitions of the visible particles resulting from a given cascade decay chain, 
and then studying the kinematic endpoints of the corresponding sorted invariant mass variables.
We shall try to keep our discussion as general as possible --- for example, in defining
the sorted invariant mass variables, we shall have in mind the generic diagram in Fig.~\ref{fig:gendecay} 
instead of the specific SUSY-inspired example of Fig.~\ref{fig:chain}.
In the case of Fig.~\ref{fig:gendecay}, a heavy resonance $D$ decays to 
$N$ indistinguishable visible particles (denoted by solid black lines) and 
one invisible particle $A$ (denoted by a red dashed line). 
We first form the set 
\beq
{\cal S}_n^N \equiv \left\{ m_{v_{i_1}v_{i_2}\ldots v_{i_n}}\right\}
\label{set}
\eeq
of all possible $n$-body invariant mass combinations $m_{v_{i_1}v_{i_2}\ldots v_{i_n}}$
for a given $n\in \left[ 2, N \right]$. The total number $C^N_n$ of elements in the 
set ${\cal S}_n^N$ depends on the choice of $n$ and is given by the binomial coefficient
\beq
C^N_n = \begin{pmatrix} N\\ n \end{pmatrix} = \frac{N!}{n!(N-n)!}.
\label{eq:binomial}
\eeq
We can now uniquely and unambiguously label the members of the set ${\cal S}_n^N$ 
by defining {\em sorted}\footnote{In what follows, we shall sometimes equivalently refer
to (\ref{eq:defsort}) as ``ranked" variables.} 
invariant mass variables in analogy to Eqs.~(\ref{mord1}-\ref{mord3}):
\beq
m_{(n,r)}\equiv {\max}_r \left\{ {\cal S}_n^N \right\}.
\label{eq:defsort}
\eeq
Using Eq.~(\ref{eq:binomial}), it is easy to see that for a given $N$, there are a total of 
\beq
\sum_{n=2}^N C^N_n  = 2^N-N-1
\label{eq:count}
\eeq 
such variables.
The physical meaning of the variable $m_{(n,r)}$ is that it is the $r$-th largest among
all possible $n$-body invariant mass combinations of visible particles in Fig.~\ref{fig:gendecay}. 
From the general definition (\ref{eq:defsort}) it is easy to make contact with the previously 
considered $N=3$ case of Fig.~\ref{fig:chain}: in the notation of (\ref{eq:defsort}), 
the variable set (\ref{IM123}) corresponds to
\beq
\left\{ m_{(2,1)}, m_{(2,2)}, m_{(2,3)}, m_{(3,1)}  \right\}.
\label{IM1234}
\eeq

In introducing the more general variables (\ref{eq:defsort}) we are motivated by several factors:
\begin{enumerate}
\item Often the visible particles in the cascade decay are indistinguishable, 
e.g.~they are all QCD jets. This can easily be the case even with the SUSY 
example of Fig.~\ref{fig:chain}, whenever the second-lightest neutralino (particle $C$)
decays predominantly hadronically to 2 jets and the lightest neutralino (particle $A$).
Another well motivated SUSY example is a squark-gluino-neutralino decay chain where 
again all three visible particles are jets. In such scenarios, {\em a priori} there is no way
to single out any particular jet, and the set (\ref{IM1234}) is the only one which makes physics sense.
\item In a purely off-shell scenario, where particle $D$ decays directly to particle $A$ 
plus $N$ visible particles, it is not possible to assign any specific order to the
decay products, even if they are distinguishable experimentally.
\item Even when the decay of particle $D$ proceeds through intermediate
narrow resonances, so that the visible decay products are emitted in some well-defined
order, this true order is unknown to the experimentalist, and can only be hypothesized.
In general, alternative theory models, with alternative orderings of the same 
final state particles, are also possible. Therefore, {\em assuming} a specific ordering
throughout the analysis is dangerous and may lead to wrong conclusions.
\item Finally, by staying clear of any theory bias, and considering the most general
case of Fig.~\ref{fig:gendecay}, we will be able to derive  (see Section~\ref{sec:off} below) 
the necessary relations which must be obeyed by the kinematic endpoints of the variables 
(\ref{eq:defsort}) in the case of a pure off-shell decay (i.e., with no intermediate 
resonances).  Any observed deviations from those predictions will signal the presence of 
new particles in addition to the mother $D$ and daughter $A$. Furthermore, 
as illustrated in Fig.~\ref{fig:endpoints} below, the measured relations among the kinematic 
endpoints are indicative of the particular on-shell event topology at hand.
\end{enumerate}

In this paper we begin the investigation of the mathematical properties of the 
variables (\ref{eq:defsort}), and, in particular, their kinematic endpoints.
In Section~\ref{sec:off} we focus on the general case of an off-shell $1\to N+1$ 
decay as depicted in Fig.~\ref{fig:gendecay}. We derive the formulas for the kinematic 
endpoints of the sorted invariant mass variables (\ref{eq:defsort}) in terms of the 
relevant physical mass parameter, the mass difference $m_D-m_A$. For $N>2$, 
the number of variables given by (\ref{eq:count}) already exceeds the number of input parameters,
which implies certain specific relations among the measured kinematic endpoints.
Among the main results of Section~\ref{sec:off} is the derivation of these relations,
as they provide a stringent test of the offshellness of the decay topology.\footnote{In 
the case of $N=2$, the number of available kinematic endpoints
is equal to the number of input parameters, thus testing for intermediate resonances 
is much more challenging. However, it can still be done by studying the {\em shape} of the invariant 
mass distribution $m_{v_1v_2}$ \cite{Cho:2012er}.}

Having dealt with the general case of arbitrary $N$ in Section~\ref{sec:off}, 
in the following sections we return to the SUSY-motivated case of $N=3$.
We shall similarly study the dependence of the sorted invariant mass endpoints 
on the physical mass parameters, in the presence of intermediate on-shell resonances.
The relevant special cases are discussed in Sections~\ref{sec:111}, \ref{sec:210}
and \ref{sec:120}. Section~\ref{sec:conc} is reserved for our summary and conclusions.

\section{The pure off-shell case of $(N+1)$-body decay}
\label{sec:off}

Consider the decay of a heavy resonance $D$ into $N$ massless visible particles
and one massive invisible particle $A$, as shown in Fig.~\ref{fig:gendecay}: 
\beq
D \rightarrow v_1 v_2 \cdots v_N A.
\label{DtoA}
\eeq
In this section we shall assume that the decay (\ref{DtoA}) proceeds in one step, 
i.e., {\em exactly} as depicted in Fig.~\ref{fig:gendecay}.
In other words, any virtual particles hiding behind the circular blob in Fig.~\ref{fig:gendecay} are sufficiently
heavy and can be integrated out to give rise to an effective contact interaction as shown in the figure. 

As mentioned in the Introduction, we treat all visible particles in the final state as {\it indistinguishable},
so that we do not know {\it a priori} the sequence in which the visible particles are emitted. 
This motivates us to consider the {\em sorted} invariant mass variables $m_{(n,r)}$ defined in Eq.~(\ref{eq:defsort}).
In what follows, we shall refer to the first index $n$ as the ``order" of the variable, 
while the second index $r$ will denote its ``rank". Obviously, the order $n$ can be chosen 
to be any integer from 2 to $N$; for completeness we shall consider all possible values of $n$,
i.e., we shall construct two-body, three-body, etc. invariant masses of visible particles.
For a given order $n$, the rank $r$ then takes values from 1 to $C^N_n$.

Our main goal in this section is to provide the analytic expressions for the kinematic endpoints
$m_{(n,r)}^{max}$ in terms of the physical masses $m_D$ and $m_A$. In all results below, 
we shall always factor out the parent mass $m_D$ and write the formulas in terms of the dimensionless
squared mass ratios\footnote{Note the following transitive and inversion properties
$$
R_{ik}=R_{ij}\cdot R_{jk}, \quad 
R_{ij}=R_{ji}^{-1}.
$$}  
\bea
R_{ij}\equiv \frac{m_i^2}{m_j^2}, \;\;\;\lbrace i, j \rbrace \in \lbrace D,C,B,A \rbrace,
\label{Rijdef}
\eea
where by assumption the particle masses obey the hierarchy
\beq
m_D>m_C>m_B>m_A.
\eeq
In the case of the pure off-shell process (\ref{DtoA}), the only two masses entering the problem are
$m_D$ and $m_A$, thus our results will be functions of $R_{AD}<1$.

\subsection{Invariant masses of order $n=2$}
\label{sec:offn2}

We first discuss the sorted invariant masses of order 2 (i.e., the two-body invariant masses),
for which a useful sum rule can be derived as follows. Using $4$-momentum conservation
for the reaction (\ref{DtoA})\footnote{From here on, we simplify our notation as
$p_i\equiv p_{v_i}$, $m_{ij}\equiv m_{v_iv_j}$, etc.}
\beq
p_D = p_1+\cdots+p_N + p_A,
\label{momcons}
\eeq
we can write
\bea
(p_D-p_A)^2=(p_1+\cdots+p_N)^2=\sum_{i=1}^{N} p_i^2 +  2\sum_{\substack{i,j=1\\i<j}}^N p_i\cdot p_j ,
\label{summ2sq}
\eea
Since the visible particles are assumed massless, $p_i^2=0$, and furthermore, 
$2p_i\cdot p_j$ is simply $(p_i+p_j)^2=m_{ij}^2$ so that the above relation can be rewritten as
\bea
(p_D-p_A)^2=\sum_{r=1}^{C_2^N} m_{(2,r)}^2.
\label{eq:offeq}
\eea
The left-hand side in this equation may be interpreted as the ``total available invariant mass squared"
which is allocated to $m_{(2,r)}^2$'s in a given event. Since Eq.~(\ref{eq:offeq}) is Lorentz-invariant,
we can evaluate its left-hand side in any frame. It is convenient to choose the rest frame of particle $D$, where
\bea
(p_D-p_A)^2=m_D^2+m_A^2-2m_DE_A.
\label{totalIM}
\eea
We are interested in kinematic endpoints, i.e., the cases in which a particular variable $m_{(2,r)}$ is maximized. 
Eq.~(\ref{eq:offeq}) suggests that in order to maximize an individual variable $m_{(2,r)}$, we must necessarily 
maximize the total quantity (\ref{totalIM}) as well. It is easy to see that (\ref{totalIM}) is maximized when 
$A$ is produced at rest and its energy $E_A=m_A$. We therefore conclude that {\em for an event which
yields a kinematic endpoint of $m_{(2,r)}$ for some value of $r$}, the following sum rule holds
\bea
\sum_{r} m_{(2,r)}^2=(m_D-m_A)^2.
\label{sr2r}
\eea
We are now in position to derive the formulas for the kinematic endpoints $m_{(2,r)}^{max}$ for various $r$.

\begin{figure}[t]
\centering
\includegraphics[width=14.5cm]{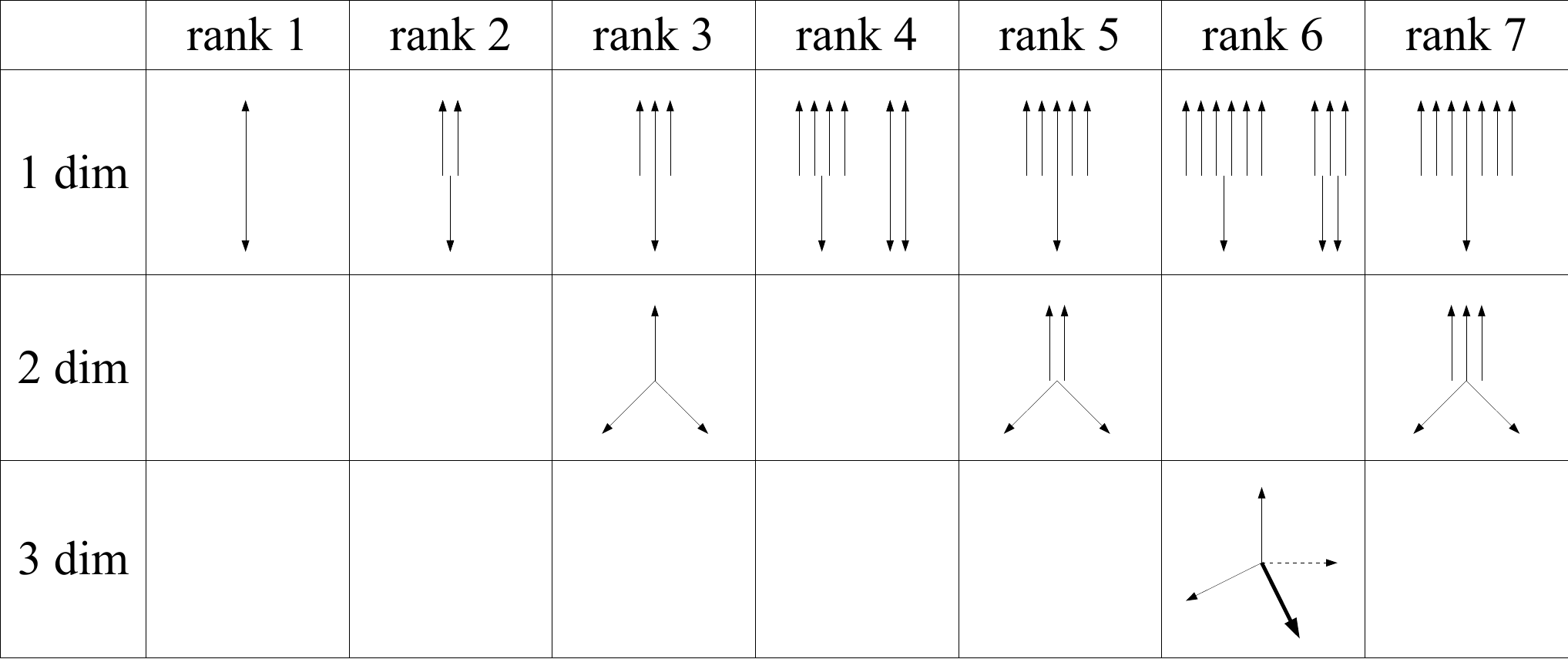}
\caption{\label{fig:table} Possible geometric orientations of the visible particle momenta in the extreme events
corresponding to the kinematic endpoints of the $m_{(2,r)}$ variables for $r\le 7$.
The lengths of the arrows are {\em not} indicative of the relative magnitudes of the momenta.
The remaining visible particles not depicted by arrows are taken to be at rest.
The configurations in the first row are collinear, the configurations in the second row are planar,
while the configurations in the third row are three-dimensional.}
\end{figure}

{\bf Rank $r=1$.} The largest possible value of $m_{(2,1)}$ is obtained when all other invariant mass combinations are
vanishing, i.e., for events in which 
\beq
m_{(2,1)} = m_{(2,1)}^{max} \quad {\rm and} \quad
m_{(2,r)} = 0, \quad {\rm for}\quad r=2,\ldots, C^N_2.
\label{conf21}
\eeq
The momentum configuration of such an event is shown in Fig.~\ref{fig:table} ---
two of the visible particles are exactly back-to-back, while the remaining visible particles,
together with the invisible particle $A$, are all at rest. The $m_{(2,1)}^{max}$ endpoint is now obtained by substituting 
(\ref{conf21}) into (\ref{sr2r}):
\bea
m_{(2,1)}^{\max}=m_D-m_A=m_D\, (1-\sqrt{R_{AD}}).
\label{m21max}
\eea

{\bf Rank $r=2$.} According to Eq.~(\ref{sr2r}), in order to maximize $m_{(2,2)}$, we need to minimize
both $m_{(2,1)}$ and $m_{(2,r)}$ for $r\ge 3$. However, by definition $m_{(2,1)}$ cannot be less than $m_{(2,2)}$,
thus for events giving the largest possible value of $m_{(2,2)}$ we expect to have
\beq
m_{(2,1)} = m_{(2,2)} = m_{(2,2)}^{max} \quad {\rm and} \quad
m_{(2,r)} = 0, \quad {\rm for}\quad r=3,\ldots, C^N_2.
\label{conf22}
\eeq
The momentum configurations of such events are also collinear, as shown in Fig.~\ref{fig:table}.
Now there are {\em three} visible particles with non-zero momenta, two of them having equal 
momenta and recoiling against the third. From (\ref{conf22}) and (\ref{sr2r}) we obtain
\bea
m_{(2,2)}^{\max}=\frac{1}{\sqrt{2}}\, (m_D-m_A)= \frac{1}{\sqrt{2}}\, m_D\, (1-\sqrt{R_{AD}}).
\label{m22max}
\eea

{\bf Higher ranks ($r>2$).} Proceeding analogously, one might {\it na\"{i}vely} expect that 
for an arbitrary rank $r$, Eqs.~(\ref{m21max}) and (\ref{m22max}) generalize to
\bea
m_{(2,r)}^{\max}\stackrel{?}{=}\frac{1}{\sqrt{r}}\, (m_D-m_A)=\frac{1}{\sqrt{r}}\,m_D\,(1-\sqrt{R_{AD}}).
\label{eq:upbound}
\eea 
However, one has to be careful to check whether physical momentum configurations exist such that
\beq
m_{(2,1)} = m_{(2,2)} = \ldots = m_{(2,r)} = m_{(2,r)}^{max} \quad {\rm and} \quad
m_{(2,i)} = 0, \quad {\rm for}\quad i=r+1,\ldots, C^N_2.
\label{conf2r}
\eeq
This check is not aways trivial. As a concrete example, consider the rank 10 variable $m_{(2,10)}$ in the case 
of $N=5$ total visible particles. It is not difficult to see that in three spatial dimensions, there are 
no allowed momentum configurations which would give the required case with 
$m_{(2,1)}=m_{(2,2)}=\cdots=m_{(2,10)}$. As a result, in this case the conjectured endpoint 
\beq
m_{(2,10)}^{\max}\stackrel{?}{=}\frac{1}{\sqrt{10}}\, (m_D-m_A)
\label{m510conjecture}
\eeq
will not be saturated, and the true endpoint will appear at slightly lower values of $m_{(2,10)}$.
However, in such situations where the general formula (\ref{eq:upbound}) happens to overestimate the kinematic endpoint,
it is nevertheless possible to obtain the correct answer by inspecting the candidate momentum configurations
for the visible particles in the rest frame of the mother particle $D$. 

\begin{figure}[t]
\centering
\includegraphics[height=5.5cm]{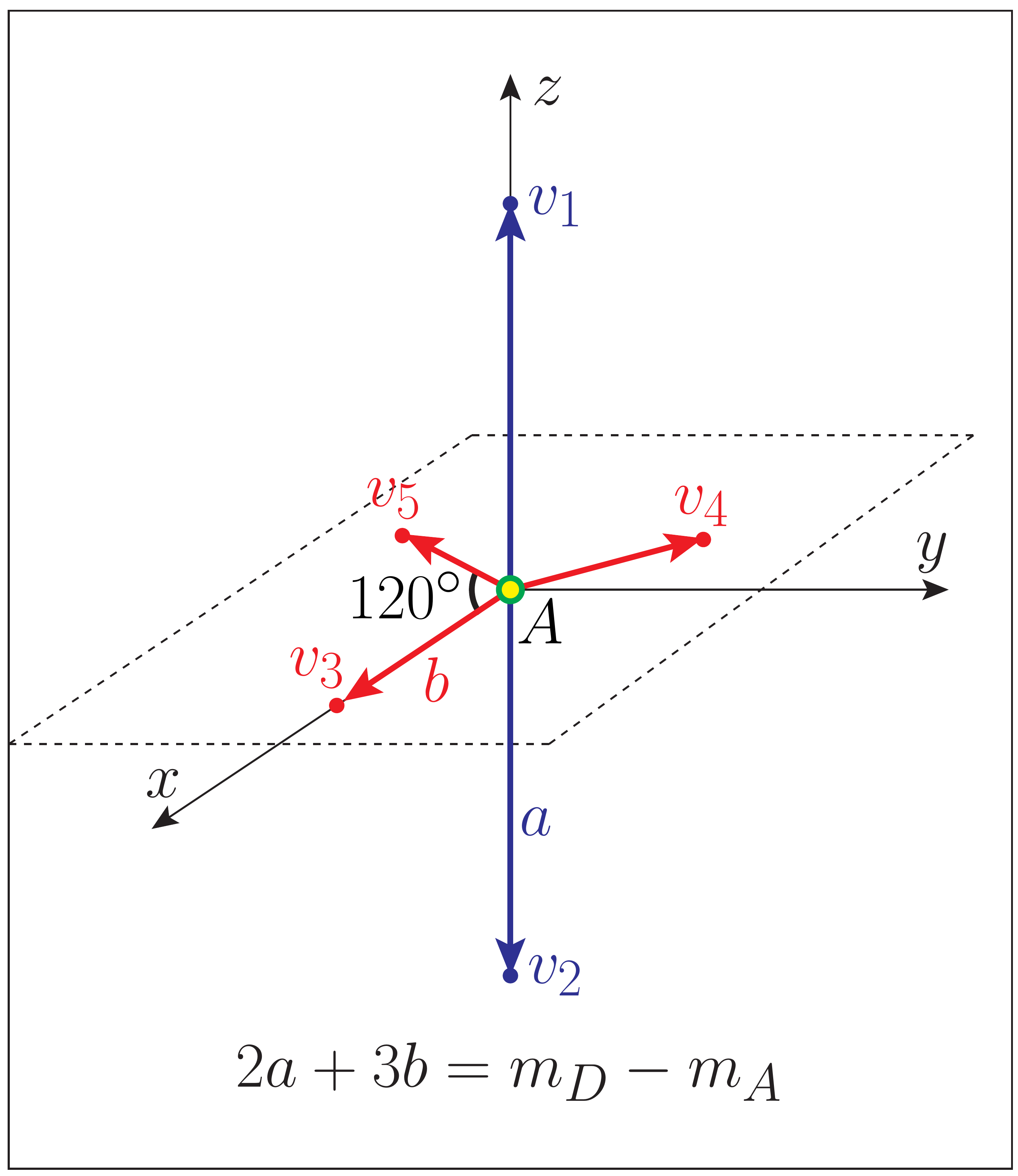}~~~~~
\includegraphics[height=5.5cm]{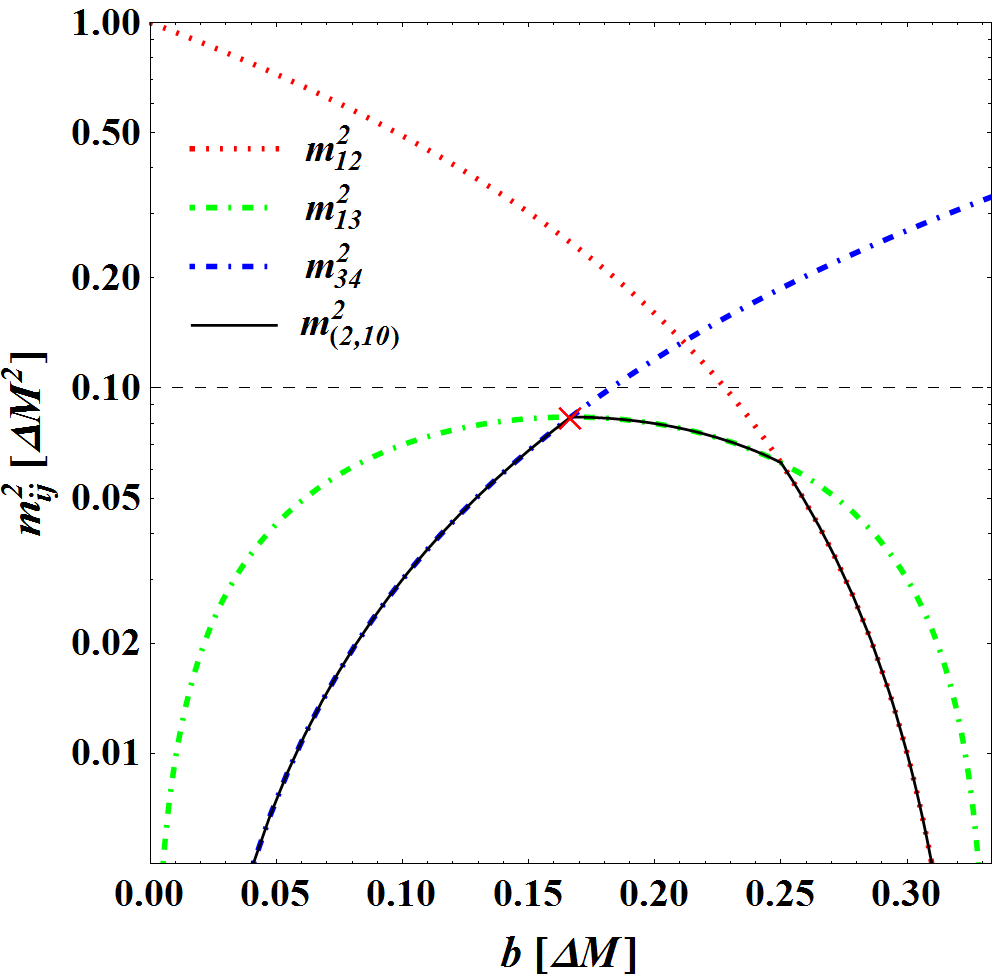}
\caption{\label{fig:conf1} Left: a candidate momentum configuration of the $N=5$ visible particles in the $D$ rest frame
for an event contributing to the $m_{(2,10)}$ endpoint. Each of the blue (red) momentum vectors has magnitude $a$ ($b$), 
and the vectors themselves are arranged in the shape of a triangular bipyramid. Energy conservation implies the relation
$2a+3b=m_D-m_A$. Right: The three different two-body invariant masses squared as a function of $b$ 
(measured in units of $\Delta M=m_D-m_A$): $m_{12}^2$ (red dotted line);
$m_{13}^2=m_{14}^2=m_{15}^2=m_{23}^2=m_{24}^2=m_{25}^2$ (green dot-dashed line);
$m_{34}^2=m_{45}^2=m_{35}^2$ (blue dot-dashed line).
The black solid line corresponds to $m_{(2,10)}^2$, which peaks at $b=\Delta M/6$, leading to the exact bound
(\ref{m5_10_exact}).}
\end{figure}
\begin{figure}[t]
\centering
\includegraphics[height=5.5cm]{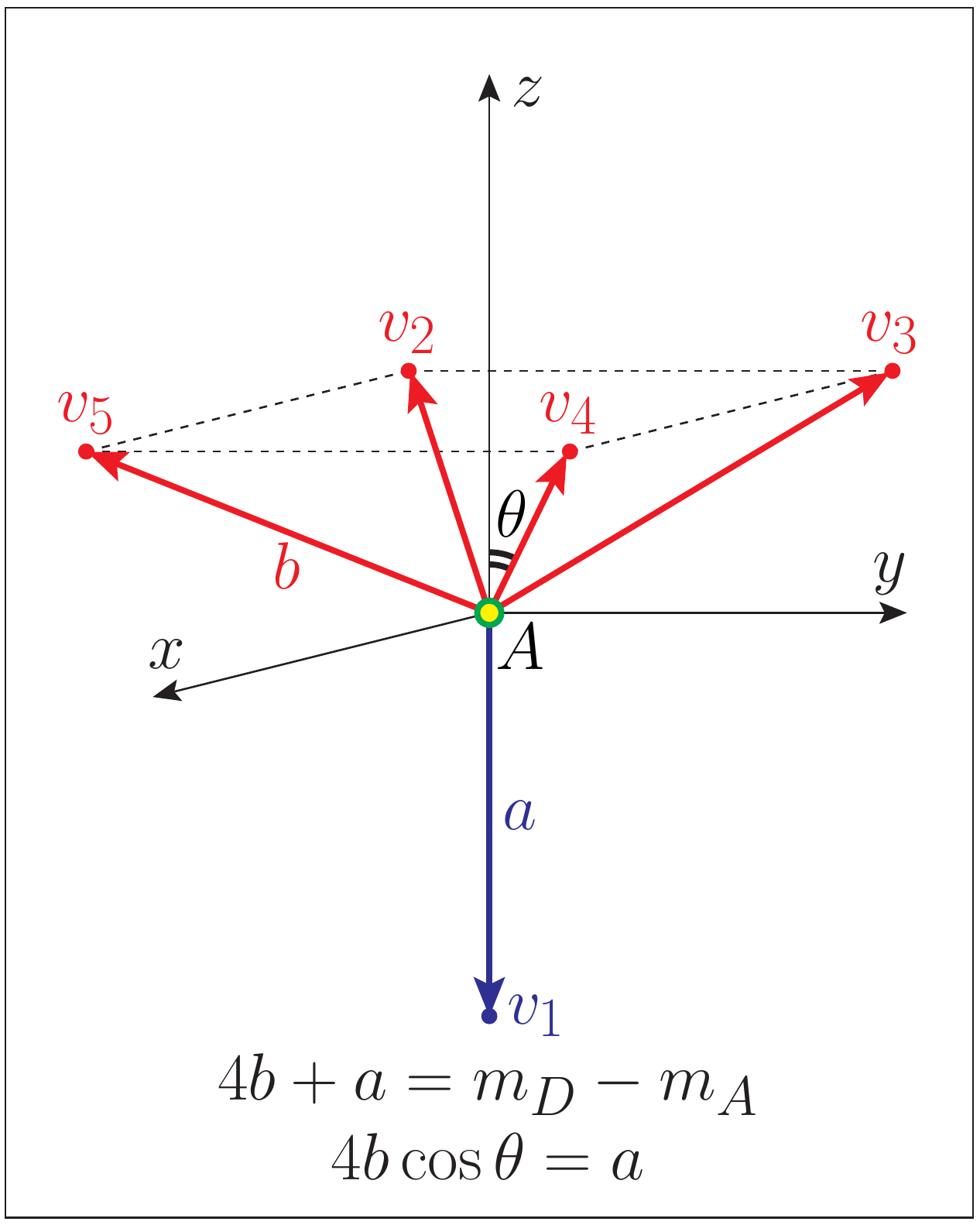}~~~~~
\includegraphics[height=5.5cm]{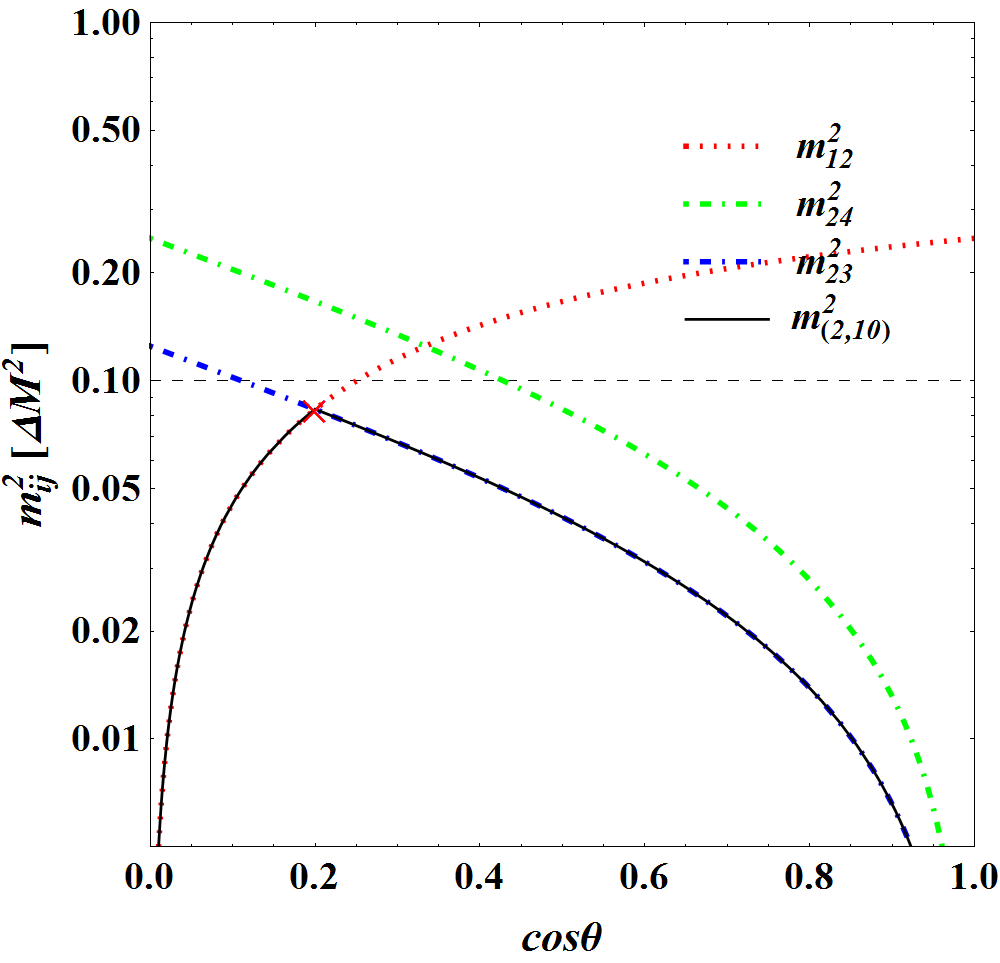}
\caption{\label{fig:conf2} The same as Fig.~\ref{fig:conf1}, but for an alternative, pyramid-like, momentum configuration.
Energy conservation implies $4b+a=m_D-m_A$, while momentum conservation along $z$ requires
$4b\cos\theta = a$, leaving only one independent degree of freedom, which in the right panel is chosen as $\cos\theta$. 
In this configuration, we have 
$m_{12}^2=m_{13}^2=m_{14}^2=m_{15}^2$ (red dotted line);
$m_{24}^2=m_{35}^2$ (green dot-dashed line); and
$m_{23}^2=m_{34}^2=m_{45}^2=m_{25}^2$ (blue dot-dashed line).
The maximal value of $m_{(2,10)}^2$ is obtained for $\cos\theta=1/12$, once again leading to the bound (\ref{m5_10_exact}).}
\end{figure}

Let us illustrate the procedure with the above example of $N=5$ visible particles, and try to obtain the exact
upper bound for $m_{(2,10)}$. For this purpose, we look for momentum configurations, in which the visible particles
$v_i$ are maximally ``spread out" in the rest frame of the mother particle $D$, while the massive 
invisible particle $A$ is still at rest. The main idea is that the desired configurations will exhibit a certain level of symmetry,
as demonstrated in Figs.~\ref{fig:conf1} and \ref{fig:conf2}. The left panels in the figures depict the momentum configurations 
of the visible particles in the $D$ rest frame, whereby momenta in blue (red) have the same magnitude $a$ ($b$).
Energy and momentum conservation imply certain relations among the magnitudes and the relative orientation 
of the momentum vectors as shown. In each case, the remaining sole degree of freedom can be varied in order to
find the maximum of the ranked variable $m_{(2,10)}$ as
\beq
N=5:\quad m_{(2,10)}^{\max}=\frac{1}{\sqrt{12}}\, (m_D-m_A).
\label{m5_10_exact}
\eeq
As expected, this bound is tighter than the na\"{i}ve expectation (\ref{m510conjecture}).

One can similarly analyze the case of $N=6$, where we find three competing momentum configurations:
a pentagonal pyramid, a square bipyramid, and a triangular prism wedge. The latter provides the 
true maximum value of the $m_{(2,15)}$ ranked variable:
\beq
N=6:\quad m_{(2,15)}^{\max}=\sqrt{\frac{253}{4536}}\, (m_D-m_A) <\frac{1}{\sqrt{15}}\, (m_D-m_A),
\label{m6_15_exact}
\eeq
while the bipyramid-like configuration provides the maximum of $m_{(2,14)}$:
\beq
N=6:\quad m_{(2,14)}^{\max}=\frac{1}{\sqrt{9\sqrt{3}}}\, (m_D-m_A) <\frac{1}{\sqrt{14}}\, (m_D-m_A).
\label{m6_14_exact}
\eeq

In summary, for large enough ranks $r$ (as in the examples just considered), 
the upper bound provided by Eq.~(\ref{eq:upbound}) will not be saturated, and the
kinematic endpoint $m_{(2,r)}^{max}$ will be found at somewhat lower values.
The lowest value of $r$ at which the prediction (\ref{eq:upbound}) begins to deviate 
from the true answer, in general depends on the number of visible particles $N$.
We have checked that Eq.~(\ref{eq:upbound}) can be trusted up to the following rank 
\beq
r'=
\left\{
\begin{array}{l}
6k^2 \hspace{2.5cm} \hbox{for }N=4k, \\
6k^2+3k \hspace{1.6cm} \hbox{for }N=4k+1, \\
6k^2+6k+1 \hspace{1cm} \hbox{for }N=4k+2, \\
6k^2+9k+3 \hspace{1cm} \hbox{for }N=4k+3,
\end{array}\right.
\label{rprime}
\eeq      
where $k$ is a non-zero integer. For ranks $r$ higher than the rank $r'$ given by Eq.~(\ref{rprime}),
the expression (\ref{eq:upbound}) provides simply an upper bound on the kinematic endpoint $m_{(2,r)}^{\max}$.

\subsection{Invariant masses of order $n>2$}
\label{sec:offn3}

Once we consider more than two particles at a time, the situation becomes more complicated.
As a concrete example, let us take $N=4$ visible particles and investigate the third order ($n=3$) variables
$m_{123}$, $m_{124}$, $m_{134}$, and $m_{234}$. Momentum conservation (\ref{momcons})
now leads to the following relation (compare to Eqs.~(\ref{summ2sq}) and (\ref{eq:offeq}))
\beq
\sum_{r=1}^{C^4_3=4} m_{(3,r)}^2=2\times 2\sum_{\substack{i,j=1\\i<j}}^{N=4} p_i\cdot p_j=2\, (p_D-p_A)^2.
\label{sumijm3r}
\eeq
As before, the kinematic endpoints are attained when particle $A$ is produced at rest in the rest frame 
of particle $D$, so that the above equation reduces to the following analogue of Eq.~(\ref{sr2r})
\bea
\sum_{r=1}^{4} m_{(3,r)}^2=2\, (m_D-m_A)^2=2\, m_D^2\, (1-\sqrt{R_{AD}})^2.
\label{sr3r}
\eea

{\bf Rank $r=1$.} There are two types of events which determine the endpoint of $m_{(3,1)}$.
The first type of events have two visible particles moving back-to-back in the $D$ rest frame, 
while the other two visible particles are at rest. In this configuration, we have
\beq
m_{(3,1)} = m_{(3,2)} = m_{(3,1)}^{max} \left(  = m_{(3,2)}^{max}\right) \quad {\rm and} \quad
m_{(3,3)} = m_{(3,4)} = 0.
\label{conf32}
\eeq
In the other configuration, three visible particles with equal energies are moving in a plane 
at $120^\circ$ with respect to each other, while the fourth visible particle is at rest.
This in turn implies that 
\beq
m_{(3,1)} = m_{(3,1)}^{max} \quad {\rm and} \quad
m_{(3,2)} = m_{(3,3)} = m_{(3,4)} = \frac{1}{\sqrt{3}}\, m_{(3,1)}.
\label{conf31}
\eeq
Using either Eq.~(\ref{conf32}) or Eq.~(\ref{conf31}) in the sum rule (\ref{sr3r}), we find
\beq
m_{(3,1)}^{\max}=m_D-m_A=m_D\, (1-\sqrt{R_{AD}}).
\label{m31max}
\eeq

{\bf Rank $r=2$.} The maximal value for $m_{(3,2)}$ is obtained for 
the momentum configuration given by (\ref{conf32}), thus the endpoint is the same as (\ref{m31max}):
\beq
m_{(3,2)}^{\max}=m_D-m_A=m_D\, (1-\sqrt{R_{AD}}).
\label{m32max}
\eeq

{\bf Ranks $r=3$ and $r=4$.} For the third- and fourth-ranked three-body invariant masses, 
we apply the reasoning from Section~\ref{sec:offn2} to obtain similar expressions. 
Our final answer for the case of $N=4$ visible particles is thus
\beq
m_{(3,r)}^{\max}=
\left\{
\begin{array}{l}
m_D\, (1-\sqrt{R_{AD}}) \hspace{2.0cm} \hbox{for }r\leq2 \\
\\
\sqrt{\frac{2}{r}}\, m_D\, (1-\sqrt{R_{AD}}) \hspace{1.4cm} \hbox{for }r\geq2. 
\end{array}\right.
\label{m3rmaxN4}
\eeq       

Having worked out this simple example, we can now generalize (\ref{m3rmaxN4}) to higher orders $n>3$. 
Suppose that there are $N$ visible particles as usual, and we consider invariant mass combinations of order $n$. 
Fixing two visible particles, say $i$ and $j$, the term $2p_i\cdot p_j$ 
appears in $C^{N-2}_{n-2}$   
invariant mass variables out of the total possible number $C^N_n$.
We have already seen that summing $2p_i\cdot p_j$ over all possible pairs of indices $i$ and $j$
is related to a sum over all possible $n$-body invariant mass combinations\footnote{See, e.g., 
the special case in Eq.~(\ref{sumijm3r}).}:
\beq
\sum_r m_{(n,r)}^2= C^{N-2}_{n-2}\times (p_D-p_A)^2.
\eeq
The kinematic endpoints that we are interested in are obtained when the right-hand-side 
of this relation is maximized (by virtue of particle $A$ being at rest in the $D$ rest frame):
\beq
\sum_r m_{(n,r)}^2= C^{N-2}_{n-2}\times (m_D-m_A)^2.
\eeq
Retracing the steps which led to Eq.~(\ref{m3rmaxN4}), we get
\beq
m_{(n,r)}^{\max}=
\left\{
\begin{array}{l}
m_D\, (1-\sqrt{R_{AD}}) \hspace{2.0cm} \hbox{for }r\leq C^{N-2}_{n-2}, \\
\\
\sqrt{
\frac{C^{N-2}_{n-2}}{r} }
\, m_D\,(1-\sqrt{R_{AD}}) \hspace{0.8cm} \hbox{for }r\geq C^{N-2}_{n-2}. 
\end{array}\right.
\label{mnrmax}
\eeq       

As already discussed at the end of Section~\ref{sec:offn2}, this formula 
provides the exact maximum only up to some rank, i.e., for sufficiently 
high ranks, it only gives an upper bound. However, even with such high ranks, the
true endpoint will still be proportional to the mass difference
$m_D-m_A=m_D(1-\sqrt{R_{AD}})$, only with a pre-factor which 
is somewhat smaller than $\sqrt{\frac{C^{N-2}_{n-2}}{r} }$.

\subsection{Testing for off-shellness}

Armed with the general result (\ref{mnrmax}), one can now design a 
specific test to verify that the decay topology is indeed a purely off-shell 
one as hypothesized in Fig.~\ref{fig:gendecay}. The main observation 
is that in a purely off-shell topology the kinematic endpoints of all 
invariant mass variables are functions of a single degree of freedom, 
$m_D-m_A$. This implies certain relationships, or ``sum rules" for short, 
among the kinematic endpoints. These sum rules are quantitatively predicted 
by Eq.~(\ref{mnrmax}). Note that 
by introducing the sorted variables (\ref{eq:defsort}), we are considering 
the largest possible number of invariant mass variables, and therefore we
obtain the largest possible number of sum rules.\footnote{Recall from 
Eq.~(\ref{eq:count}) that the number of sorted variables is $2^N-N-1$.
Therefore the total number of sum rules in the purely off-shell case is 
$2^N-N-2$.}

For illustration, let us consider the simplest case of $N=3$ visible particles,
as in the SUSY-like decay chain of Fig.~\ref{fig:chain}. There are 
$2^3-3-1=4$ sorted variables given by (\ref{IM1234}), and one unknown 
degree of freedom, $m_D-m_A$, which leaves us with three sum rules. Therefore,
if this were a purely off-shell process, the following relations must hold
\bea
\frac{ m_{(3,1)}^{max} } { m_{(2,1)}^{max} } &=& 1, \label{sr1}\\ [2mm]
\frac{ m_{(2,2)}^{max} } { m_{(2,1)}^{max} } &=& \frac{1}{\sqrt{2}}, \label{sr2} \\ [2mm]
\frac{ m_{(2,3)}^{max} } { m_{(2,1)}^{max} } &=& \frac{1}{\sqrt{3}}. \label{sr3}
\eea
The violation of one or more of these relations would indicate one of two things ---
either the presence of intermediate on-shell resonances, or 
some sort of momentum-dependent couplings (form-factors).

\begin{figure}[t]
\centering
\includegraphics[width=8cm]{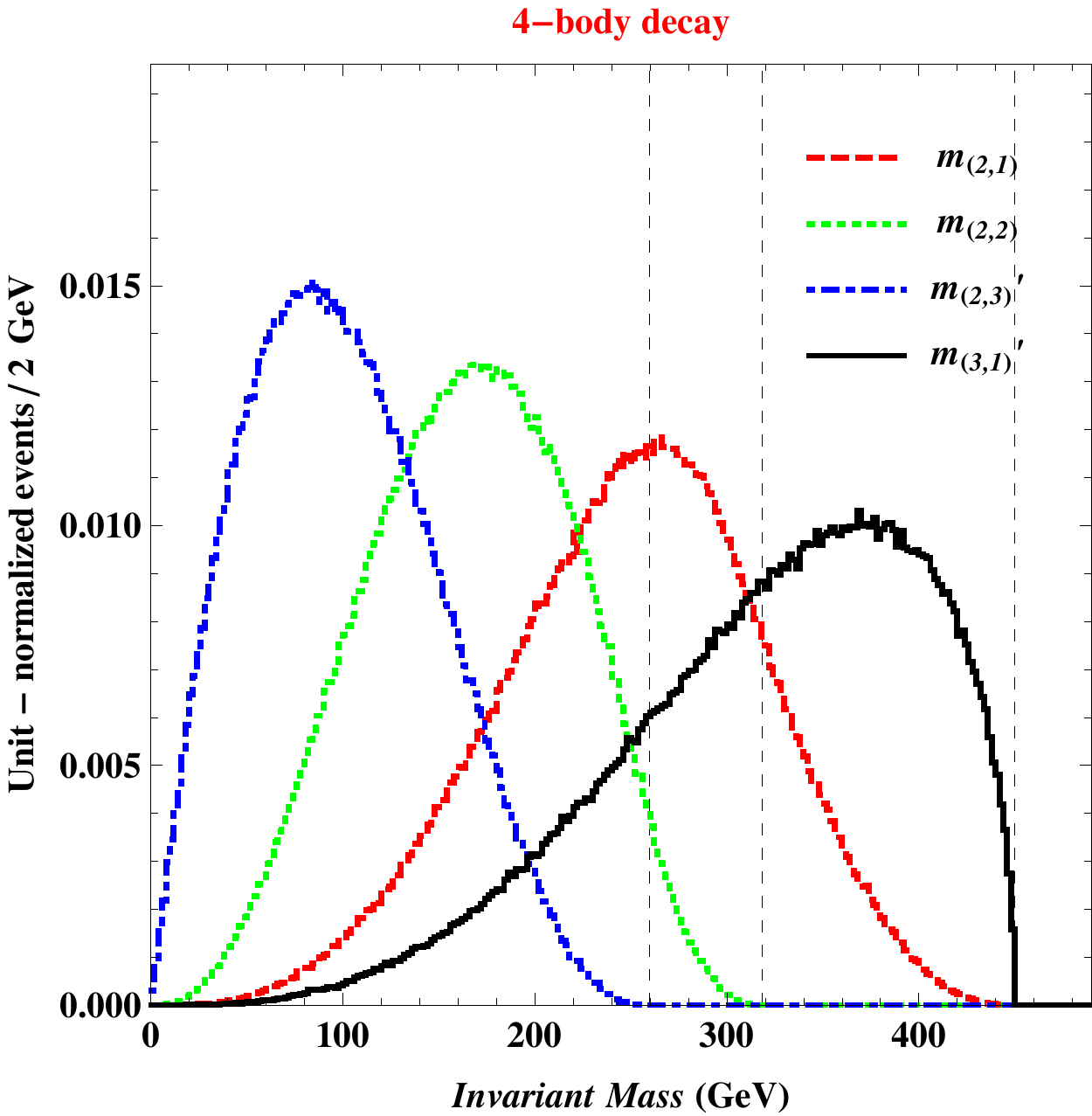}
\caption{\label{fig:4body} The distributions of the sorted variables 
$m_{(2,1)}$ (red dashed line),
$m_{(2,2)}$ (green dotted line),
$m_{(2,3)}$ (blue dot-dashed line) and
$m_{(3,1)}$ (black solid line), for a 4-body decay ($N=3$ in Eq.~(\ref{DtoA}))
with $m_D=500$ GeV and $m_A=50$ GeV.
The kinematic endpoints predicted by (\ref{mnrmax}) are
marked with the vertical black dashed lines and are
located at 450 GeV, 318 GeV, 260 GeV and 450 GeV, correspondingly.
}
\end{figure}

The relationships (\ref{sr1}-\ref{sr3}) are illustrated in Fig.~\ref{fig:4body}.
We consider a four-body pure off-shell decay, i.e., the reaction (\ref{DtoA}) 
with $N=3$, and plot the distributions of the four relevant sorted invariant mass
variables: $m_{(2,1)}$ (red dashed line), $m_{(2,2)}$ (green dotted line),
$m_{(2,3)}$ (blue dot-dashed line) and $m_{(3,1)}$ (black solid line).
Eq.~(\ref{mnrmax}) predicts that their respective endpoints will be located 
at 450 GeV, 318 GeV, 260 GeV and 450 GeV. Fig.~\ref{fig:4body}
shows that the endpoint structure for $m_{(3,1)}$ can be identified very well 
and the value of the endpoint clearly agrees with the theoretical prediction.
The two-body invariant mass distributions for $m_{(2,i)}$
also saturate the theoretical bounds (\ref{mnrmax}). 
However, we note that those distributions 
are relatively shallow near their upper kinematic endpoints \cite{BK,Cho:2012er,Giudice:2011ib},
which might make the experimental extraction of those endpoints rather 
challenging \cite{Lester:2006yw}.\footnote{In principle, one should be able to 
benefit from the knowledge of the asymptotic behavior of the distribution near the endpoint,
which, however, is only known for the cases of $m_{(2,1)}$ and $m_{(3,1)}$ \cite{BK}.}

\section{The pure on-shell decay topology with $N=3$ visible particles}
\label{sec:111}

\begin{figure}[t]
\centering
\includegraphics[width=12cm]{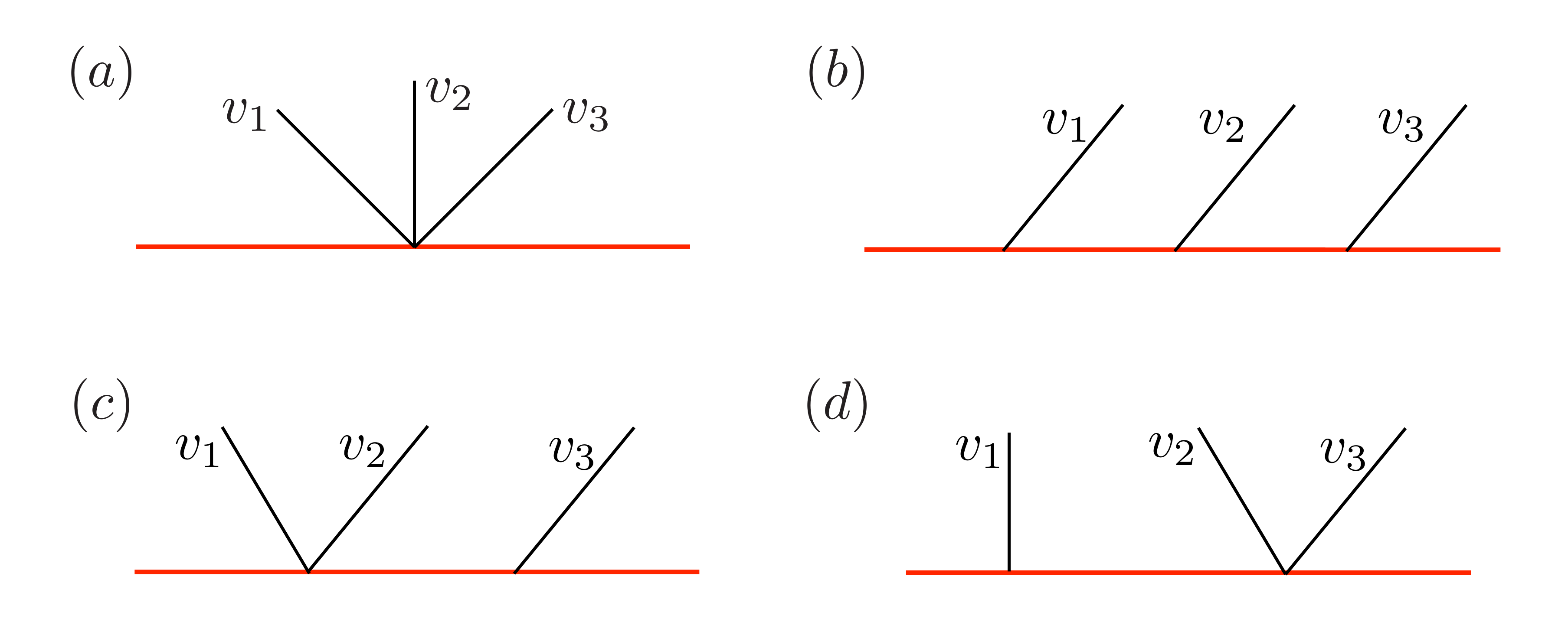}
\caption{\label{fig:4decays} The four possible decay topologies with $N=3$ massless 
visible particles $v_1$, $v_2$ and $v_3$.
}
\end{figure}

Having considered the purely off-shell case in complete generality in the previous section, 
we now turn our attention to decay topologies with intermediate on-shell resonances.
Unfortunately, the general analysis for an arbitrary number of visible particles $N$
gets quite complicated, which is why in this and the subsequent sections we shall
limit our discussion to the most interesting case of $N=3$, as in the 
SUSY-like decay chain from Fig.~\ref{fig:chain}. In particular, we shall focus on the 
four possibilities depicted in Fig.~\ref{fig:4decays}. 
The event topology of Fig.~\ref{fig:4decays}(a) is simply a special case of a 
purely off-shell decay already considered in the previous section.
The event topology of Fig.~\ref{fig:4decays}(b) is the typical SUSY scenario from 
Fig.~\ref{fig:chain} and will be the main subject of this section. It involves a 
sequence of three 2-body decays, where each decay produces one visible particle.
We shall sometimes refer to the diagram of Fig.~\ref{fig:4decays}(b) as a $(1,1,1)$ decay topology.
The event topologies of Figs.~\ref{fig:4decays}(c) and \ref{fig:4decays}(d)  
are ``hybrid" event topologies in the sense that they involve both a two-body and a three-body decay.
Correspondingly, the diagram of Fig.~\ref{fig:4decays}(c) will be referred to as a
$(2,1)$ topology and will be studied in the next Section~\ref{sec:210}, while
the diagram of Fig.~\ref{fig:4decays}(d) will be labelled as a
$(1,2)$ topology and will be considered in Section~\ref{sec:120}.

\subsection{The phase space structure of the $(1,1,1)$ decay topology}

\begin{figure}[t] 
\begin{center}
\includegraphics[width=14cm]{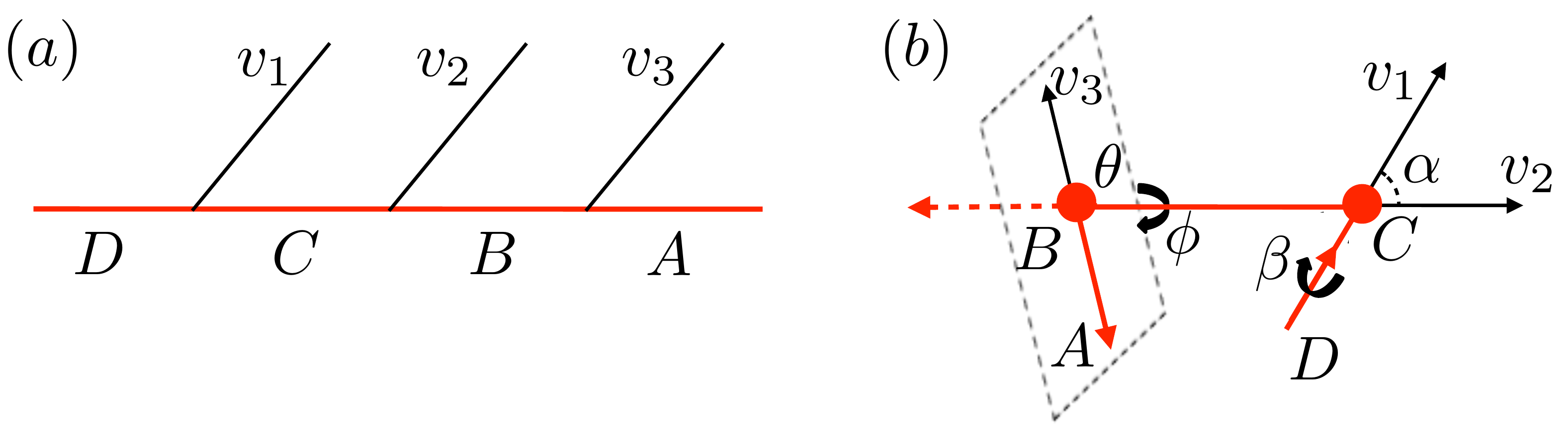} 
\caption{\label{fig:type3} 
(a) The $(1,1,1)$ decay topology of a heavy particle $D$ into a lighter particle $A$ and three 
massless visible particles $v_1$, $v_2$ and $v_3$.  
(b) The relevant kinematic variables:
$\alpha$ is the angle between the momenta of $v_1$ and $v_2$ in the rest frame of particle $C$, 
$\theta$ is the polar angle of the $v_3$ momentum with respect to the direction of $C$ in the $B$ rest frame, 
and $\phi$ is the azimuthal angle between the planes defined by the momenta of a) $v_1$ and $v_2$ 
and b) $v_2$ and $v_3$, in the $C$ rest frame. } 
\end{center}
\end{figure}

Before discussing the properties of the ranked variables (\ref{eq:defsort}),
it is instructive to review the properties of the allowed phase space
in terms of the original variables (\ref{IMbasic}) \cite{Costanzo:2009mq,CLTASI}. 
The relevant kinematic variables for the $(1,1,1)$ decay of a heavy particle $D$ 
into an invisible particle $A$ and visible particles $v_1$, $v_2$ and $v_3$ are 
depicted in Fig.~\ref{fig:type3}. We note that the $(1,1,1)$ decay is most conveniently
described in the rest frame of particle $C$ as shown in Fig.~\ref{fig:type3}(b). 
Na\"{i}vely, the total number of degrees of freedom is four, but one of them 
(here the overall azimuthal angle $\beta$) 
can be neglected taking into account the azimuthal symmetry of this phase space.
The remaining three degrees of freedom are:  the polar angle $\alpha$ of the momentum of $v_2$ 
with respect to the direction of $v_1$, the polar angle $\theta$ of the momentum of $v_3$ with respect 
to the direction of $C$ in the rest frame of particle $B$, and the azimuthal angle $\phi$ 
between the planes defined by $(v_1,v_2)$ and $(v_2,v_3)$ in the $C$ rest frame.

This phase space can be equivalently described in terms of the invariant mass variables
\beq
\left\{ m_{v_1v_2}, m_{v_1v_3}, m_{v_2v_3} \right\}.
\label{IMv1v2v3}
\eeq
However, to simplify notation, from here on we shall work with the dimensionless variables
\bea
x &\equiv& \frac{m^2_{v_2v_3}}{m_D^2} = \frac{(p_2+p_3)^2}{m_D^2},  \label{xdef} \\ [2mm]
y &\equiv& \frac{m^2_{v_1v_2}}{m_D^2} = \frac{(p_1+p_2)^2}{m_D^2}, \label{ydef}\\ [2mm]
z &\equiv& \frac{m^2_{v_1v_3}}{m_D^2} = \frac{(p_1+p_3)^2}{m_D^2}, \label{zdef}
\eea
instead of the dimensionful set (\ref{IMv1v2v3}). In terms of the SUSY-like decay chain of Fig.~\ref{fig:chain},
the variable $x$ corresponds to $m^2_{\ell\ell}$, the variable $y$ is the analogue of
$m^2_{j\ell_n}$, while $z$ represents $m^2_{j\ell_f}$. 

The three angular degrees of freedom $\theta$, $\alpha$ and $\phi$ from Fig.~\ref{fig:type3}(b)
can be mapped to the three dimensionless invariant mass variables
(\ref{xdef}-\ref{zdef}) as follows: 
\bea
x&=& \frac{x_{0}}{2}\left(1-\cos\theta\right) ,  \label{xmap}\\
y&=&\frac{y_{0}}{2}\left(1-\cos\alpha\right) ,  \label{ymap}\\
z&=&\frac{z_{0} \,e^{-\eta}}{2} \left\{ \cosh\eta-\cos\theta \sinh\eta-\cos\alpha(\cos\theta \cosh\eta-\sinh\eta)
-\cos\phi \sin\theta\sin\alpha\right\}, ~~~ \label{zmap}
\label{eq:xyz}
\eea
where
\bea
x_{0} &\equiv& x_{max}=R_{CD} (1-R_{AB})(1-R_{BC}), \label{x0def}\\
y_{0} &\equiv& y_{max}=(1-R_{BC})(1-R_{CD}) , \label{y0def} \\
z_{0} &\equiv& z_{max}=(1-R_{AB})(1-R_{CD}), \label{z0def}
\eea
and
\beq
\eta  \equiv \ln\left(\frac{m_C}{m_B}\right).
\eeq

\begin{figure}[t] 
\begin{center}
\includegraphics[width=15cm]{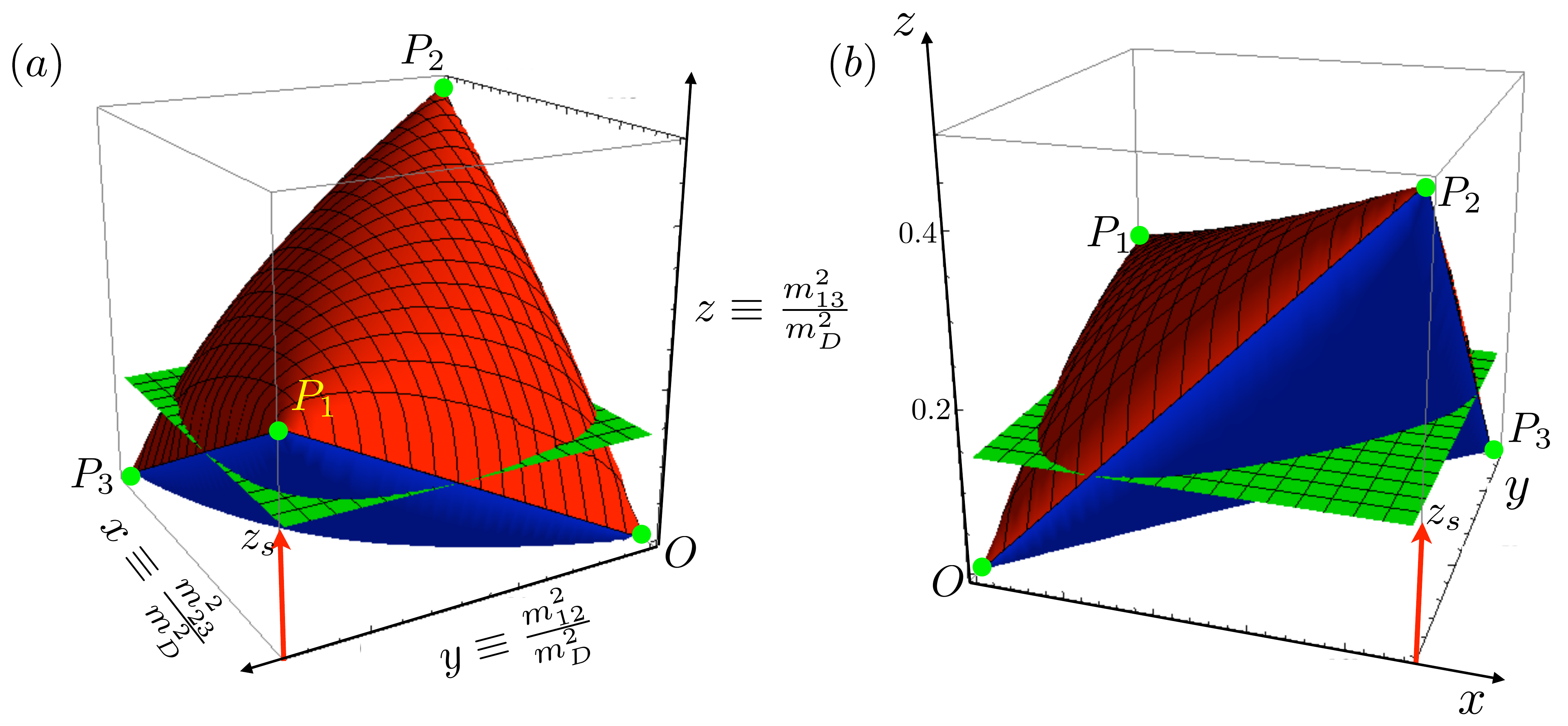}
\end{center}
\caption{\label{fig:fullphase}  Two different views of the allowed phase space for a
$(1,1,1)$ cascade decay chain, in terms of the dimensionless variables (\ref{xdef}-\ref{zdef}). 
The boundary surface is made of two separate sheets --- the top one, colored red, 
is given by Eq.~(\ref{zplus}), while the bottom (blue) sheet is given by Eq.~(\ref{zminus}).  
The three points $\left\{P_1,\, P_2,\, P_3\right\}$ determine the shape of the surface. 
In order to study the ranked invariant mass distributions, we perform a CT-like scan
at a fixed $z$ value $z_s$, as shown by the green plane.  For this illustration,
the mass spectrum has been chosen as $\left(m_D,m_C,m_B,m_A\right)  = \left(1000,700,500,100\right)$ GeV.} 
\end{figure} 

The allowed phase space spanned by (\ref{xmap}-\ref{zmap}) is shown in Fig.~\ref{fig:fullphase}.
Its shape has been likened to that of a samosa \cite{CLTASI} and can be parametrized by two
functions, $z^{+}(x,y)$ for the top surface (colored in red) and $z^{-}(x,y)$ for the bottom surface (colored in blue). 
In order to find the explicit form of $z^{\pm}(x,y)$, we note that the angle $\phi$ enters only 
the definition of $z$ in Eq.~(\ref{zdef}). Thus the extreme values of $z$ (for a fixed $x$ and $y$)
are found for the extreme values of $\phi$, namely, $\phi=0$ and $\phi=\pi$ \cite{Costanzo:2009mq}:
\bea
z^{+}(x,y) &=& z_{0} \left[ \sqrt{\frac{x}{x_{0}} \left(1-\frac{y}{y_{0}}\right)}
+e^{-\eta} \sqrt{\frac{y}{y_{0}} \left(1-\frac{x}{x_{0}}\right)} \right]^2 \textrm{ for } \phi=\pi, \label{zplus}\\
z^{-}(x,y) &=& z_{0} \left[ \sqrt{\frac{x}{x_{0}} \left(1-\frac{y}{y_{0}}\right)}
-e^{-\eta} \sqrt{\frac{y}{y_{0}} \left(1-\frac{x}{x_{0}}\right)}\right]^2 \textrm{ for } \phi=0. \label{zminus}
\eea
The exact shape of the ``samosa" is determined by the location of the four ``corner points" 
which in Fig.~\ref{fig:fullphase} are denoted as $\{O,\,P_1,\,P_2,\,P_3\}$ 
\bea
(\theta=0,\alpha=0) &\rightarrow&    O=\left(0,\,0,\,0\right), \\
(\theta=0,\alpha=\pi) &\rightarrow&P_1 = \left(0,\,y_{0},\,e^{-2\eta}\,z_0 \right) \equiv (0,y_0,z_c), \label{P1coor}\\
(\theta=\pi,\alpha=0) &\rightarrow&P_2 = \left(x_{0},\, 0,\, z_{0} \right),  \label{P2coor}\\
(\theta=\pi,\alpha=\pi) &\rightarrow&P_3 = \left(x_{0},\,y_{0}, \,0\right). \label{P3coor}
\eea

\subsection{Computer tomography of the allowed phase space}

In order to analyze the allowed phase space from Fig.~\ref{fig:fullphase} in terms of the sorted variables
(\ref{eq:defsort}), we need to rank the variables $x$, $y$ and $z$ among themselves. 
We do this by performing a computerized tomography (CT) scan in which the relevant cross 
sectional CT images are obtained at a fixed value $z=z_s$ along the $z$-axis 
(see the green plane in Fig.~\ref{fig:fullphase}). 
In this CT scan process the point $P_1$ plays a special role because it divides the 
obtained images into two groups. Whenever the scan image is taken ``below" $P_1$, 
i.e., at a value of $z_s$ smaller than the $z$ component of $P_1$
\beq
z_c \equiv e^{-2 \eta}\, z_0 = z_0\, R_{BC},
\eeq
the boundary of the image consists of two segments obtained from setting $z^+(x,y)=z_s$,
interspersed with another two segments given by  $z^-(x,y)=z_s$ (see Fig.~\ref{fig:fullphase}).
On the other hand, when the image is taken ``above" $P_1$, i.e., when $z_s>z_c$,
the corresponding image boundary is made up of only one segment from each 
surface (top and bottom). 

The basic procedure of ranking $x$, $y$ and $z$ with the CT scan method is the following. 
For a fixed $z=z_s$, the intersection of the green plane shown in Fig.~\ref{fig:fullphase}
with the interior of the samosa determines the allowed values $x(z_s)$ and $y(z_s)$ at this 
particular value of $z=z_s$. We then sort $x(z_s)$ and $y(z_s)$ and find their respective maxima:
\bea 
r_1\equiv\max\left\{\max\left[x(z_s),\;y(z_s)\right]\right\}, \quad
r_2\equiv \max\left\{\min\left[x(z_s),\;y(z_s) \right]\right\}. \label{eq:defr1r2}
\eea
Once this is done, we need to compare the thus obtained values of $r_1$ and $r_2$ to 
the value of $z$, so we sort again $ \lbrace r_1,r_2,z_s \rbrace$ by magnitude for all 
possible $r_1$ and $r_2$ to obtain the ``local'' maxima of the sorted invariant masses
at a given $z_s$\footnote{We remind the reader that we are using the notation
${\max}_r$ to indicate the $r$-th largest among a given set of elements.
The index $r$ in Eq.~(\ref{m2rzs}) is thus the rank index which in this case 
takes values $r=1,2,3$.}:
\beq
\left. \left( m_{(2,r)}^{\max}\right)^2\right|_{z_s} = m_D^2\cdot \max_{r_1,r_2}{}_{r}\left[r_1,r_2,z_s \right].
\label{m2rzs}
\eeq
Finally, we find the ``global'' maxima of the sorted invariant mass variables 
by maximizing (\ref{m2rzs}) for all allowed $z_s$:
\bea
m_{(2,\text{r})}^{\max}=\max_{z_s} \left\lbrace \left.m_{(2,\text{r})}^{\max} \right|_{z_s} \right\rbrace.
\label{m2rscan}
\eea 
In evaluating $r_2$, it is convenient to fold the $(x,y)$ plane along the $x=y$ line \cite{Burns:2009zi}.
This motivates us to treat separately the cases of $x_0 < y_0$ and $x_0 \ge y_0$.  

\subsubsection{CT scans for the case of $x_0 < y_0$}

\begin{figure}[t] 
\begin{center}
\includegraphics[width=14.5cm]{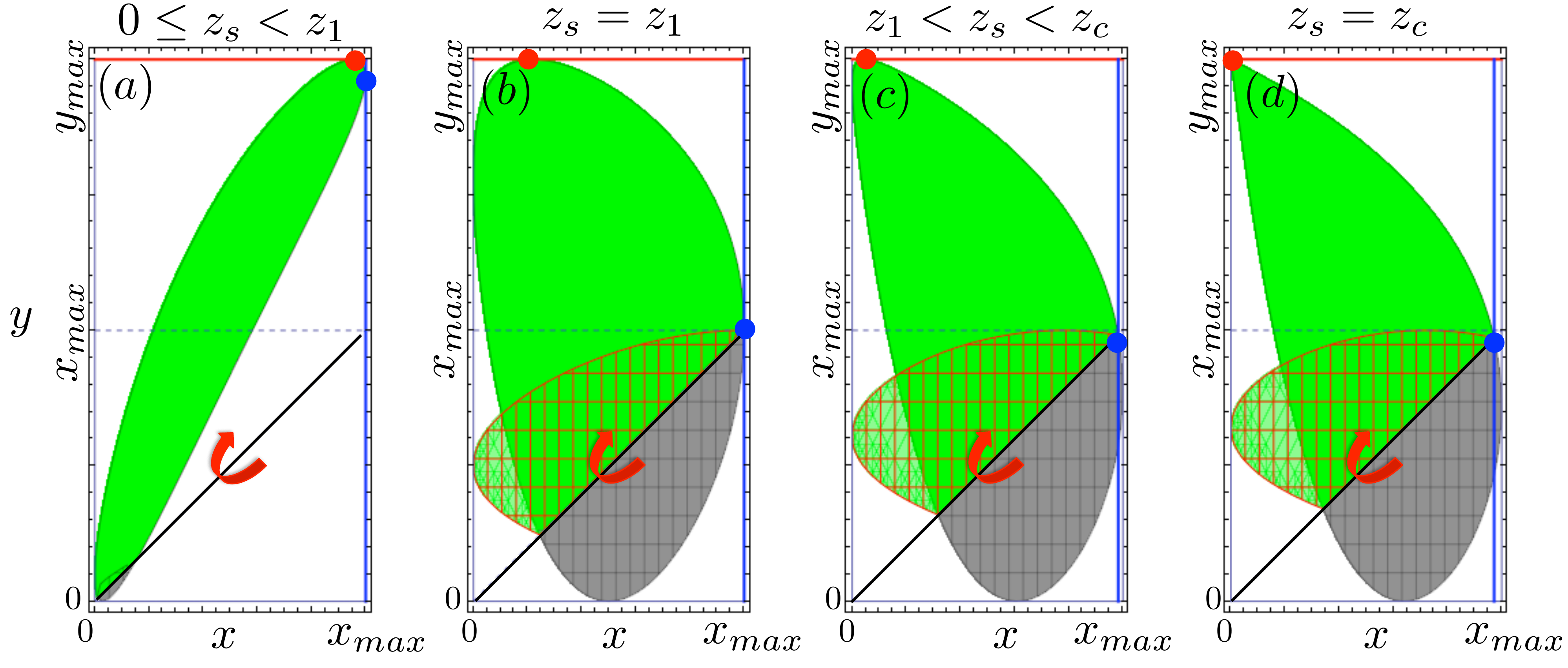}
\includegraphics[width=14.5cm]{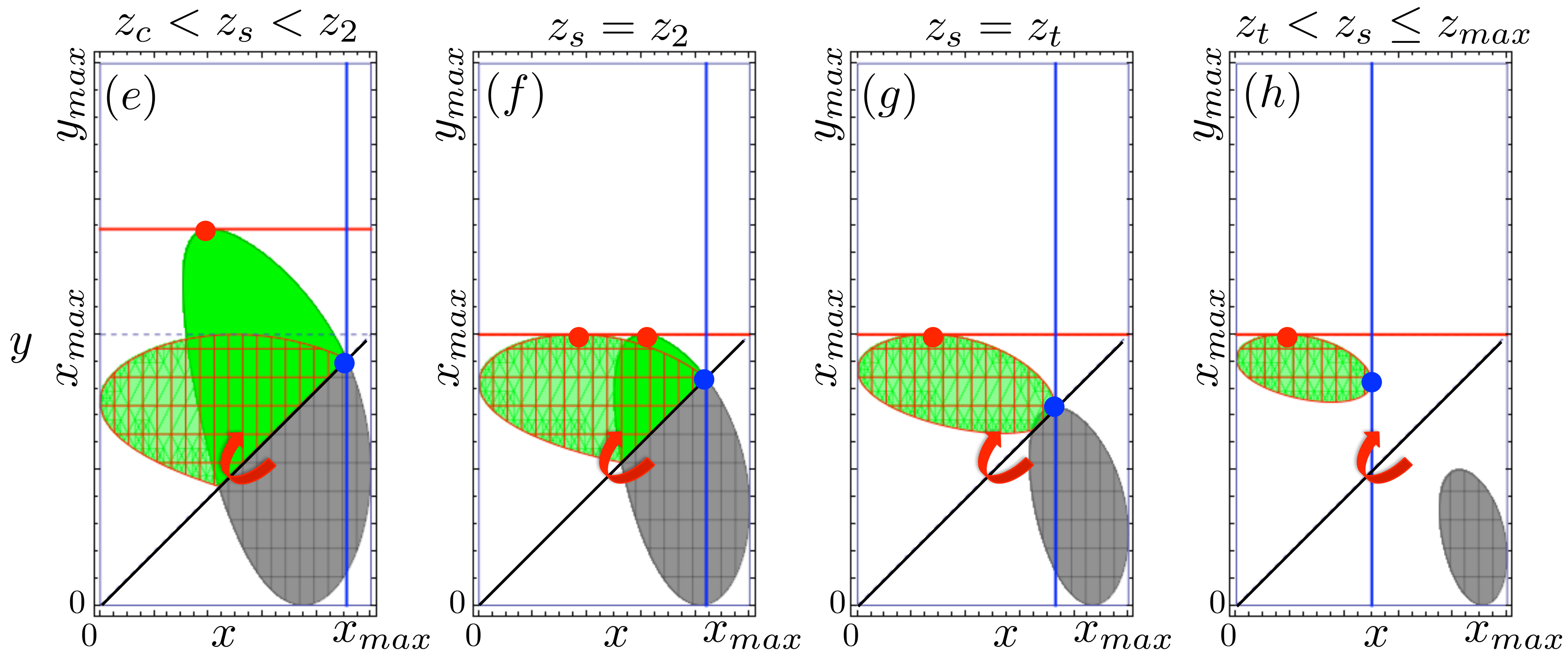}
\end{center}
\caption{\label{fig:caseI}  For the case of $x_0 < y_0$, we show the eight characteristic CT images obtained
at different fixed values of $z_s$. Each CT image typically consists of a green region, in which $x<y$, and a grey region, in which $x>y$.
In order to rank $x$ and $y$, we fold each CT image along the diagonal line $x=y$, thus mapping each grey region onto a
corresponding green hatched region. The extremal values of $r_1$ and $r_2$ are then found by examining the two green regions 
(with and without a grid hatch). In each panel, the red (blue) dot indicates the location of the point giving the value of $r_1$ ($r_2$). } 
\end{figure}

We first discuss the case of $x_0 < y_0$, i.e., when the range of possible $y$ values is larger than the range
of possible $x$ values. According to Eqs.~(\ref{x0def}) and (\ref{y0def}), this occurs for mass spectra obeying the relation
\beq
x_0<y_0 \Longleftrightarrow R_{CD}<\frac{1}{2-R_{AB}}. \label{x0lessy0}
\eeq
Fig.~\ref{fig:caseI} shows the eight characteristic shapes of the CT images obtained at different fixed values of $z_s$.
Each CT image typically consists of a green region, in which $x<y$, and a grey region, in which $x>y$.
In order to rank $x$ and $y$, we fold each CT image along the diagonal line $x=y$, 
mapping the grey region onto a corresponding green hatched region. 
The variables (\ref{eq:defr1r2}) are then found by extremizing over the two green regions 
(with and without a grid hatch). In each panel, the red dot indicates the location of the point 
within the green regions which has the largest $y$ coordinate, giving the value of $r_1$.
Similarly, the blue dot indicates the location of the point within the green regions which has the 
largest $x$ coordinate, thus defining the value of $r_2$.

As demonstrated in Fig.~\ref{fig:caseI}, there exist four special intermediate values of the scanning 
coordinate $z_s$, namely $z_s=\{z_1,z_c,z_2,z_t\}$. At those values of $z_s$, 
one of the $r_i$ variables, either $r_1$ or $r_2$, when considered as a function of $z_s$, 
exhibits some interesting behavior.
\begin{figure}[t] 
\begin{center}
\includegraphics[width=14cm]{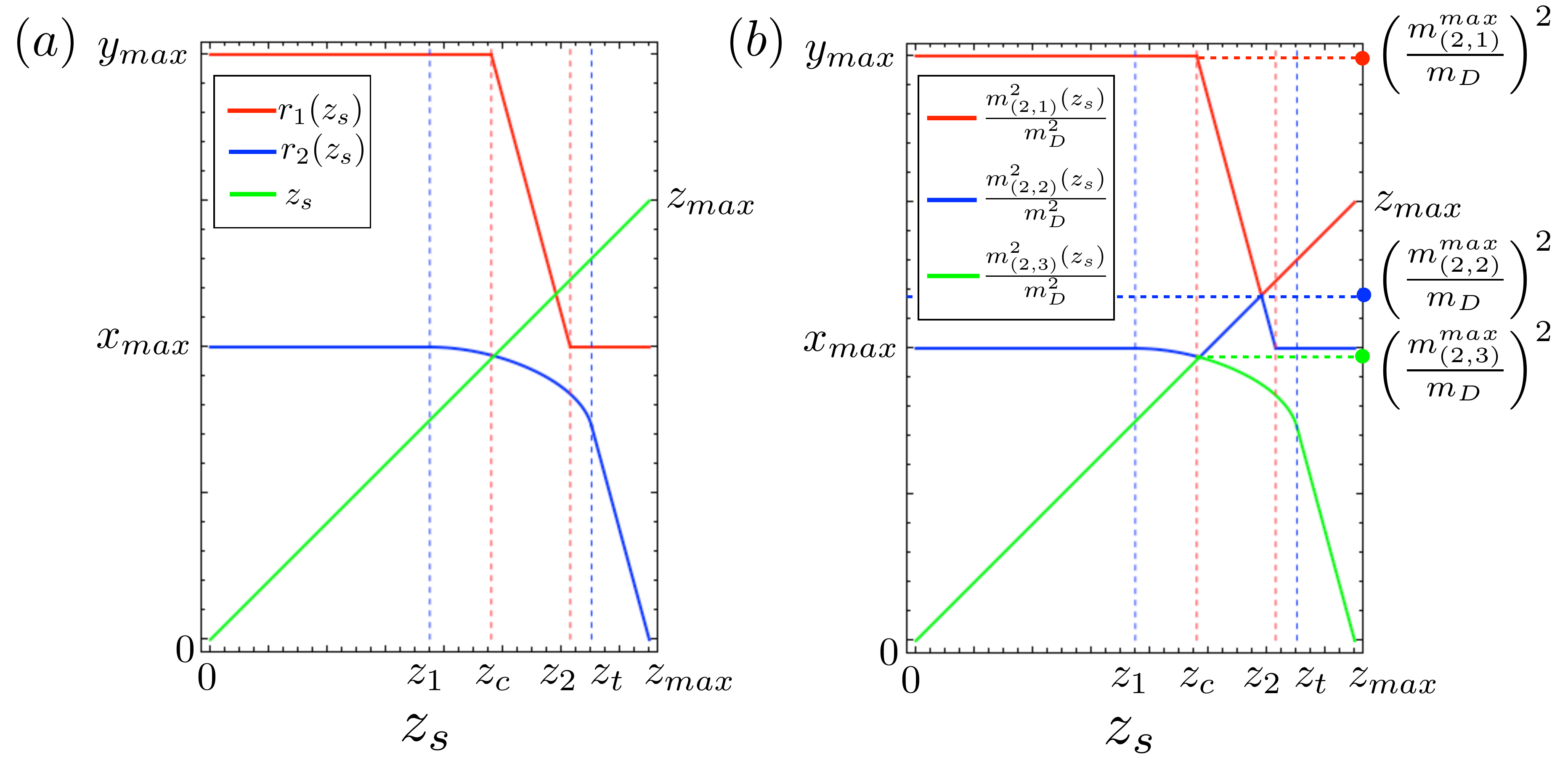} 
\end{center}
\caption{\label{fig:caseISCAN}  
(a) The functional dependence of $r_1(z_s)$ (red line) and $r_2(z_s)$ (blue line), for the case of 
$x_0<y_0$ depicted in Fig.~\ref{fig:caseI}.
(b) The corresponding sorted invariant masses $m_{(2,1)}^2(z_s)$ (red line),
$m_{(2,2)}^2(z_s)$ (blue line) and $m_{(2,3)}^2(z_s)$ (green line).
} 
\end{figure} 
This is depicted more clearly in Fig.~\ref{fig:caseISCAN}(a), where we track the functional dependence 
of $r_1(z_s)$ and $r_2(z_s)$. At first, for very small values of $z_s$ (panels (a) and (b) in Fig.~\ref{fig:caseI}), 
both $r_1(z_s)$ and $r_2(z_s)$ stay constant at $r_1(z_s)=y_0$ and $r_2(z_s)=x_0$\footnote{Recall 
that in this subsection we have assumed that $y_0>x_0$}. As the value of $z_s$ is being increased, the blue point
marking the location of $r_2$ is lowered until it eventually reaches the diagonal line of $x=y$. 
This occurs at a special value of $z_s\equiv z_1$ such that
\beq
z_s = z_1 \equiv z^\pm (x=x_0,y=x_0) = z_0 \left(1-\frac{x_0}{y_0}\right).
\eeq
As we continue to increase $z_s$ beyond $z_1$ and up to $z_c$
(panels (c) and (d) in Fig.~\ref{fig:caseI}), two effects take place.
First, the value of $r_2$ is not given by $x_0$ any more, but is obtained 
from the folding along the $x=y$ line. The functional dependence $r_2(z_s)$
is thus given implicitly by the equation
\beq
z^+(r_2,r_2)=z_s.
\label{r2implicit}
\eeq
Second, the red point in Fig.~\ref{fig:caseI}(a-d) indicating the value of $r_1$ moves to the left, 
until it eventually reaches the point $(x,y)=(0,y_0)$, where the $z$ coordinate is given by
\beq
z_s = z_c \equiv z^\pm(x=0,y=y_0) = z_0\, e^{-2\eta}.
\eeq
Comparing to (\ref{P1coor}), we see that the scan depicted in Fig.~\ref{fig:caseI}(d)
is taken through point $P_1$ in Fig.~\ref{fig:fullphase}, and from that point on, the value of $r_1$ will begin to decrease with $z_s$.
Indeed, as the value of $z_s$ increases further from $z_c$ to $z_2$ (panels (e) and (f) in Fig.~\ref{fig:caseI}),
the green region shrinks and $r_1(z_s)$ decreases linearly with $z_s$ as
\beq
r_1(z_s) = \frac{y_0}{2}\left(1+\coth\eta\right)\left(1-\frac{z_s}{z_0}\right),
\label{eq:tangenty} 
\eeq
until $r_1(z_s)$ reaches the value $x_0$. This occurs at a value of $z_s=z_2$ 
such that $r_1(z_2)=x_0$. Inverting (\ref{eq:tangenty}) and solving for $z_2$, we obtain
\bea 
z_2 = z_0\left[ 1-\frac{x_0}{y_0}\left(1-e^{-2\, \eta}\right)\right] .
\label{eq:z2}
\eea
Finally, for $z_s > z_2$, $r_1(z_s)$ again becomes constant at $r_1=x_0$. Thus the complete 
functional behavior of $r_1(z_s)$ is given by
\bea
r_1(z_s)=\left\{
\begin{array}{l l}
y_0 &\hbox{for }0\leq z_s \leq z_c; \\
\frac{y_0}{2}\left(1+\coth\eta\right) \left(1-\frac{z_s}{z_0}\right)&\hbox{for } z_c < z_s \leq z_2; \\
x_0 &\hbox{for } z_2 < z_s \leq z_0,
\end{array} \right.  
\label{r1zsfunction}
\eea
as illustrated by the red solid line in Fig.~\ref{fig:caseISCAN}(a).

In order to complete the discussion of Figs.~\ref{fig:caseI} and \ref{fig:caseISCAN}, we again turn our
focus to $r_2(z_s)$, which was given implicitly by (\ref{r2implicit}). However, this is true only as long as the 
CT images are crossed by the diagonal line $x=y$. Eventually, as $z_s$ approaches its maximal value $z_0$,
the CT image is confined to the region\footnote{Recall that the ``tip" 
of the samosa is located at point $P_2$ in Fig.~\ref{fig:fullphase}, whose coordinates are
given by Eq.~(\ref{P2coor}).} with $x\sim x_0$ and $y\sim 0$ and may not extend all the way 
up to the $x=y$ line, as depicted in Fig.~\ref{fig:caseI}(h). As shown in Fig.~\ref{fig:caseI}(g),
the image becomes ``disconnected" from the diagonal line $x=y$ at the point
\beq
z_s = z_t =  \frac{z_0}{2} \left(1-\frac{x_0}{y_0} +\sqrt{\left(1-\frac{x_0}{y_0}\right)^2+4\, \frac{x_0}{y_0}\, e^{-2\, \eta}}\right). 
\label{eq:zt}
\eeq
From that point on, for $z_s>z_t$, the value of $r_2$ is determined by the rightmost point of the 
hatched green CT image, i.e., the blue point in Fig.~\ref{fig:caseI}(h).
Thus we find that for $z_s>z_t$, the functional dependence of $r_2(z_s)$ is given by
\beq
r_2(z_s) = \frac{y_0}{2}\left(1+\coth\eta\right)\left(1-\frac{z_s}{z_0}\right).
\label{r2slope}
\eeq
Interestingly, this is the same function as (\ref{eq:tangenty}). This can also be seen by inspecting Fig.~\ref{fig:caseISCAN}(a),
where the blue and red slanted straight segments are lined up.

In conclusion, we comment on the relative ordering of the special points $\{z_1,z_c,z_2,z_t\}$.
First, the definition (\ref{eq:zt}) can be rewritten as
\beq
z_t = \frac{1}{2}
\left[ z_2-\frac{x_0}{y_0}\, z_c
+ \sqrt{ \left(  z_2+\frac{x_0}{y_0}\, z_c\right)^2  + 4\, \frac{x_0}{y_0}\, z_c (z_0-z_2)\, }
\right],
\eeq
which makes it evident that $z_t>z_2$. Furthermore, the definition (\ref{eq:z2}) can be rewritten as
\begin{subequations}
\label{z2hierarchy}
\begin{eqnarray}
z_2 &=& z_1 + \frac{x_0}{y_0}\, z_c  \\
       &=& z_c + \left(1-\frac{x_0}{y_0}\right)\left(1- e^{-2\eta}\right)\, z_0.
\end{eqnarray}
\end{subequations}
Since $e^{-2\eta}=R_{BC}\le 1$ by definition and $x_0<y_0$ by the assumption of this subsection,
Eqs.~(\ref{z2hierarchy}) imply that $z_2$ is larger than both $z_1$ and $z_c$. Finally, the relative size of
$z_c$ and $z_1$ is not predetermined, but depends on the mass spectrum. Therefore, the ordering is
\beq
\left\{z_1,z_c\right\} \le z_2 \le z_t.
\eeq

We are now in position to derive the endpoints of the ranked invariant mass variables.
We first add the straight line $z=z_s$ as the green line in Fig.~\ref{fig:caseISCAN}(a),  
and proceed to order $\{r_1(z_s), \, r_2(z_s), \, z_s\}$ for each value of $z_s$ as shown
in Fig.~\ref{fig:caseISCAN}(b), where the red, blue, and green colored curves track the values of 
the largest, the second largest, and the smallest invariant mass combination for each $z_s$. 
Therefore, the endpoint of each ranked invariant mass arises at the maximum of each colored curve. 
Since $r_1$ and $r_2$ have already been ordered among themselves in accordance with 
(\ref{eq:defr1r2}), all that is left to do now is to compare $r_1(z_s)$ and $r_2(z_s)$ to the value 
of $z_s$ itself. This leads to two interesting intersection points seen in Fig.~\ref{fig:caseISCAN}(a): 
first, the intersection point between $r_1(z_s)$ and the straight line $z_s$
\beq
r_1(z_s)=z_s  \Longleftrightarrow z_s = r_{1z} \equiv  \frac{\left(1-R_{CD}\right)\left(1-R_{AB}\right)}{2-R_{AB}},
\label{ctdef}
\eeq 
and the corresponding intersection point between $r_2(z_s)$ and the straight line $z_s$
\bea
r_2(z_s)=z_s  \Longleftrightarrow z_s = r_{2z} &\equiv&  
1-R_{CD}\left\{2+R_{BC}\left(2-R_{AB}\right)\right\}   \nonumber \\ [2mm]
&+&2\sqrt{R_{BC}R_{CD} \left\{R_{CD}\left(3-R_{AB}\right)-1\right\}}. \label{cfdef}
\eea 

Armed with these results, it is now straightforward to determine the kinematic endpoints of the 
sorted invariant mass variables on a case-by-case basis. 
We postpone the presentation of the relevant results until Section~\ref{sec:results},
after we have had the chance to also discuss the case of $x_0>y_0$, which is the subject of the next 
subsection.

\subsubsection{CT scans for the case of $x_0 > y_0$}  

We now repeat the analysis of the previous subsection for the case of $x_0>y_0$.
Note that the vertices $P_2$ and $P_3$ from Fig.~\ref{fig:fullphase} have a common
$x$ coordinate $x=x_0$ (see also Eqs.~(\ref{P2coor}) and (\ref{P3coor})).
Therefore, for any value of $z_s$, we will have the simple relation (contrast this to (\ref{r1zsfunction}))
\beq
r_1(z_s) = x_0.
\eeq
\begin{figure}[t]  
\includegraphics[width=14.5cm]{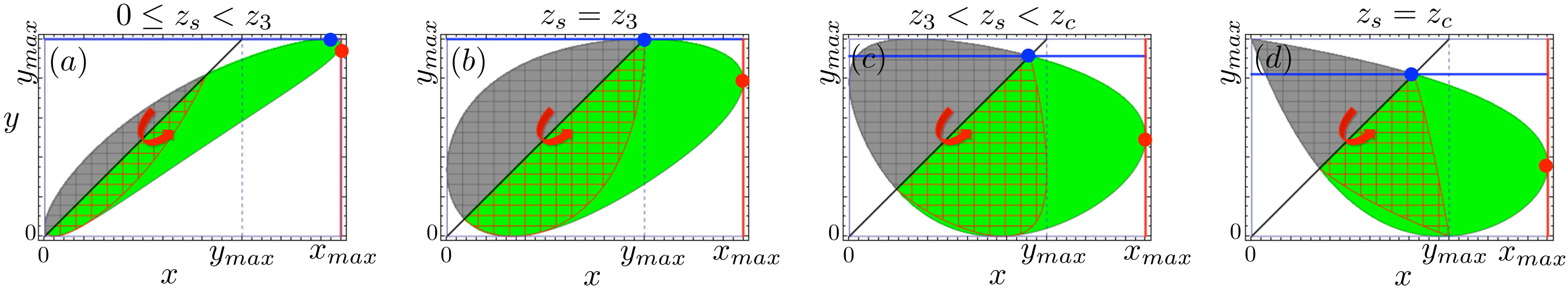}\\
\includegraphics[width=14.5cm]{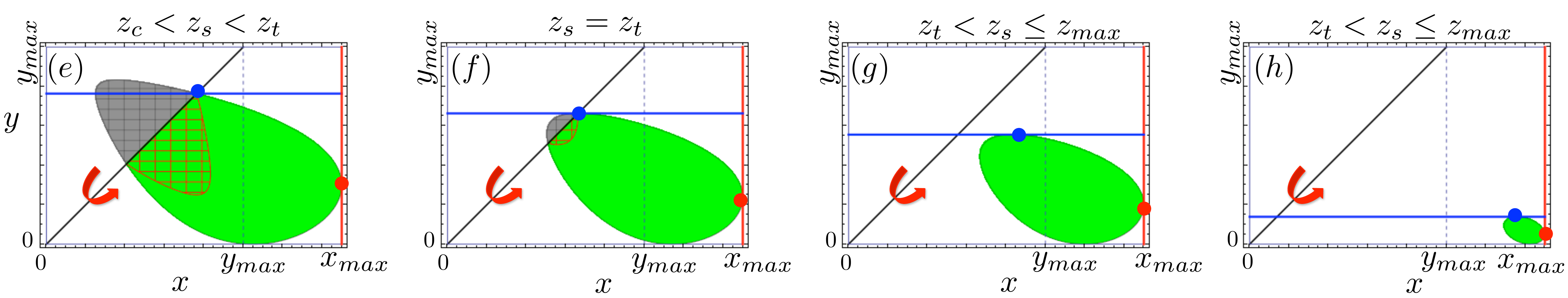} 
\caption{\label{fig:caseII}  The same as Fig.~\ref{fig:caseI}, but for the case of $x_0 > y_0$.
} 
\end{figure} 
This fact can be clearly observed in Figs.~\ref{fig:caseII} and \ref{fig:caseIISCAN},
which are the $x_0>y_0$ analogues of Figs.~\ref{fig:caseI} and \ref{fig:caseISCAN}, respectively.
On the other hand, the behavior of the function $r_2(z_s)$ is similar to the $x_0<y_0$ 
case from the previous subsection, only now the roles of $x$ and $y$ are reversed.
Again, there are two special points, $z_s=z_3$ and $z_s=z_t$, where the functional form of
$r_2(z_s)$ changes. Fig.~\ref{fig:caseII}(a,b) reveals that as we start increasing $z_s$ from 0,
the value of $r_2$ stays constant at $r_2(z_s)=y_0$. Eventually the blue point determining the 
value of $r_2$ reaches the diagonal line $x=y$. This occurs at a special value of $z_s=z_3$ given by
\bea
z_s=z_3=z^{\pm}\left(y_0,y_0\right)= z_0 \left(1-\frac{y_0}{x_0}\right) e^{-2\, \eta}.
\eea
Once $z_s$ exceeds $z_3$, the blue point begins to track the straight line of $x=y$ 
and $r_2(z_s)$ is again found from (\ref{r2implicit}). As illustrated in Fig.~\ref{fig:caseII}(c-f),
this trend continues until the contour of $z^+(x,y)=z_s$ is detached from the $x=y$ line at
the point $z_s=z_t$, where $z_t$ is again given by (\ref{eq:zt}).
Finally, for $z_s>z_t$, $r_2(z_s)$ is the maximal $y$-coordinate of the contour line
$z^+(x,y)=z_s$ which was already computed earlier in Eq.~(\ref{eq:tangenty}). 

\begin{figure}[t] 
\begin{center}
\includegraphics[width=15cm]{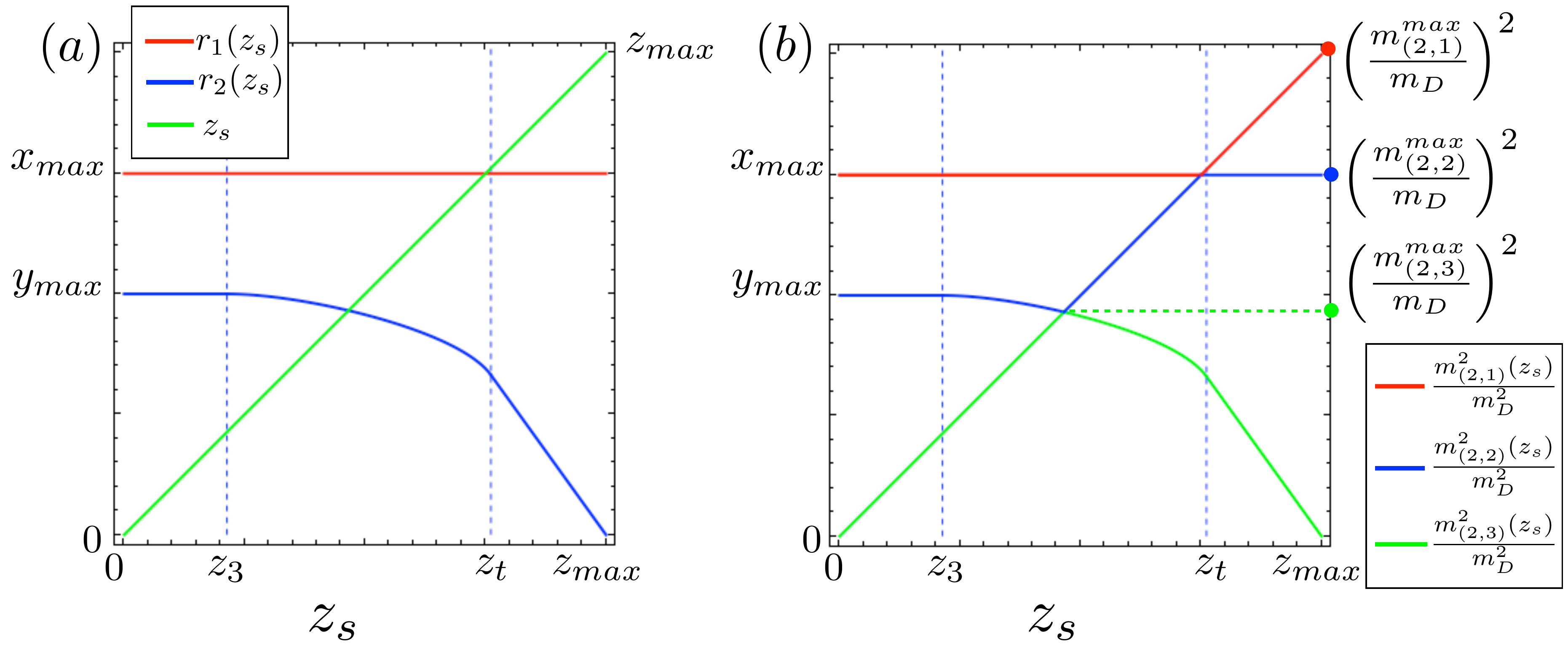} 
\end{center}
\caption{\label{fig:caseIISCAN}  The same as Fig.~\ref{fig:caseISCAN}, but for the case
$x_0>y_0$ shown in Fig.~\ref{fig:caseII}.
} 
\end{figure} 

The rest of the steps are identical to those for the previous case of $x_0<y_0$. 
Having ranked $x$ and $y$ among themselves in terms of $r_1$ and $r_2$, it remains to rank $z$ relative to $r_1$ and $r_2$,
as illustrated in Fig.~\ref{fig:caseIISCAN}(b). The relevant results are summarized in the next subsection.

\subsection{Results summary for the $(1,1,1)$ decay topology}
\label{sec:results}

In this subsection, we collect all relevant results derived from the arguments in the previous two subsections. 
We have seen that the endpoints of the ranked invariant mass variables are given in terms of the endpoints 
$x_0$, $y_0$ and $z_0$ of the original unranked variables (\ref{IMv1v2v3}), as well as the two intersection points
(\ref{ctdef}) and (\ref{cfdef}). The actual mass spectrum $\{m_D,m_C,m_B,m_A\}$ then determines 
which of these five expressions applies to which ranked variable.\footnote{In other words, the endpoint expressions 
are piece-wise defined functions of the mass parameters $\{m_D,m_C,m_B,m_A\}$.} 
Therefore, in presenting the results, we must first describe how the mass parameter space $\{m_D,m_C,m_B,m_A\}$
is partitioned into domains, and then specify the relevant endpoint formulas for each individual domain.

\begin{figure}[t]
\centering
\includegraphics[width=11.5cm]{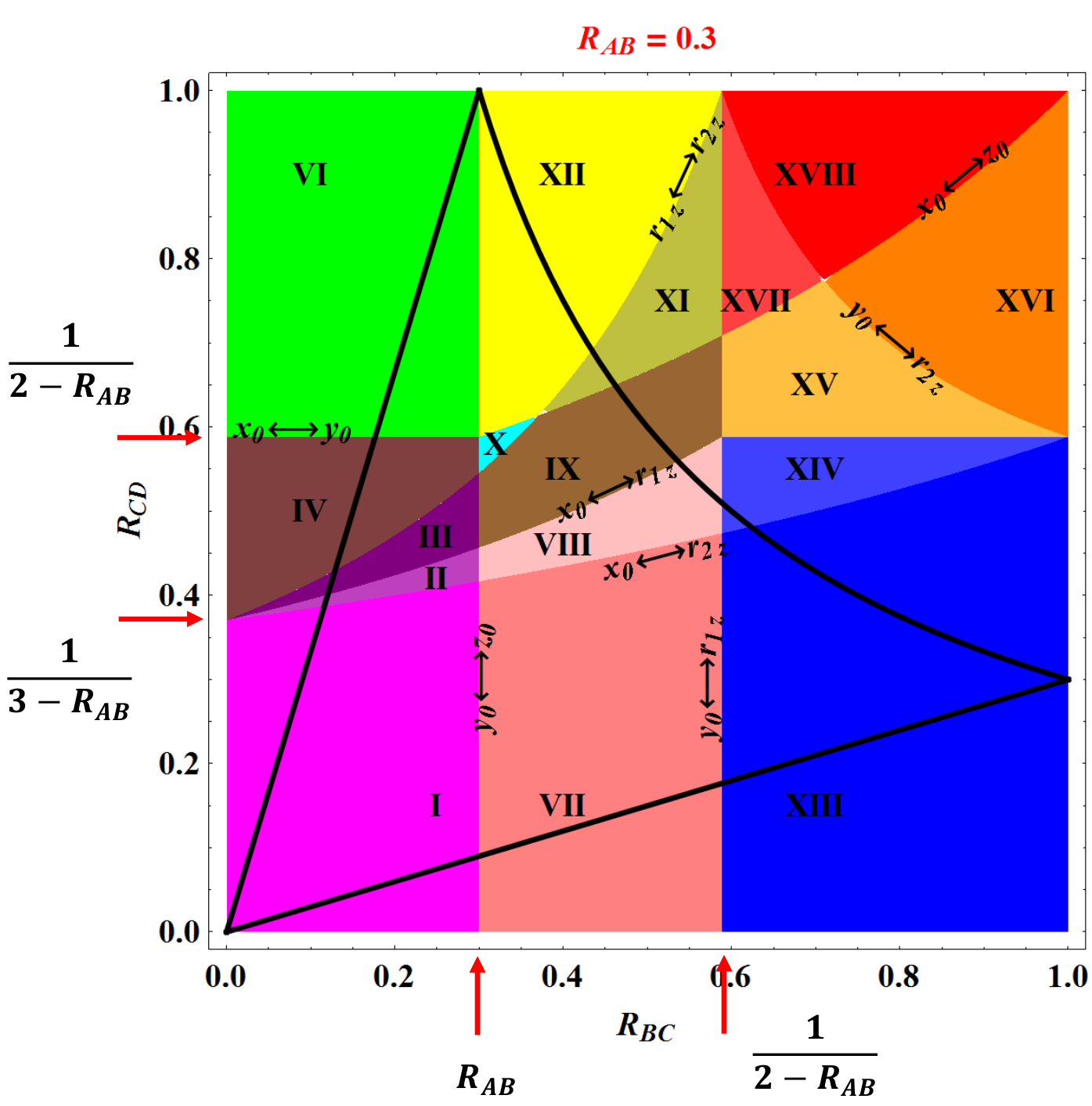}
\caption{\label{fig:regionplotonshell0.3} 
The colors indicate the relevant definition domains for the endpoint formulas (\ref{m2rendpoints})
in the $(R_{BC},R_{CD})$ plane for fixed $R_{AB}=0.3$.
Note the presence of Region $X$ (the cyan triangular-shaped region) at this relatively low value of $R_{AB}$.
The solid black curves denote the domains relevant for the $m_{(3,1)}$ endpoint formulas (\ref{T1}).
}
\end{figure}

In order to define the domains, we can factor out the overall scale, say $m_D$, and then trade the remaining
three mass variables $\{m_C,m_B,m_A\}$ for the dimensionless ratios (\ref{Rijdef}). Even though the resulting
parameter space $\{R_{AB},R_{BC},R_{CD}\}$ is three-dimensional, it turns out that all relevant domains can 
be exhibited on a suitably chosen 2-dimensional slice at constant $R_{AB}$, as illustrated in Figs.~\ref{fig:regionplotonshell0.3}
and \ref{fig:regionplotonshell0.7}.

The color coding in Figs.~\ref{fig:regionplotonshell0.3} and \ref{fig:regionplotonshell0.7} shows that there are 18 different regions
which are needed in order to define the endpoints of the sorted two-body invariant mass variables $m_{(2,r)}$. 
Sixteen of those regions are always visible, for any value of the fixed parameter $R_{AB}$.
On the other hand, Region $X$ only appears at low values of $R_{AB}$, 
$R_{AB} < (3- \sqrt{5})/2$ (see Fig.~\ref{fig:regionplotonshell0.3}),
while Region $V$ only appears at high values of $R_{AB}$, 
$R_{AB} > (3- \sqrt{5})/2$ (see Fig.~\ref{fig:regionplotonshell0.7}). In what follows,
each region will be defined by specifying a range of $R_{BC}$ values and a corresponding range for $R_{CD}$ values.
Given the geometry of Figs.~\ref{fig:regionplotonshell0.3} and \ref{fig:regionplotonshell0.7}, it is convenient to define the 
region boundaries at a fixed $R_{AB}$
by treating $R_{BC}$ as the independent variable, and $R_{CD}$ as the dependent variable, as follows:
\begin{figure}[t]
\centering
\includegraphics[width=11.5cm]{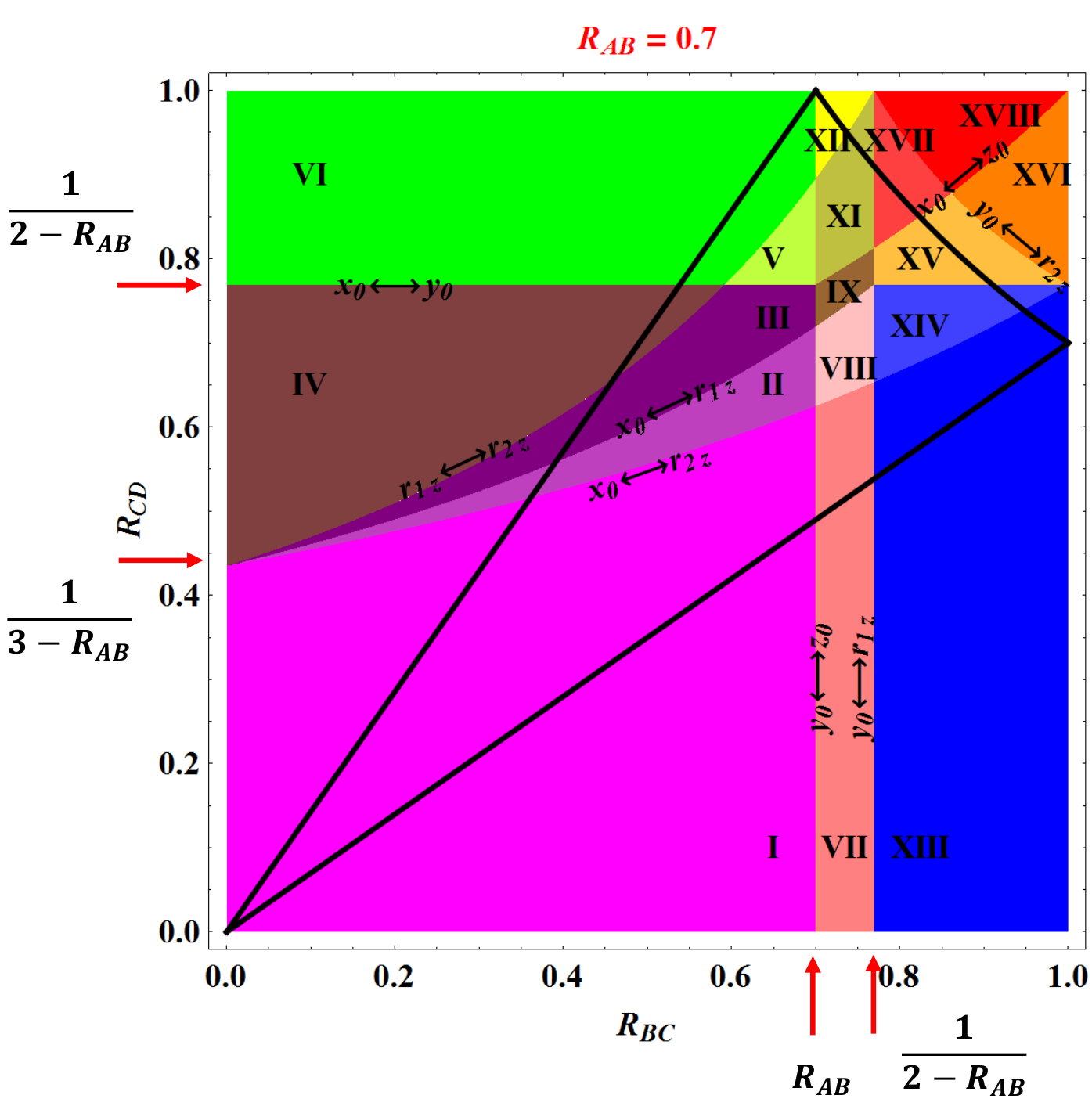}
\caption{\label{fig:regionplotonshell0.7} 
The same as  Fig.~\ref{fig:regionplotonshell0.3} but for $R_{AB}=0.7$.
Note the disappearance of Region $X$ and the emergence of Region $V$ (the yellowgreen triangular-shaped region).
}
\end{figure}
\bea
x_0\leftrightarrow r_{2z}:  \quad
R_{CD} &=& f_{x_0\leftrightarrow r_{2z}} (R_{BC}) \equiv 
\frac{1}{3-R_{AB}-R_{BC}},
\label{x0r2z}
\\ 
x_0\leftrightarrow r_{1z}:  \quad
R_{CD} &=& f_{x_0\leftrightarrow r_{1z}} (R_{BC}) \equiv 
\frac{1}{3-R_{AB} -(2-R_{AB})R_{BC}},
\label{x0r1z}
\\ 
r_{1z}\leftrightarrow r_{2z}:  \quad
R_{CD} &=& f_{r_{1z}\leftrightarrow r_{2z}} (R_{BC}) \equiv 
\frac{1}{3-R_{AB} -(2-R_{AB})^2R_{BC}},
\label{r1zr2z}
\\
x_0\leftrightarrow y_0:  \quad
R_{CD} &=& f_{x_0\leftrightarrow y_0} (R_{BC}) \equiv \frac{1}{2-R_{AB}},
\label{x0y0}
\\
x_0\leftrightarrow z_0:  \quad
R_{CD} &=& f_{x_0\leftrightarrow z_0} (R_{BC}) \equiv \frac{1}{2-R_{BC}},
\label{x0z0}
\\ 
y_0\leftrightarrow r_{2z}:  \quad
R_{CD} &=& f_{y_0\leftrightarrow r_{2z}} (R_{BC}) \equiv 
\frac{1}{3-R_{AB}-R^{-1}_{BC}}.
\label{y0r2z}
\eea
The functions $f(R_{BC})$ defining those boundaries are labelled by the 
replacement which needs to be done in the endpoint formulas (\ref{m2rendpoints}) below
as one crosses the corresponding boundary and moves from one region to the next.

In addition, Figs.~\ref{fig:regionplotonshell0.3} and \ref{fig:regionplotonshell0.7} also display 
two vertical boundaries
\bea
y_0\leftrightarrow z_0:  \quad
R_{BC} &=& f_{y_0\leftrightarrow z_0} (R_{AB}) \equiv R_{AB},
\label{y0z0}
\\ 
y_0\leftrightarrow r_{1z}:  \quad
R_{BC} &=& f_{y_0\leftrightarrow r_{1z}} (R_{AB}) \equiv \frac{1}{2-R_{AB}}.
\label{y0r1z}
\eea

\begin{table}[t]
\centering
\begin{tabular}{| c | c || c | c || c | c |}
\hline
\multicolumn{2}{|c||}{Region}      & \multicolumn{2}{c||}{ $R_{CD}$ range}    & \multicolumn{2}{c|}{ $R_{BC}$ range} \\ 
\hline
Label      & Color      &  min  &  max  & min  & max\\  \hline\hline
$I$         &  magenta  &  0  &  $f_{x_0\leftrightarrow r_{2z}}$  &  &        \\ \cline{1-4}
$II$        & orchid   & $f_{x_0\leftrightarrow r_{2z}}$    & $f_{x_0\leftrightarrow r_{1z}}$   & &   \\    \cline{1-4}
$III$         &  purple  &  $f_{x_0\leftrightarrow r_{1z}}$   &   $\min(f_{r_{1z}\leftrightarrow r_{2z}},f_{x_0\leftrightarrow y_0})$  &  0  &  $f_{y_0\leftrightarrow z_0}$      \\  \cline{1-4}
$IV$        &  brown  & $f_{r_{1z}\leftrightarrow r_{2z}}$    &  $f_{x_0\leftrightarrow y_0}$  & &   \\    \cline{1-4}
$V$         &  yellowgreen  & $f_{x_0\leftrightarrow y_0}$   & $f_{r_{1z}\leftrightarrow r_{2z}}$  &  &        \\  \cline{1-4}
$VI$        &  green  & $\max(f_{x_0\leftrightarrow y_0},f_{r_{1z}\leftrightarrow r_{2z}})$   &  1  & &   \\   \hline\hline
$VII$         & lightsalmon   &  0  & $f_{x_0\leftrightarrow r_{2z}}$   &  &        \\  \cline{1-4}
$VIII$        & pink   &  $f_{x_0\leftrightarrow r_{2z}}$   &  $f_{x_0\leftrightarrow r_{1z}}$   & &   \\    \cline{1-4}
$IX$         &  brown  & $f_{x_0\leftrightarrow r_{1z}}$    & $\min(f_{x_0\leftrightarrow z_0},f_{r_{1z}\leftrightarrow r_{2z}})$  & $f_{y_0\leftrightarrow z_0}$  &  $f_{y_0\leftrightarrow r_{1z}} $      \\  \cline{1-4}
$X$        & cyan   &  $f_{r_{1z}\leftrightarrow r_{2z}}$   &  $f_{x_0\leftrightarrow z_0}$  & &   \\    \cline{1-4}
$XI$         & darkkhaki   &  $f_{x_0\leftrightarrow z_0}$  & $f_{r_{1z}\leftrightarrow r_{2z}}$   &  &        \\  \cline{1-4}
$XII$        & yellow   & $\max(f_{r_{1z}\leftrightarrow r_{2z}},f_{x_0\leftrightarrow z_0})$    &  1  & &   \\   \hline\hline
$XIII$         &  blue  &  0  &  $f_{x_0\leftrightarrow r_{2z}}$  &  &        \\  \cline{1-4}
$XIV$        &  skyblue  & $f_{x_0\leftrightarrow r_{2z}}$    & $f_{x_0\leftrightarrow y_0}$    & &   \\    \cline{1-4}
$XV$         &  coral  &  $f_{x_0\leftrightarrow y_0}$   & $\min(f_{x_0\leftrightarrow z_0},f_{y_0\leftrightarrow r_{2z}})$  &  $f_{y_0\leftrightarrow r_{1z}} $&   1         \\  \cline{1-4}
$XVI$        &  orange  &  $f_{y_0\leftrightarrow r_{2z}}$  &   $f_{x_0\leftrightarrow z_0}$  & &   \\    \cline{1-4}
$XVII$         &  fuchsia  &  $f_{x_0\leftrightarrow z_0}$  & $f_{y_0\leftrightarrow r_{2z}}$  &  &        \\  \cline{1-4}
$XVIII$        &  red  &  $\max(f_{x_0\leftrightarrow z_0},f_{y_0\leftrightarrow r_{2z}})$  &  1  &   &    \\   \hline
\end{tabular}
\caption{\label{tab:regions}
Definition of the colored regions seen in Figs.~\ref{fig:regionplotonshell0.3} and \ref{fig:regionplotonshell0.7}. }
\end{table}

Using the definitions of the boundaries (\ref{x0r2z}-\ref{y0r1z}), the 18 colored regions\footnote{A close inspection of the 
endpoint formulas (\ref{m2rendpoints}) given below reveals that in Regions $IX$ and $XV$ the endpoints are given
by the same expressions, so those two regions can be effectively merged. The same observation applies to 
Regions $XI$ and $XVII$. Thus, strictly speaking, for the purposes of defining the endpoints of the ranked
$m_{(2,r)}$ variables, one only needs to consider 16 cases.}
seen in Figs.~\ref{fig:regionplotonshell0.3} and \ref{fig:regionplotonshell0.7} can be explicitly defined as in Table~\ref{tab:regions}.
Then, the kinematic endpoints for the sorted two-body invariant mass variables $m_{(2,r)}$ are given by
\bea
(m_{(2,1)}^{\max}, m_{(2,2)}^{\max}, m_{(2,3)}^{\max} )\hs&=&\hs
m_D \times
\l\{
\baa{l l l}
\l(\sqrt{y_0},\sqrt{r_{1z}},\sqrt{x_0}\r) &  \text{in Region I} \\
\l(\sqrt{y_0},\sqrt{r_{1z}},\sqrt{r_{2z}}\r) &  \text{in Region II} \\
\l(\sqrt{y_0},\sqrt{x_0},\sqrt{r_{2z}}\r) &  \text{in Region III} \\
\l(\sqrt{y_0},\sqrt{x_0},\sqrt{r_{1z}}\r) &  \text{in Region IV} \\
\l(\sqrt{x_0},\sqrt{y_0},\sqrt{r_{2z}}\r) &  \text{in Region V} \\
\l(\sqrt{x_0},\sqrt{y_0},\sqrt{r_{1z}}\r) &  \text{in Region VI} \\
\l(\sqrt{z_0},\sqrt{r_{1z}},\sqrt{x_0}\r) &  \text{in Region VII} \\
\l(\sqrt{z_0},\sqrt{r_{1z}},\sqrt{r_{2z}}\r) &  \text{in Region VIII} \\
\l(\sqrt{z_0},\sqrt{x_0},\sqrt{r_{2z}}\r) &  \text{in Region IX} \\
\l(\sqrt{z_0},\sqrt{x_0},\sqrt{r_{1z}}\r) &  \text{in Region X} \\
\l(\sqrt{x_0},\sqrt{z_0},\sqrt{r_{2z}}\r) &  \text{in Region XI} \\
\l(\sqrt{x_0},\sqrt{z_0},\sqrt{r_{1z}}\r) &  \text{in Region XII} \\
\l(\sqrt{z_0},\sqrt{y_0},\sqrt{x_0}\r) &  \text{in Region XIII} \\
\l(\sqrt{z_0},\sqrt{y_0},\sqrt{r_{2z}}\r) &  \text{in Region XIV} \\
\l(\sqrt{z_0},\sqrt{x_0},\sqrt{r_{2z}}\r) &  \text{in Region XV} \\
\l(\sqrt{z_0},\sqrt{x_0},\sqrt{y_0}\r) &  \text{in Region XVI} \\
\l(\sqrt{x_0},\sqrt{z_0},\sqrt{r_{2z}}\r) &  \text{in Region XVII} \\
\l(\sqrt{x_0},\sqrt{z_0},\sqrt{y_0}\r)  &  \text{in Region XVIII} \\
\eaa\r. .
\label{m2rendpoints}
\eea 

For completeness, we also list the well known endpoint formulas for the 
three-body invariant mass $m_{(3,1)}$ \cite{Allanach:2000kt,Gjelsten:2004ki}
\bea
m_{(3,1)}^{\max} \hs&=&\hs 
m_D \times
\l\{
\baa{l l}
\sqrt{(1-R_{CD})(1-R_{AB}R_{BC})}, & \text{ for }R_{CD}<R_{AB}R_{BC}; \\ [2mm]
\sqrt{(1-R_{BC})(1-R_{CD}R_{AB})},  & \text{ for }R_{BC}<R_{AB}R_{CD}; \\ [2mm]
\sqrt{(1-R_{AB})(1-R_{BC}R_{CD})}, & \text{ for }R_{AB}<R_{BC}R_{CD}; \\ [2mm]
\sqrt{(1-\sqrt{R_{AB}R_{BC}R_{CD}})}, & \text{ otherwise.}
\eaa\r.   
\label{T1}
\eea

\section{Type $(2,1)$ cascade decay chain}
\label{sec:210}

\begin{figure}[t!] 
\begin{center}
\includegraphics[width=15cm]{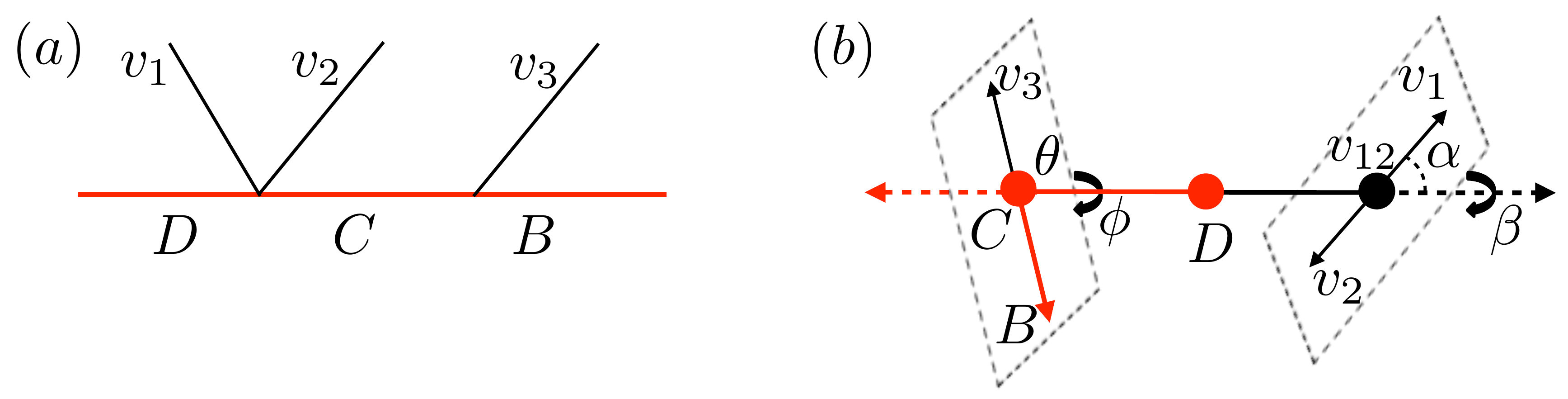} 
\caption{\label{fig:type21} (a) Type $(2,1)$ cascade decay chain. Here $v_i$, $(i=1,2,3)$, represents
a visible SM particle, while $B$, $C$, and $D$ are new physics particles. (b)
The relevant kinematics in the rest frame of particle $D$.
Here $\alpha$ is the polar angle of $v_1$ with respect to the direction of the composite system 
of $v_1\oplus v_2$, $\theta$ is the polar angle of $v_3$ with respect to the direction of $D$ 
in the rest frame of particle $C$, and $\beta$ and $\phi$ are respectively the azimuthal angles 
of $v_1$ and $v_3$ about the axis defined by particles $C$ and $D$. } 
\end{center}
\end{figure}

In this section, we proceed to analyze one of the hybrid decay topologies, namely type $(2,1)$.
The relevant decay chain is depicted in Fig.~\ref{fig:type21}(a):
a massive particle $D$ decays via a three-body decay into two massless visible particles $v_1$ and $v_2$
and an on-shell intermediate particle $C$. In turn, particle $C$ decays into a massless visible particle $v_3$ 
and an invisible particle $B$. The relevant decay configuration seen in the rest frame of particle 
$D$ is illustrated in Fig.~\ref{fig:type21}(b). Again, na\"{i}vely there exist four degrees of freedom,
however, out of the two azimuthal angles, $\beta$ and $\phi$, only their difference 
is relevant --- we can then safely parametrize it with $\phi$, and set $\beta$ to zero. 

\begin{figure}[t] 
\begin{center}
\includegraphics[width=14.5cm]{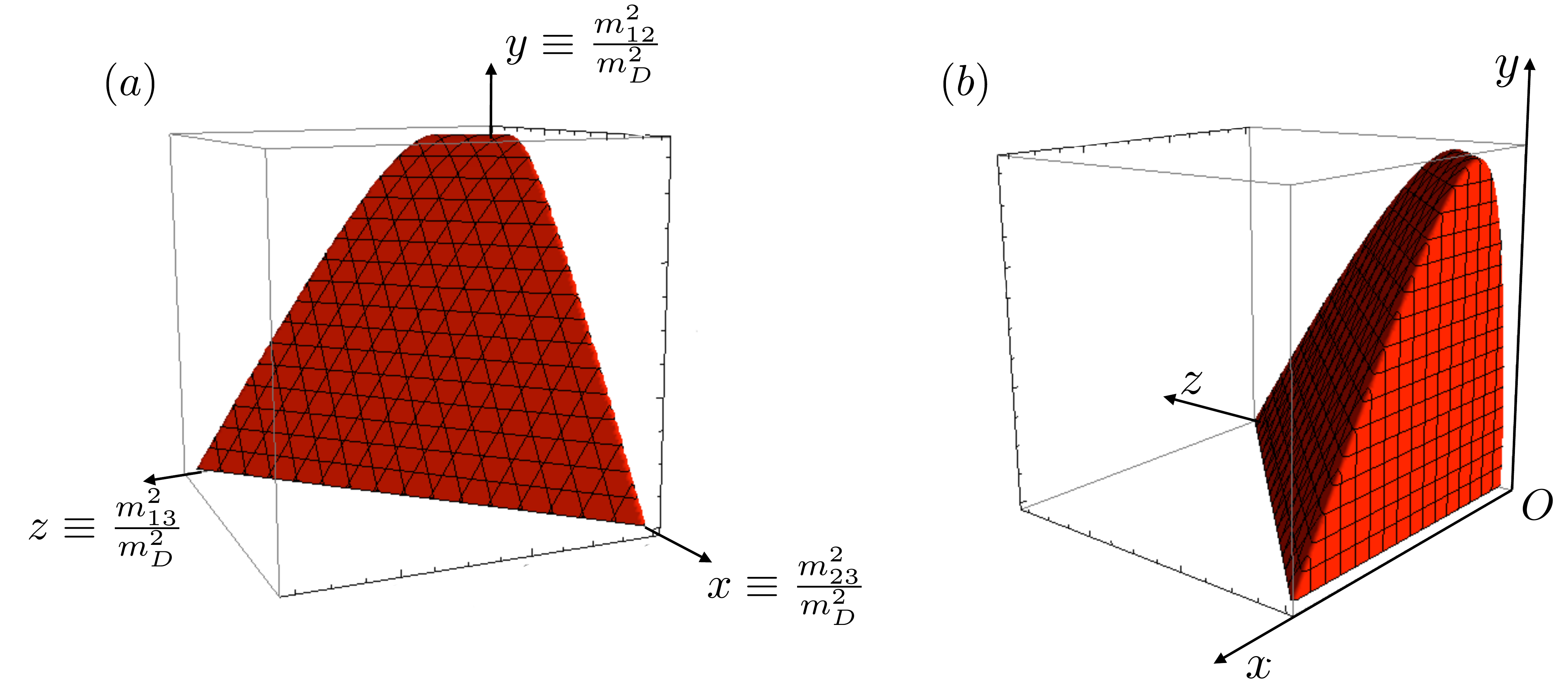}
\end{center}
\caption{\label{fig:fullphase21}  
 Two different views of the allowed phase space for a
$(2,1)$ cascade decay chain, in terms of the dimensionless variables (\ref{xdef}-\ref{zdef}). 
The boundary surface is given by Eqs.~(\ref{xzboundary}-\ref{sigma21-}).  
In order to study the ranked invariant mass distributions, here we scan
at a fixed $y$ value $y_s$.  For this illustration,
the mass spectrum has been chosen as $\left(m_D,m_C,m_B\right)  = \left(1000,700,500\right)$ GeV.
} 
\end{figure} 
 
The allowed $\{x,y,z\}$ phase space is shown in Fig.~\ref{fig:fullphase21}.  
In order to obtain the equation for the surface boundary of the allowed region, we start from 
the kinematic relation
\bea
x+z&=&1+R_{BC}R_{CD}-y-\frac{2E_B^{(D)}}{m_D} \nonumber \\
&=&\frac{1-R_{BC}}{2} \left[ 1-R_{CD}-y-\sqrt{(1+R_{CD}-y)^2-4R_{CD}}\cos\theta \right],
\label{xpz}
\eea
where the superscript on $E_B$ implies that this energy is measured in the rest frame of particle $D$. 
Notice that the symmetry $v_1 \leftrightarrow v_2$ implies that the variables $x$ and $z$ 
always enter in the combination $x+z$. Then, the equation for the boundary surface is
\beq
x+z = \Sigma^{\pm}(y),
\label{xzboundary}
\eeq
where the functions $\Sigma^{\pm}(y)$, which are the analogues of
(\ref{zplus},\ref{zminus}), are obtained from (\ref{xpz}) for the extreme values of $\cos\theta$:
\bea
\Sigma^{+}_{(x,z)}(y) &\equiv& \max_{y}(x+z)  =  \frac{1-R_{BC}}{2} \left[1-R_{CD}-y + \sqrt{\left(1+R_{CD}-y\right)^2-4 R_{CD}}\right], 
\label{sigma21+}\\
\Sigma^{-}_{(x,z)}(y) &\equiv& \min_{y}(x+z)  = \frac{1-R_{BC}}{2} \left[1-R_{CD}-y - \sqrt{\left(1+R_{CD}-y\right)^2-4 R_{CD}}\right] .
\label{sigma21-}
\eea

\begin{figure}[t] 
\begin{center}
\includegraphics[width=15cm]{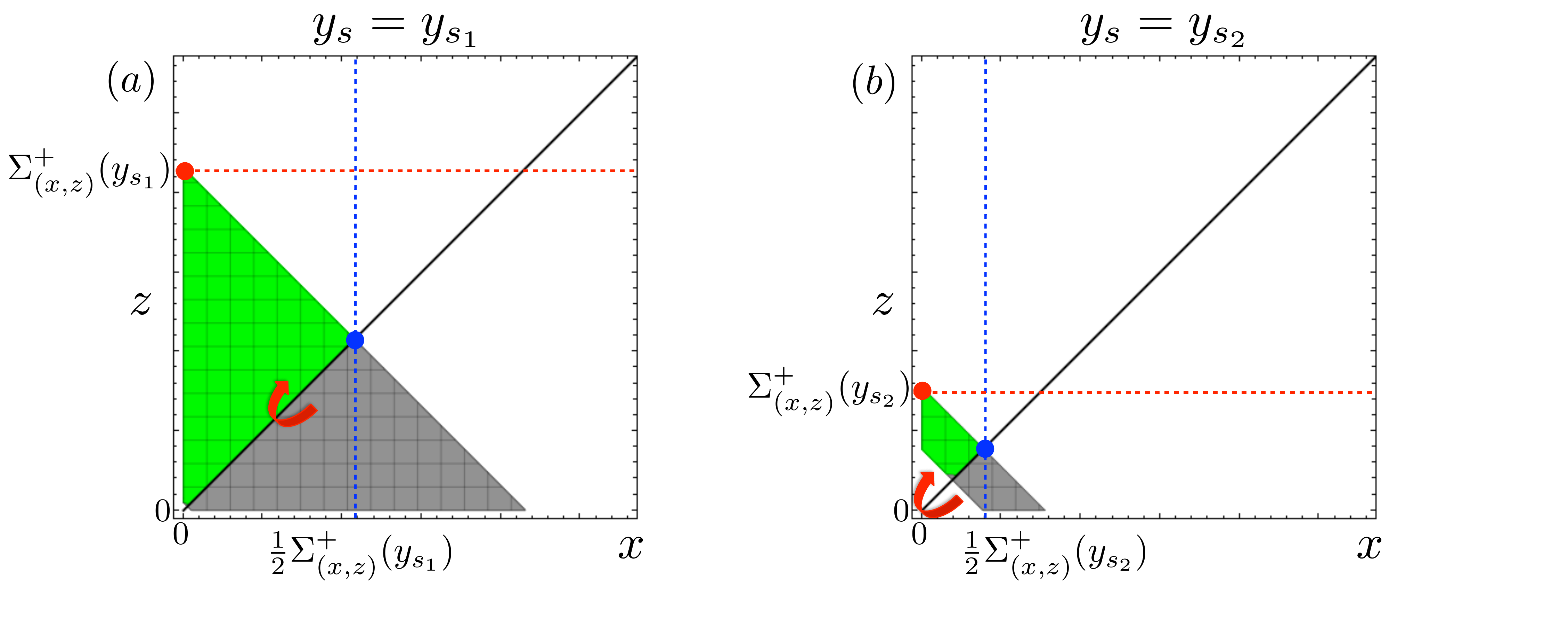}
\includegraphics[width=15cm]{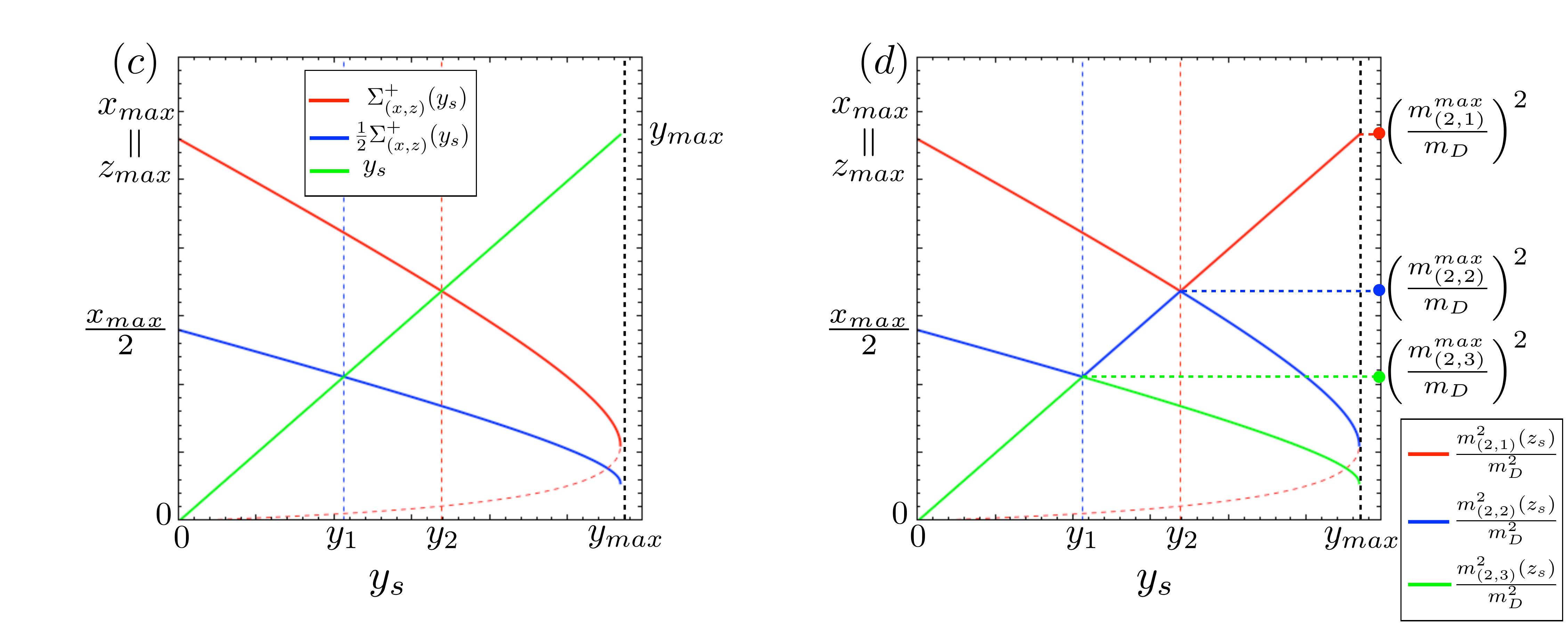}
\end{center}
\caption{\label{fig:case3}   (a,b): CT-images in the $(x,z)$ plane, for a small (left) and a large (right) value of $y_s$.  
Due to the $x\leftrightarrow z$ symmetry, the green and grey halves of the CT image are identical.
(c,d) Illustration of the ranking procedure among $x$, $y$ and $z$, in analogy to Figs.~\ref{fig:caseISCAN} and \ref{fig:caseIISCAN}.
} 
\end{figure} 
 
In order to derive the maximal values of the sorted invariant masses of order 2, 
we repeat the scanning procedure from the previous section, only now
we scan along the $y$-axis. For a given $y=y_s$, the CT image in the $(x,z)$ plane is an isosceles trapezoid,
as shown in Fig.~\ref{fig:case3}. We again rank $x(y_s)$ and $z(y_s)$ for all possible pairs of $(x,z)$ to 
obtain the ranked variables
\bea 
r_1\equiv\max\left\{\max\left[x(y_s),\;z(y_s)\right]\right\}, \quad
r_2\equiv \max\left\{\min\left[x(y_s),\;z(y_s) \right]\right\}. 
\eea
Due to the simple geometry of Fig.~\ref{fig:case3}(a,b), $r_1$ and $r_2$ can be readily computed as
\bea
r_1(y_s)&=&\Sigma^{+}_{(x,z)}(y)=\frac{1-R_{BC}}{2} \left[1-R_{CD}-y + \sqrt{\left(1+R_{CD}-y\right)^2-4 R_{CD}}\right], \\ 
r_2(y_s)&=& \frac{\Sigma^{+}_{(x,z)}(y)}{2}=\frac{1-R_{BC}}{4} \left[1-R_{CD}-y + \sqrt{\left(1+R_{CD}-y\right)^2-4 R_{CD}}\right].
\label{r1r2ys}
\eea
We then compare the variables $r_1(y_s)$, $r_2(y_s)$, and $y=y_s$ as before,
see Fig.~\ref{fig:case3}(c,d). There are two special points, denoted by $y_1$ and $y_2$, which
arise from the intersection of $r_2(y_s)$ and $r_1(y_s)$ with the straight line $y=y_s$:
\bea
\frac{1}{2} \Sigma^{+}_{(x,z)}(y_1)  = y_1~~ &\rightarrow& ~~ y_1 = \frac{1}{2} \left(1-R_{BC}\right) \left[\frac{2}{3-R_{BC}}-R_{CD}\right], \label{21_y1} \\
\Sigma^{+}_{(x,z)}(y_2)  = y_2~~ &\rightarrow&~~ y_2 =   \left(1-R_{BC}\right) \left[\frac{1}{2-R_{BC}}-R_{CD}\right]. \label{21_y2}
\eea
The kinematic endpoints for the sorted two-body invariant masses 
$m_{(2,r)}$ will be given in terms of the endpoints $x_0=z_0$ and $y_0$
of the original unsorted variables and the two special points $y_1$ and $y_2$ given by
(\ref{21_y1}) and (\ref{21_y2}). Fig.~\ref{fig:case3}(d) shows an example where 
$m_{(2,1)}^{max}=m_D \sqrt{y_0}$,
$m_{(2,2)}^{max}=m_D \sqrt{y_2}$ and
$m_{(2,3)}^{max}=m_D \sqrt{y_1}$.
The relevant formulas for the general case are collected in Section~\ref{results21}.
   
\begin{figure}[t] 
\begin{center}
\includegraphics[width=16cm]{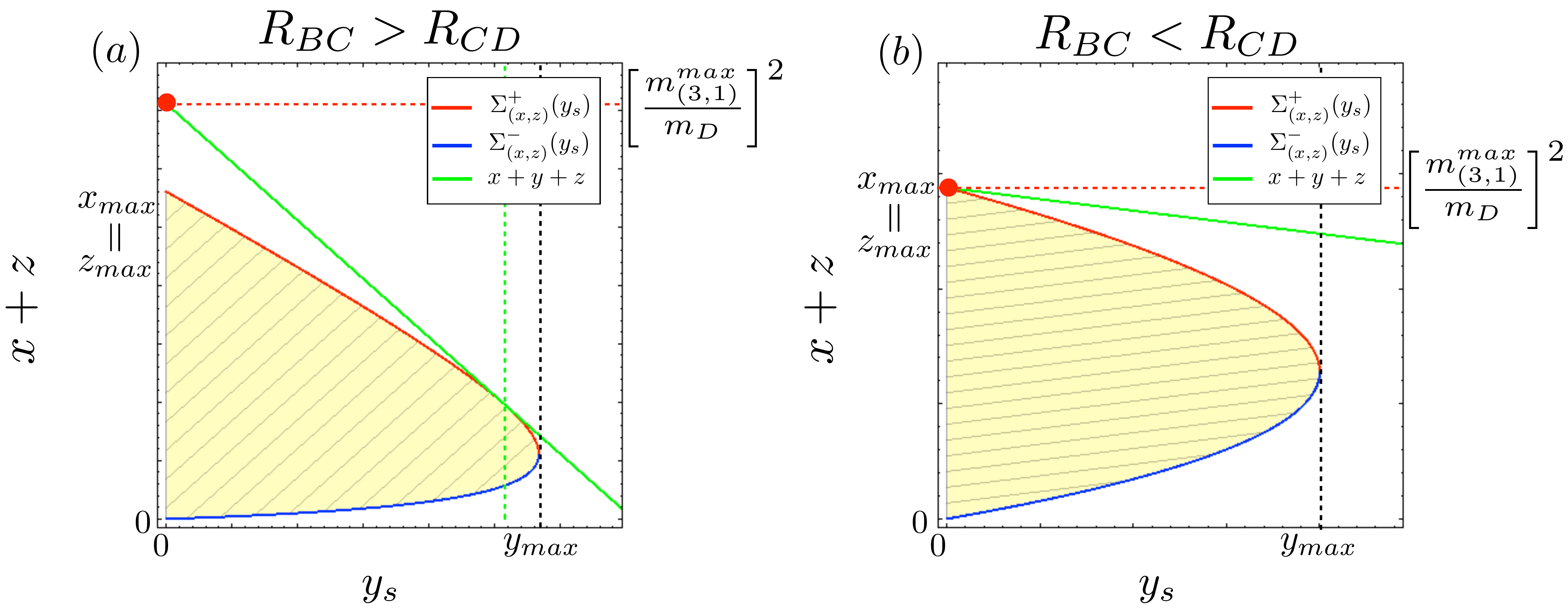}
\end{center}
\caption{\label{fig:case21qll}   The allowed phase space (yellow-shaded region) of a Type $(2,1)$ decay chain in the 
$(y,x+z)$ plane. The kinematic boundary is given by $\Sigma^{\pm}_{(x,z)}(y_s)$ for a given $y_s$.
The two panels show the two distinct cases for $m_{(3,1)}^{max}$: (a) $R_{BC}>R_{CD}$ and (b) $R_{BC}<R_{CD}$.} 
\end{figure}

In conclusion of this subsection, we discuss the derivation of the endpoint of the $m_{(3,1)}$ variable 
for the case of a type $(2,1)$ decay topology. Since in supersymmetry models one typically gets 
either $(1,1,1)$ or $(1,2)$ decay topologies, this case has not drawn much attention in the existing literature.
It is most convenient to analyze the decay kinematics in the plane of $(x+z)$ versus $y$, as
shown in Fig.~\ref{fig:case21qll}. Since all visible particles are assumed massless, we have the identity 
\beq
\frac{m_{(3,1)}^2}{m_D^2}=x+y+z\equiv k,
\label{kdef}
\eeq
which describes a straight line with a slope $-1$ and intercept $k$: 
\bea
x+z = -y+k.
\label{eq:tanline}
\eea

Given Eq.~(\ref{kdef}), in order to maximize $m_{(3,1)}$, we need to maximize
the intercept, $k$, with respect to the full allowed phase space of $(y,x+z)$,
keeping the slope fixed at $-1$. 
As shown in Fig.~\ref{fig:case21qll}, the boundary of the allowed phase space 
is given by $\Sigma_{(x,z)}^{\pm}(y)$, and we need to consider two distinct cases,
illustrated in panels (a) and (b).
Note that $\Sigma_{(x,z)}^+(y)$ is a monotonically decreasing function of $y$, whose (negative) slope 
increases in magnitude and approaches $-\infty$ at $y=y_{\max}$. 
Therefore, if the slope at $y=0$ is larger than $-1$, then Eq.~(\ref{eq:tanline}) 
will appear as a tangential line on the curve of $\Sigma_{(x,z)}^+(y)$
as shown in Fig.~\ref{fig:case21qll}(a), and the resulting $k$ will give rise 
to the maximum $m_{(3,1)}$. If, however, the slope at $y=0$ is already smaller than $-1$, then no such
tangential line is possible, and the line giving the maximum will pass through $(y,x+z)=(0,\Sigma^+_{(x,z)}(0))$,
as shown in Fig.~\ref{fig:case21qll}(b). A simple algebra results in the following kinematic endpoints:
\bea
k_{\max}=\l\{
\baa{ll}
(1-\sqrt{R_{BC}R_{CD}})^2, & R_{CD}<R_{BC}; \\ [2mm]
(1-R_{BC})(1-R_{CD}), & \text{otherwise.}
\eaa
\r.
\eea

\subsection{Results summary for the $(2,1)$ decay topology}
\label{results21}

The endpoints of the $m_{(2,r)}$ and $m_{(3,1)}$ sorted invariant mass variables are 
given in terms of the nominal endpoints
\bea
x_0 &\equiv& x_{max} = (1-R_{BC})(1-R_{CD}), \\ [2mm]
y_0 &\equiv& y_{max} = (1-\sqrt{R_{CD}})^2,
\eea
and the intersection points $y_1$ and $y_2$ given by (\ref{21_y1}) and (\ref{21_y2}).
The endpoint formulas are again piecewise-defined functions, and the boundaries of the
defining regions are given by functions $R_{CD}=f(R_{BC})$ in analogy to (\ref{x0r2z}-\ref{y0r2z}):
\bea
x_0\leftrightarrow y_0:  \quad
R_{CD} &=& f_{x_0\leftrightarrow y_0} (R_{BC}) \equiv 
\frac{R^2_{BC}}{(2-R_{BC})^2},
\label{21_x0y0}
\\ 
y_2\leftrightarrow x_0/2:  \quad
R_{CD} &=& f_{y_2\leftrightarrow x_0/2} (R_{BC}) \equiv 
\frac{R_{BC}}{(2-R_{BC})},
\label{21_y2x02}
\\ 
y_0\leftrightarrow y_1:  \quad
R_{CD} &=& f_{y_0\leftrightarrow y_1} (R_{BC}) \equiv 
\frac{4}{(3-R_{BC})^2}.
\label{21_y0y1}
\eea

\begin{figure}[t]
\centering
\includegraphics[width=9cm]{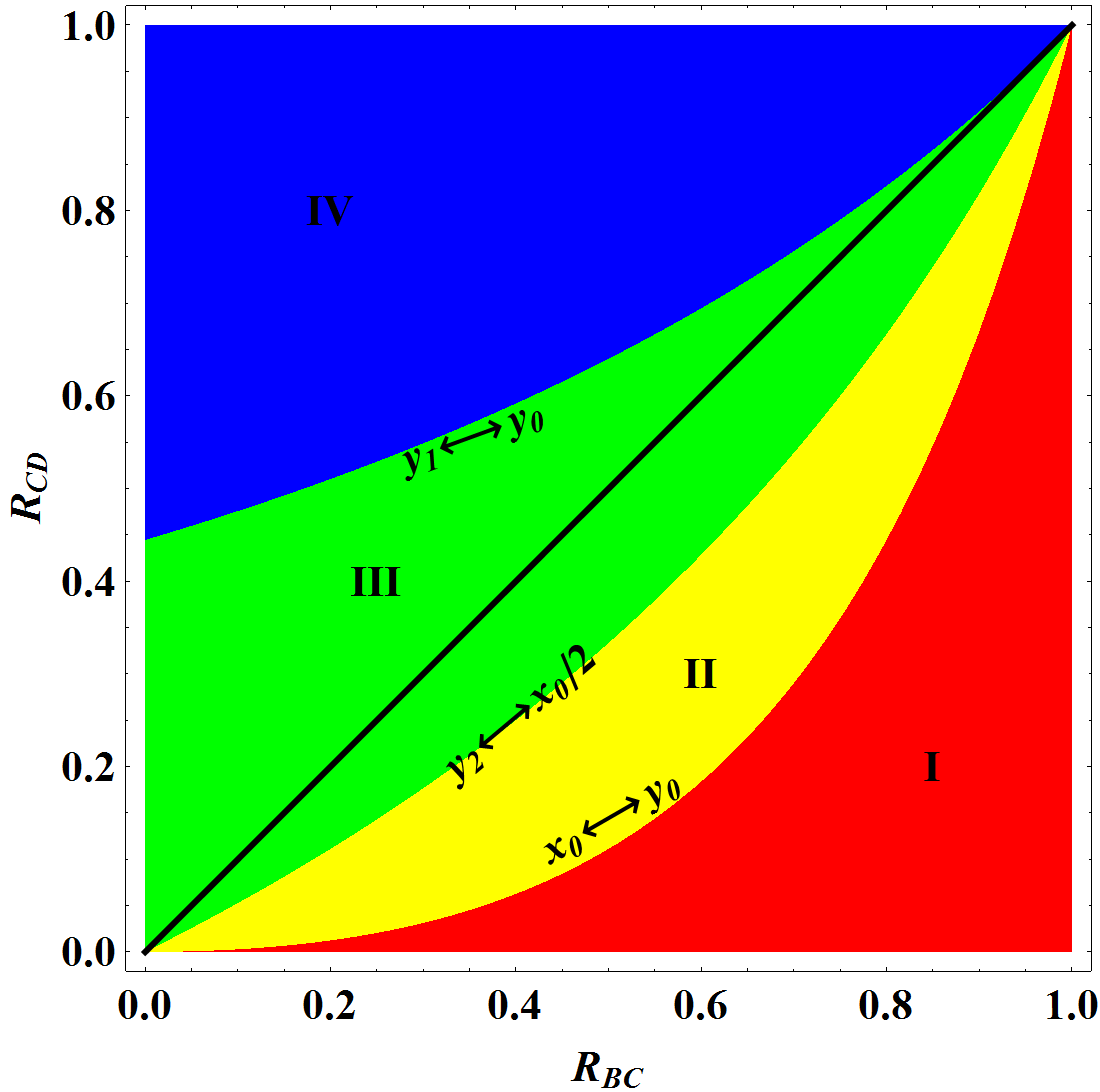}
\caption{\label{fig:regionplot21} The same as Figs.~\ref{fig:regionplotonshell0.3}
and \ref{fig:regionplotonshell0.7}, but for the Type (2,1) decay topology discussed in Section \ref{sec:210}.}
\end{figure}

The boundaries (\ref{21_x0y0}-\ref{21_y0y1}) divide the $(R_{BC},R_{CD})$ parameter space of the 
$(2,1)$ decay topology into the 4 color-coded regions shown in Fig.~\ref{fig:regionplot21}.
Then, the kinematic endpoints of the ranked two-body invariant mass distributions are given by
\bea
(m_{(2,1)}^{\max}, m_{(2,2)}^{\max}, m_{(2,3)}^{\max} )\hs&=&\hs
m_D \times
\l\{
\baa{l l l}
\l(\sqrt{y_0},\sqrt{y_2},\sqrt{y_1}\r) &  \text{in Region I;} \\ [2mm]
\l(\sqrt{x_0},\sqrt{y_2},\sqrt{y_1}\r) &  \text{in Region II;} \\ [2mm]
\l(\sqrt{x_0},\sqrt{x_0/2},\sqrt{y_1}\r) &  \text{in Region III;} \\ [2mm]
\l(\sqrt{x_0},\sqrt{x_0/2},\sqrt{y_0}\r) &  \text{in Region IV;} 
\eaa\r.  
\eea
while the endpoint of the three-body invariant mass is 
\bea
m_{(3,1)}^{max}\hs&=&\hs 
m_D \times
\l\{
\baa{l l }
1-\sqrt{R_{BC}R_{CD}} & \quad \text{for }R_{CD} < R_{BC}; \\ [2mm]
\sqrt{(1-R_{BC})(1-R_{CD})} & \quad \text{otherwise.}
\eaa\r. 
\eea

 \section{Type $(1,2)$ cascade decay chain}
 \label{sec:120}

\begin{figure}[t] 
\begin{center}
\includegraphics[width=15cm]{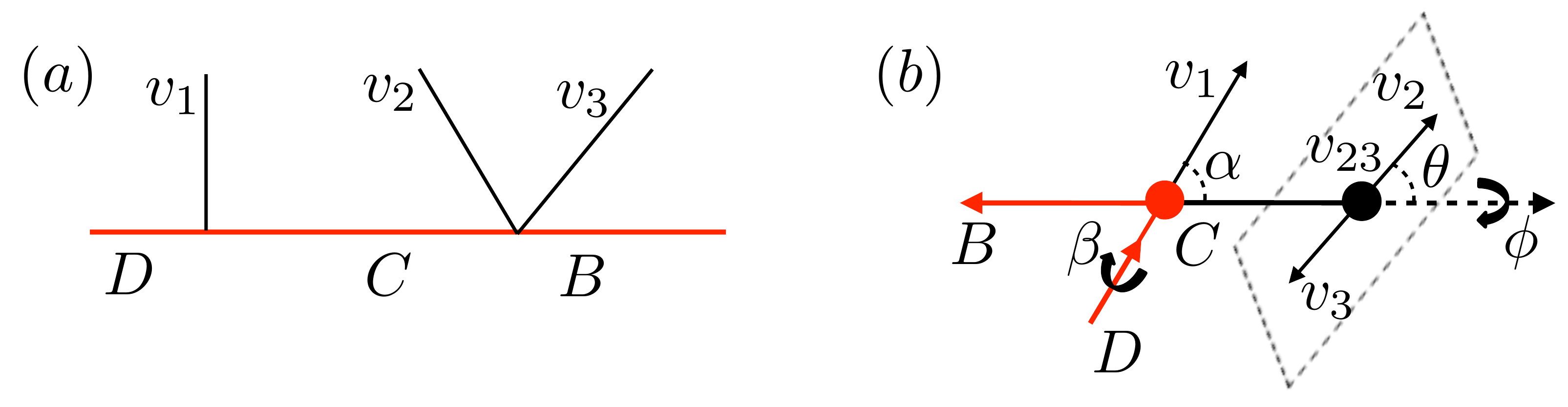} 
\caption{\label{fig:type12} (a) Type $(1,2)$ cascade decay chain. (b) The relevant kinematics in the rest frame of particle $C$.
Here $\alpha$ ($\theta$) is the polar angle of $v_1$ ($v_2$) with respect to the direction of the composite system of $v_2\oplus v_3$, 
and $\beta$ and $\phi$ are the azimuthal angles of $v_1$ and $v_2$ about the axis defined by particles $B$ and $C$. } 
\end{center}
\end{figure}
 
In this section, we analyze the decay topology of type $(1,2)$, which is depicted in Fig.~\ref{fig:type12}(a).
First, a massive particle $D$ decays into a visible particle $v_1$ along with an on-shell intermediate particle $C$, 
and in turn, particle $C$ decays into two visible particles $v_2$ and $v_3$ and an invisible particle 
$B$ via a three-body decay. It is convenient to analyze the kinematics in the rest frame of particle $C$, 
as illustrated in Fig.~\ref{fig:type12}(b). As one might expect, most of the analysis leading to the final formulae will be similar 
to that of the preceding section.

\begin{figure}[t] 
\begin{center}
\includegraphics[width=14.5cm]{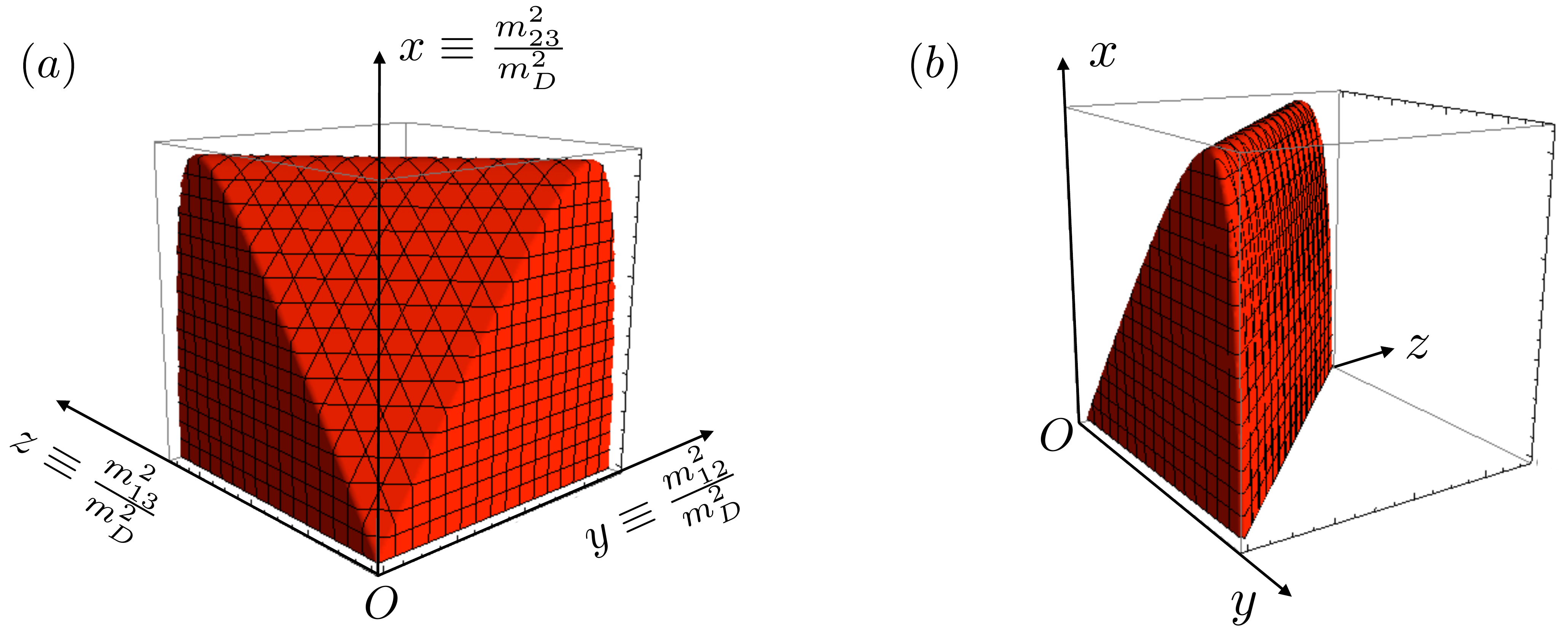}
\end{center}
\caption{\label{fig:fullphase12}  The same as Fig.~\ref{fig:fullphase12}, but for a
$(1,2)$ cascade decay chain. The boundary surface is now given by 
(\ref{yzboundary}-\ref{12Sigma-}) and the scanning is done at a fixed value for $x$.} 
\end{figure} 

The allowed $\l\{x,y,z\r\}$ phase space is illustrated in Fig.~\ref{fig:fullphase12}.
Notice the $y\leftrightarrow z$ symmetry which follows from the $v_2 \leftrightarrow v_3$ symmetry.
The boundary of the allowed region can be derived from a kinematic relation analogous to (\ref{xpz}):
\bea
y+z=\frac{1-R_{CD}}{2} \left[1-R_{BC}+\frac{x}{R_{CD}} - \sqrt{\left(1+R_{BC}-\frac{x}{R_{CD}}\right)^2-4 R_{BC}}\cos\theta \right].
\eea
The boundary equation is obtained from here by taking $\cos\theta=\pm1$. One finds
\beq
y+z = \Sigma^{\pm}(x),
\label{yzboundary}
\eeq
where
\bea
\Sigma^{+}_{(y,z)}(x) &=& \frac{1-R_{CD}}{2} \left[1-R_{BC}+\frac{x}{R_{CD}} + \sqrt{\left(1+R_{BC}-\frac{x}{R_{CD}}\right)^2-4 R_{BC}}\right], 
\label{12Sigma+}\\
\Sigma^{-}_{(y,z)}(x) &=&  \frac{1-R_{CD}}{2} \left[1-R_{BC}+\frac{x}{R_{CD}} - \sqrt{\left(1+R_{BC}-\frac{x}{R_{CD}}\right)^2-4 R_{BC}}\right].
\label{12Sigma-}
\eea

\begin{figure}[t] 
\begin{center}
\includegraphics[width=15cm]{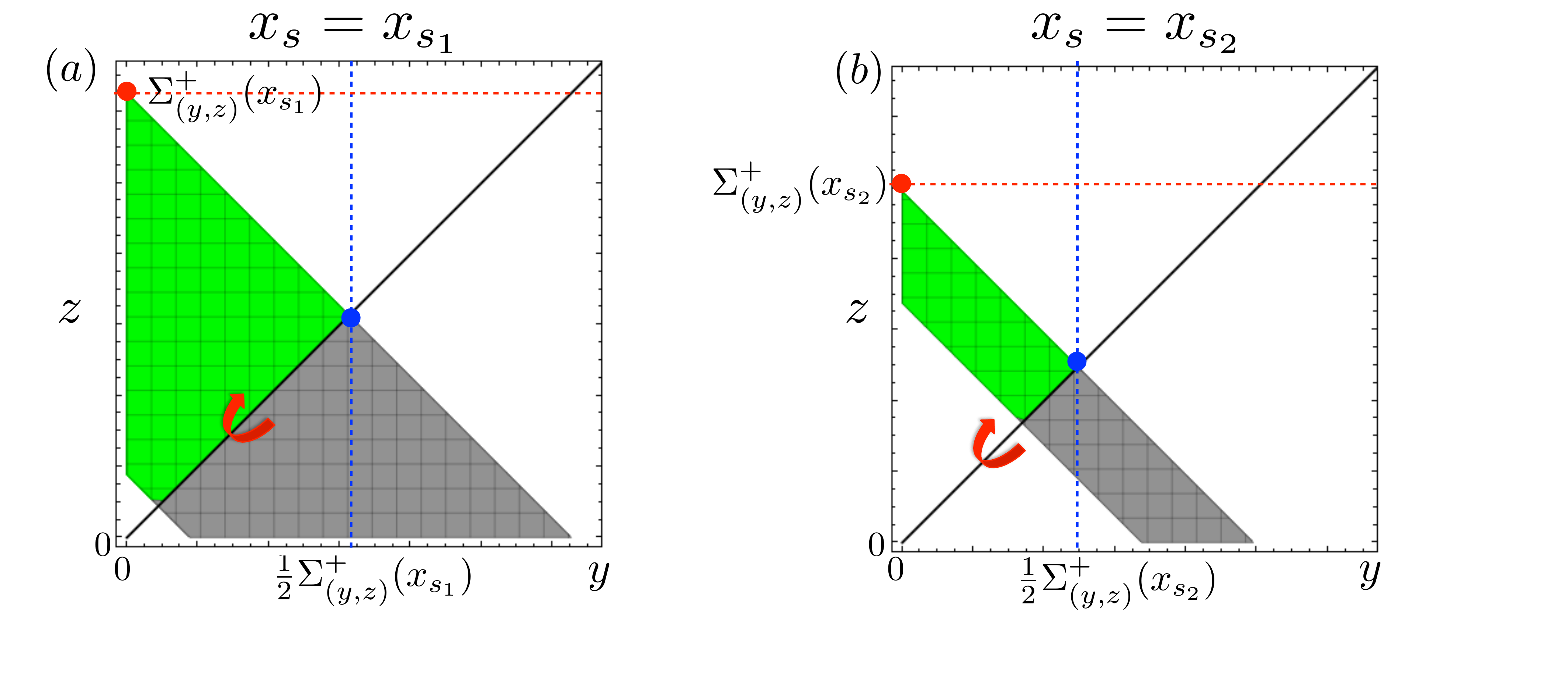}
\includegraphics[width=15cm]{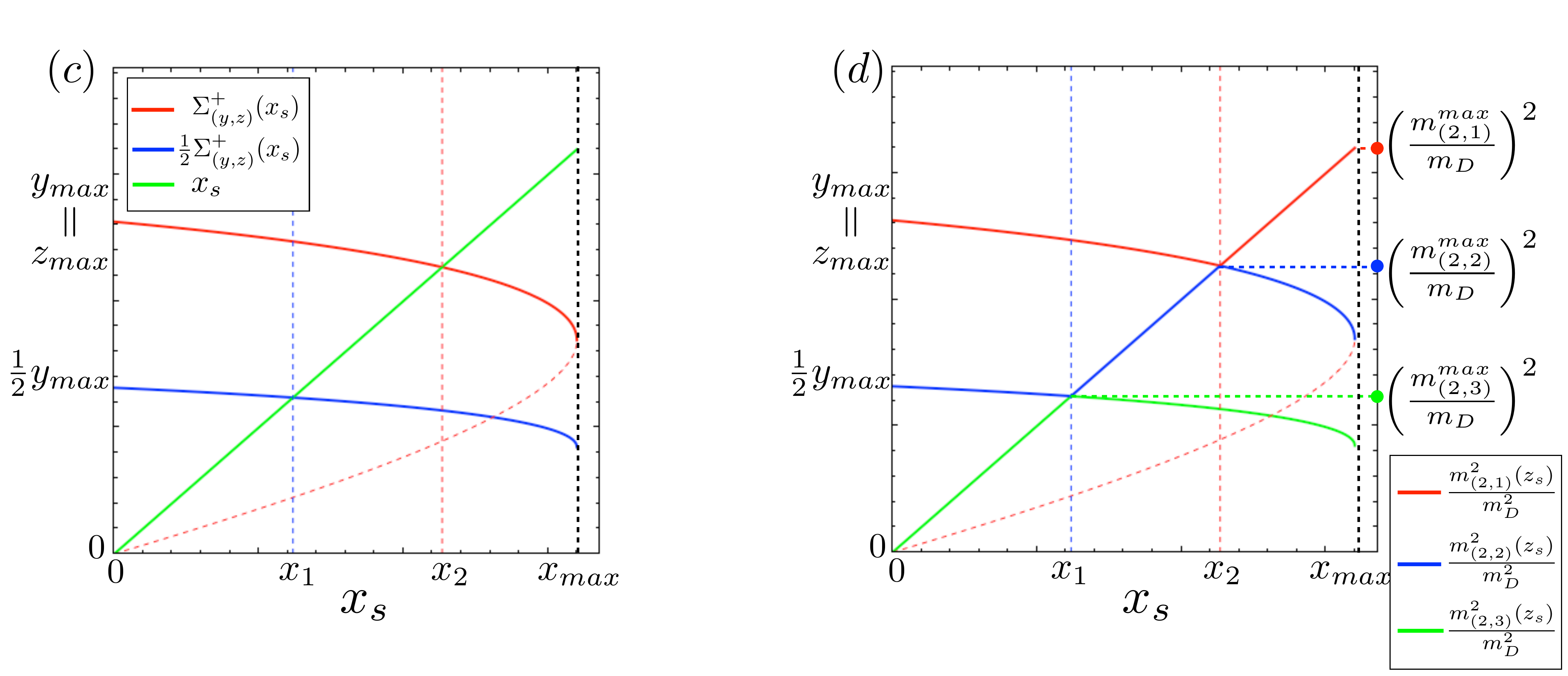}
\end{center}
\caption{\label{fig:case3p}  (a,b): CT-images in the $(y,z)$ plane at two fixed values for $x_s$, with
$x_{s_1}<x_{s_2}$. Due to the $y\leftrightarrow z$ symmetry, the green and grey halves of the CT image are identical.
(c,d) Illustration of the ranking procedure among $x$, $y$ and $z$, in analogy to Figs.~\ref{fig:caseISCAN}, \ref{fig:caseIISCAN}
and \ref{fig:case3}(c,d).
} 
\end{figure}

In order to find the largest two-body invariant masses, we again scan the allowed phase space shown in Fig.~\ref{fig:fullphase12},
this time at fixed values for $x=x_s$. The obtained images are again isosceles trapezoids with bases of length $\Sigma^{+}_{(y,z)}(x_s)$
and $\Sigma^{-}_{(y,z)}(x_s)$, as shown in Fig.~\ref{fig:case3p}(a,b).
We then rank $y(x_s)$ and $z(x_s)$ in analogy to (\ref{eq:defr1r2}) and (\ref{r1r2ys})
\bea 
r_1\equiv\max\left\{\max\left[y(x_s),\;z(x_s)\right]\right\}, \quad
r_2\equiv \max\left\{\min\left[y(x_s),\;z(x_s) \right]\right\}. 
\eea
%
Just like the previous case of type $(2,1)$ topology, due to the symmetric structure of the phase space, 
the corresponding $r_1$ and $r_2$ are simple to evaluate --- they are given by 
$\Sigma_{(y,z)}^+(x_s)$ and $\Sigma_{(y,z)}^+(x_s)/2$, respectively. 

Finally, the ranking procedure among $r_1(x_s)$, $r_2(x_s)$ and $x_s$ again introduces 
two intersection points, $x_1$ and $x_2$, which arise from the crossing of $r_2(x_s)$ with $x=x_s$,
and $r_1(x_s)$ with $x=x_s$, respectively (see Fig.~\ref{fig:case3p}(c,d)):
\bea
\frac{1}{2} \Sigma^{+}_{(y,z)}(x_1)  = x_1~~ &\rightarrow&~~ x_1 = \frac{1-R_{CD}}{2}  \left[\frac{1-R_{CD}(3-2R_{BC})}{1-3R_{CD}}\right],
\label{x1def} \\
\Sigma^{+}_{(y,z)}(x_2)  = x_2~~ &\rightarrow&~~ x_2 =    \frac{1-R_{CD}}{1-2 R_{CD}}\left[1-R_{CD}(2-R_{BC})\right].
\label{x2def}
\eea
The endpoints of the ranked two-body invariant masses $m_{(2,r)}$ will be given in terms 
of $y_0(=z_0)$, $x_0$, $y_0/2$, $x_1$, or $x_2$, depending on the mass spectrum.
Fig.~\ref{fig:case3p}$(c,d)$ illustrates a specific example where 
$m_{(2,1)}^{max}=m_D \sqrt{x_0}$,
$m_{(2,2)}^{max}=m_D \sqrt{x_2}$ and
$m_{(2,3)}^{max}=m_D \sqrt{x_1}$.
The relevant formulas for the general case are collected in Section~\ref{results12}.

Finally, we discuss the three-body invariant mass $m_{(3,1)}$. 
The decay topology $(1,2)$ is very common in supersymmetry,
e.g., in the decay $\tilde q\to q\tilde\chi^0_2$ of a squark $\tilde q$ to the second-to-lightest neutralino 
$\tilde\chi^0_2$, which in turn decays by a three-body decay to the lightest neutralino $\tilde\chi^0_1$
and a couple of jets or leptons: $\tilde\chi^0_2 \to q\bar{q} \tilde\chi^0_1$ or 
$\tilde\chi^0_2 \to \ell^+\ell^- \tilde\chi^0_1$. The expression for the endpoint $m_{(3,1)}^{\max}$ 
is already known (see, e.g.~\cite{Lester:2006cf}) and here we shall simply re-derive it using the method 
from Section~\ref{sec:210}.

\begin{figure}[t] 
\begin{center}
\includegraphics[width=16cm]{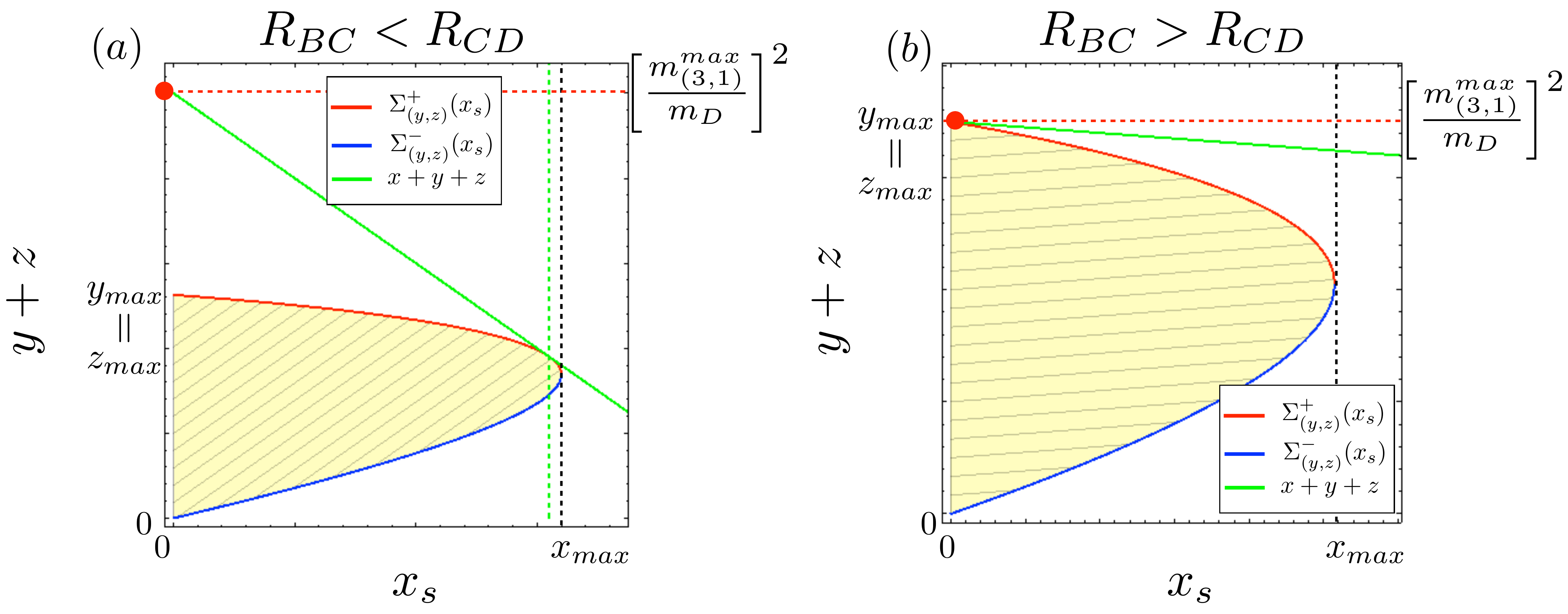}
\end{center}
\caption{\label{fig:case12qll}   
The same as Fig.~\ref{fig:case21qll}, but for a type $(1,2)$ decay topology, where
the relevant phase space is best viewed in the $(x,y+z)$ plane.
} 
\end{figure}

Again, it is convenient to study the variable of interest 
\beq
m_{(3,1)}^2 = x + y + z \equiv k
\label{kdef2}
\eeq
in the plane of $(y+z)$ versus $x$, as depicted in Fig.~\ref{fig:case12qll}, where
the allowed region is shaded in yellow. In this plane, the relation (\ref{kdef2})
again represents a straight line 
\bea
y+z=-x+k. \label{eq:tanline2}
\eea 
with constant negative slope $-1$ and intercept $k$. Just like in the previous section, the task is to find the 
point on the phase space boundary which maximizes the intercept, $k$, for a fixed slope $-1$.
As the two panels of Fig.~\ref{fig:case12qll} show, one again has to consider two cases, depending on whether the 
slope of the boundary curve $\Sigma^+_{(y,z)}(x)$ at $x=0$ is larger or smaller than $-1$. 
In the former case, shown in Fig.~\ref{fig:case12qll}(a), the endpoint $m_{(3,1)}^{\max}$ is obtained from
a line tangential to the boundary, while in the latter case the endpoint is simply given by $y_0$:
\bea
m_{(3,1)}^{max}\hs&=&\hs 
m_D \times
\l\{
\baa{ll}
(1-\sqrt{R_{BC}R_{CD}}) & R_{CD}> R_{BC}, \\ [2mm]
\sqrt{(1-R_{CD})(1-R_{BC})} & \text{otherwise.}
\eaa
\r.
\eea

\subsection{Results summary for the $(1,2)$ decay topology}
\label{results12}

\begin{figure}[t]
\centering
\includegraphics[width=9cm]{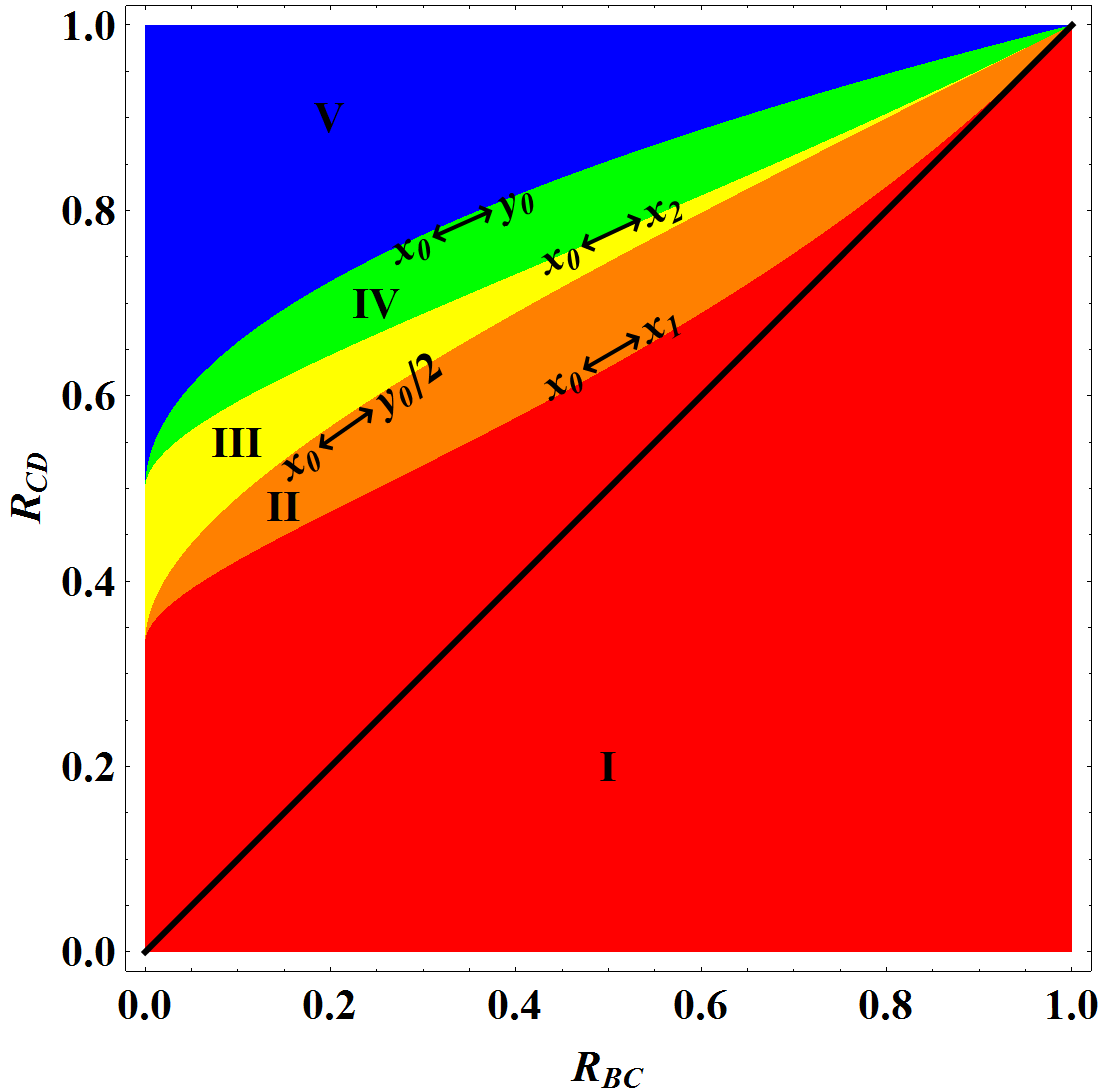}
\caption{\label{fig:regionplot12} The same as Fig.~\ref{fig:regionplot21}, but for the decay topology of Type $(1,2)$.}
\end{figure}

The endpoints of the $m_{(2,r)}$ and $m_{(3,1)}$ sorted invariant mass variables are 
given in terms of the nominal endpoints for the $(1,2)$ decay topology
\bea
x_0 &\equiv& x_{max}  = R_{CD}\left(1-\sqrt{R_{BC}}\right)^2 , \\ [2mm]
y_0 &\equiv& y_{max}  = z_0 \left(\equiv z_{max}\right)  = (1-R_{CD})(1-R_{BC}),
\eea
and the intersection points $x_1$ and $x_2$ given by (\ref{x1def}) and (\ref{x2def}).
Again, the formulas are piecewise-defined functions whose relevant domains are illustrated in Fig.~\ref{fig:regionplot12}.
In analogy to (\ref{21_x0y0}-\ref{21_y0y1}) the boundaries of the colored regions in Fig.~\ref{fig:regionplot12} 
are defined in terms of the functions $R_{CD}=f(R_{BC})$ as follows:
\bea
x_0\leftrightarrow x_1:  \quad
R_{CD} &=& f_{x_0\leftrightarrow x_1} (R_{BC}) \equiv 
\frac{1}{3-2\sqrt{R_{BC}}},
\label{12_x0x1}
\\ 
x_0\leftrightarrow y_0/2:  \quad
R_{CD} &=& f_{x_0\leftrightarrow y_0/2} (R_{BC}) \equiv 
\frac{1+\sqrt{R_{BC}}}{3-\sqrt{R_{BC}}},
\label{12_x0y02}
\\ 
x_0\leftrightarrow x_2:  \quad
R_{CD} &=& f_{x_0\leftrightarrow x_2} (R_{BC}) \equiv 
\frac{1}{2-\sqrt{R_{BC}}},
\label{12_x0x2}
\\ 
x_0\leftrightarrow y_0:  \quad
R_{CD} &=& f_{x_0\leftrightarrow y_0} (R_{BC}) \equiv 
\frac{1+\sqrt{R_{BC}}}{2}.
\label{12_x0y_0}
\eea
Unlike the case of the type $(2,1)$ decay topology discussed in Fig.~\ref{fig:regionplot21}, 
in Fig.~\ref{fig:regionplot12} we now get 5 different regions.\footnote{The additional region III arises
due to the possibility of having $\frac{1}{2} \Sigma^{+}_{(y,z)}(x_s=0) < x_0< \Sigma^{+}_{(y,z)}(x_s=x_0)$.
The analogous case for a type $(2,1)$ decay topology is impossible, due to the relation 
$\frac{1}{2} \Sigma^{+}_{(x,z)}(y_s=0) > \Sigma^{+}_{(x,z)}(y_s=y_0)$,
as seen in Fig.~\ref{fig:case3}(c,d).}

The kinematic endpoints of the sorted two-body invariant mass distributions are given by
\bea
(m_{(2,1)}^{\max}, m_{(2,2)}^{\max}, m_{(2,3)}^{\max} )\hs&=&\hs
m_D \times
\l\{
\baa{l l l}
\l(\sqrt{y_0},\sqrt{y_0/2},\sqrt{x_0}\r) &  \text{in Region I;} \\ [2mm]
\l(\sqrt{y_0},\sqrt{y_0/2},\sqrt{x_1}\r) &  \text{in Region II;} \\ [2mm]
\l(\sqrt{y_0},\sqrt{x_0},\sqrt{x_1}\r) &  \text{in Region III;} \\ [2mm]
\l(\sqrt{y_0},\sqrt{x_2},\sqrt{x_1}\r) &  \text{in Region IV;} \\ [2mm]
\l(\sqrt{x_0},\sqrt{x_2},\sqrt{x_1}\r) &  \text{in Region V;} 
\eaa\r.  
\eea
while the endpoint of the three-body invariant mass is 
\bea
m_{(3,1)}^{max}\hs&=&\hs 
m_D \times
\l\{
\baa{l l }
1-\sqrt{R_{BC}R_{CD}} & \quad \text{for }R_{CD} > R_{BC}; \\ [2mm]
\sqrt{(1-R_{CD})(1-R_{BC})} & \quad \text{otherwise.}
\eaa\r. 
\eea

\section{Conclusions and outlook}
\label{sec:conc}

The dark matter problem greatly motivates the search for semi-invisibly decaying resonances in Run II of the LHC.
After the discovery of such particles, their masses will most likely have to be measured using the classic 
kinematic endpoint techniques. In fact, such techniques can already be usefully applied in the current data --- for example, 
following the procedure outlined in \cite{Burns:2008va}, the CMS collaboration has published an analysis of simultaneous extraction of the 
top, $W$ and neutrino masses from the measurement of kinematic endpoints in the $t\bar{t}$ dilepton system 
\cite{Chatrchyan:2013boa}.

\begin{figure}[t] 
\begin{center}
\includegraphics[width=7.5cm]{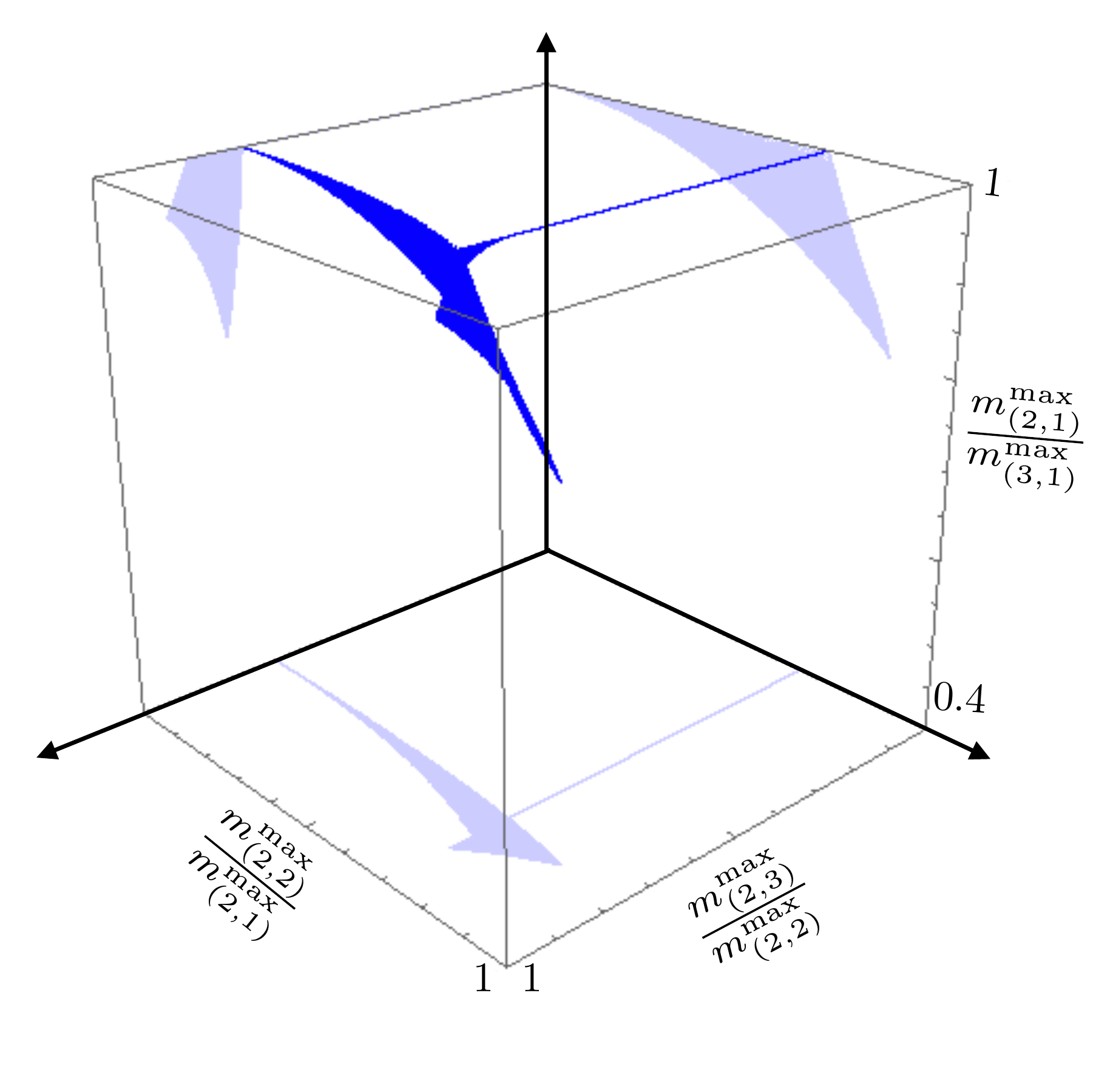}
\includegraphics[width=7.5cm]{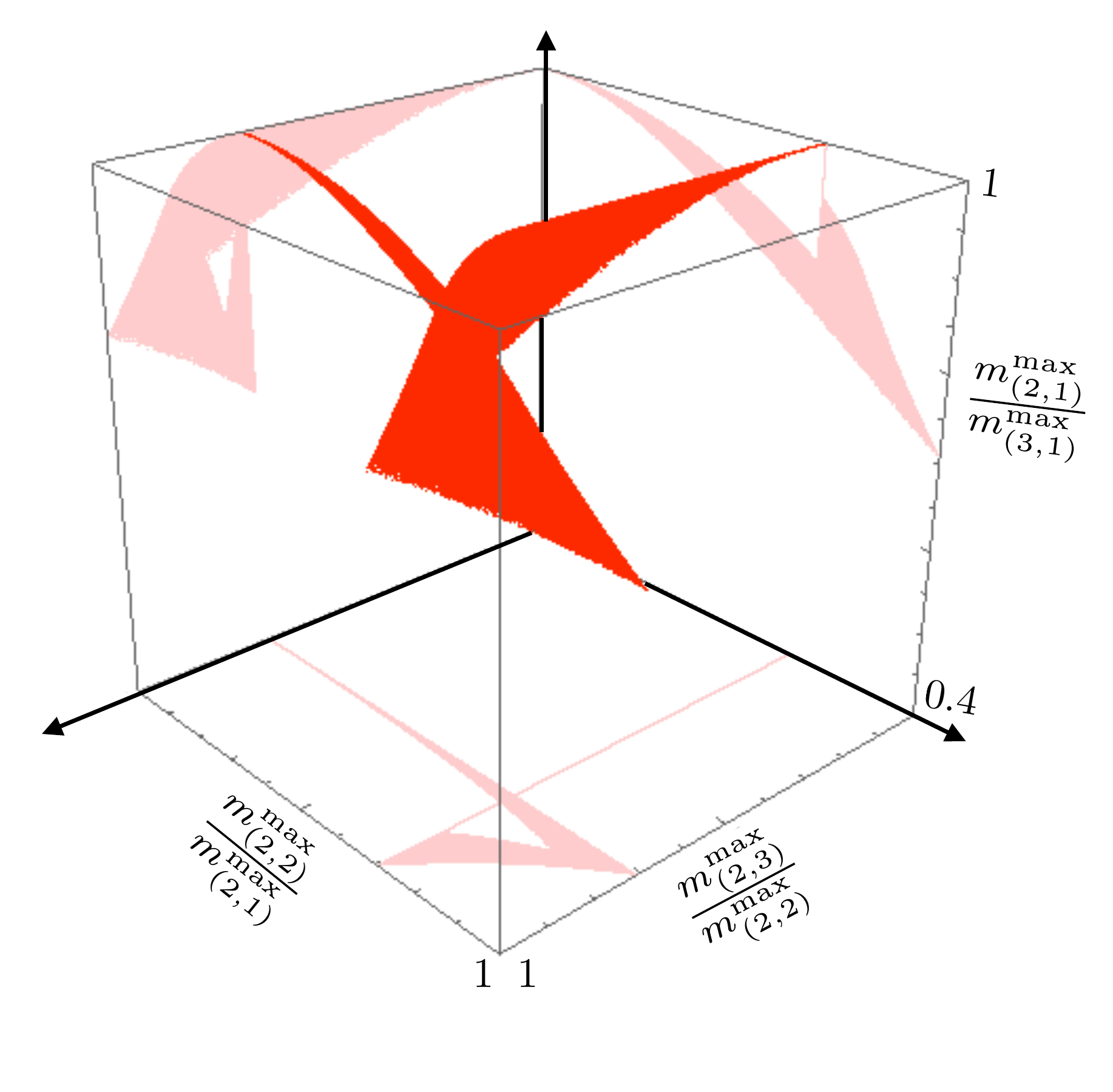}
\end{center}
\caption{\label{fig:endpoints}   
The correlations among the endpoint ratios
$x=\frac{m^{max}_{(2,3)}}{m^{max}_{(2,2)}}$,
$y=\frac{m^{max}_{(2,2)}}{m^{max}_{(2,1)}}$ and
$z=\frac{m^{max}_{(2,1)}}{m^{max}_{(3,1)}}$,
for the case of 
the type $(2,1)$ event topology from Fig.~\ref{fig:type21}(a) (left panel) 
and the type $(1,2)$ event topology from Fig.~\ref{fig:type12}(a) (right panel).
Here the lighter shaded images on each of the three planes $(x,\,y,\,0.4)$, $(0,\,y,\,z)$ and $(x,\,0,\,z)$
are the corresponding projections of the $3D$ surface
$(x,y,z)=\left(x(R_{BC},R_{CD}),y(R_{BC},R_{CD}),z(R_{BC},R_{CD})\right)$ which was
obtained by scanning over the allowed values of $R_{BC}\in (0,1)$ and $R_{CD}\in (0,1)$.
Note that the range of the $z\,$axis is from 0.4 to 1.
} 
\end{figure}

In this paper, we revisited the classic method for mass determination via kinematic endpoints, where one studies the 
invariant mass distributions of suitable collections of visible decay products, and extracts their upper kinematic endpoints.
We generalized the existing studies on the subject in several ways:
\begin{itemize}
\item We shied away from making any assumptions about the structure of the decay topology, 
and considered the invariant masses of all possible sets of visible decay products. This led us to the 
introduction in Eq.~(\ref{eq:defsort}) of the sorted invariant mass variables $m_{(n,r)}$, where we consider 
all possible partitions of $n$ visible particles, and then rank the resulting invariant masses in order.
The so defined sorted invariant mass variables allow us to study SUSY-like decay chains in a fully model-independent way.
\item In Section~\ref{sec:off} we considered a completely general semi-invisible decay with no intermediate resonances,
where a heavy particle $D$ decays directly to an arbitrary number $N$ of massless SM particles and a single massive NP particle $A$.
For this very general case, we derived the corresponding formulas for the endpoints of the sorted invariant mass variables,
Eq.~(\ref{mnrmax}). The importance of those results lies in the fact that they allow the experimenter to test for the presence of intermediate 
on-shell resonances between particles $D$ and $A$ --- in the absence of such resonances, the ratios of all endpoints are 
uniquely predicted by Eq.~(\ref{mnrmax}). Any measured deviation from those ratios will signal the presence of some other new
intermediate particles.
\item In the second half of the paper, i.e.~Sections~\ref{sec:111}, \ref{sec:210} and \ref{sec:120}, we considered 
the SUSY-motivated case of $N=3$, and focused on the three possible event topologies with one or two intermediate on-shell particles.
Once again, we derived the corresponding formulas for the kinematic endpoints of the sorted invariant mass variables in terms of the 
physical mass spectrum. Each possible event topology predicts certain correlations among the 
observed endpoints, as illustrated in Fig.~\ref{fig:endpoints}. 
\end{itemize}

In conclusion, we are hoping that the model-independent approach to the kinematic endpoint method
presented in this paper will soon be tested in real data after a new physics discovery.
At the same time, the results presented here may provide useful mathematical insights to 
researches interested in phase space kinematics.

\acknowledgments
We would like to thank W. S. Cho for collaboration at an early stage of this project. MP is particularly grateful to C.\,Lester for the detailed understanding of a phase space in a SUSY cascade decay chain and a Mathematica code to produce Fig.~\ref{fig:fullphase}.
DK would like to thank K.\,Agashe for supporting and general advice during the initial stage of this project. 
DK acknowledges support by LHC-TI postdoctoral fellowship under grant NSF-PHY-0969510.
DK and KM are supported by DOE Grant No.\,DE-SC0010296.
MP is supported by IBS under the project code, IBS-R018-D1.


\begin{thebibliography}{99}

\bibitem{Chatrchyan:2012jja} 
  S.~Chatrchyan {\it et al.}  [CMS Collaboration],
  ``Study of the Mass and Spin-Parity of the Higgs Boson Candidate Via Its Decays to Z Boson Pairs,''
  Phys.\ Rev.\ Lett.\  {\bf 110}, 081803 (2013)
  [arXiv:1212.6639 [hep-ex]].
  
\bibitem{Aad:2013xqa} 
  G.~Aad {\it et al.}  [ATLAS Collaboration],
  ``Evidence for the spin-0 nature of the Higgs boson using ATLAS data,''
  Phys.\ Lett.\ B {\bf 726}, 120 (2013)
  [arXiv:1307.1432 [hep-ex]].
  
\bibitem{Chatrchyan:2013mxa} 
  S.~Chatrchyan {\it et al.}  [CMS Collaboration],
  ``Measurement of the properties of a Higgs boson in the four-lepton final state,''
  Phys.\ Rev.\ D {\bf 89}, 092007 (2014)
  [arXiv:1312.5353 [hep-ex]].

\bibitem{Arrenberg:2013rzp} 
  S.~Arrenberg, H.~Baer, V.~Barger, L.~Baudis, D.~Bauer, J.~Buckley, M.~Cahill-Rowley and R.~Cotta {\it et al.},
  ``Working Group Report: Dark Matter Complementarity,''
  arXiv:1310.8621 [hep-ph].
    
\bibitem{Bartl:1986hp} 
  A.~Bartl, H.~Fraas and W.~Majerotto,
  ``Production and Decay of Neutralinos in e+ e- Annihilation,''
  Nucl.\ Phys.\ B {\bf 278}, 1 (1986).

\bibitem{Birkedal:2004xn} 
  A.~Birkedal, K.~Matchev and M.~Perelstein,
  ``Dark matter at colliders: A Model independent approach,''
  Phys.\ Rev.\ D {\bf 70}, 077701 (2004)
  [hep-ph/0403004].

\bibitem{Feng:2005gj} 
  J.~L.~Feng, S.~Su and F.~Takayama,
  ``Lower limit on dark matter production at the large hadron collider,''
  Phys.\ Rev.\ Lett.\  {\bf 96}, 151802 (2006)
  [hep-ph/0503117].
      
\bibitem{Appelquist:2000nn} 
  T.~Appelquist, H.~C.~Cheng and B.~A.~Dobrescu,
  ``Bounds on universal extra dimensions,''
  Phys.\ Rev.\ D {\bf 64}, 035002 (2001)
  [hep-ph/0012100].

\bibitem{Cheng:2002ab} 
  H.~C.~Cheng, K.~T.~Matchev and M.~Schmaltz,
  ``Bosonic supersymmetry? Getting fooled at the CERN LHC,''
  Phys.\ Rev.\ D {\bf 66}, 056006 (2002)
  [hep-ph/0205314].

\bibitem{Barr:2010zj} 
  A.~J.~Barr and C.~G.~Lester,
  ``A Review of the Mass Measurement Techniques proposed for the Large Hadron Collider,''
  J.\ Phys.\ G {\bf 37}, 123001 (2010)
  [arXiv:1004.2732 [hep-ph]].

\bibitem{Wang:2008sw} 
  L.~T.~Wang and I.~Yavin,
  ``A Review of Spin Determination at the LHC,''
  Int.\ J.\ Mod.\ Phys.\ A {\bf 23}, 4647 (2008)
  [arXiv:0802.2726 [hep-ph]].

\bibitem{Allanach:2000kt} 
  B.~C.~Allanach, C.~G.~Lester, M.~A.~Parker and B.~R.~Webber,
  ``Measuring sparticle masses in nonuniversal string inspired models at the LHC,''
  JHEP {\bf 0009}, 004 (2000)
  [hep-ph/0007009].

\bibitem{Gjelsten:2004ki} 
  B.~K.~Gjelsten, D.~J.~Miller and P.~Osland,
  ``Measurement of SUSY masses via cascade decays for SPS 1a,''
  JHEP {\bf 0412}, 003 (2004)
  [hep-ph/0410303].
  
\bibitem{Burns:2009zi} 
  M.~Burns, K.~T.~Matchev and M.~Park,
  ``Using kinematic boundary lines for particle mass measurements and disambiguation in SUSY-like events with missing energy,''
  JHEP {\bf 0905}, 094 (2009)
  [arXiv:0903.4371 [hep-ph]].

\bibitem{MP}
F.~Moortgat and L.~Pape, CMS Physics TDR, Vol. II, Report No. CERN-LHCC-2006, Chap.~13.4, p.~410.

\bibitem{Matsumoto:2006ws} 
  S.~Matsumoto, M.~M.~Nojiri and D.~Nomura,
  ``Hunting for the Top Partner in the Littlest Higgs Model with T-parity at the CERN LHC,''
  Phys.\ Rev.\ D {\bf 75}, 055006 (2007)
  [hep-ph/0612249].

\bibitem{Rajaraman:2010hy} 
  A.~Rajaraman and F.~Yu,
  ``A New Method for Resolving Combinatorial Ambiguities at Hadron Colliders,''
  Phys.\ Lett.\ B {\bf 700}, 126 (2011)
  [arXiv:1009.2751 [hep-ph]].

\bibitem{Bai:2010hd} 
  Y.~Bai and H.~C.~Cheng,
  ``Identifying Dark Matter Event Topologies at the LHC,''
  JHEP {\bf 1106}, 021 (2011)
  [arXiv:1012.1863 [hep-ph]].
      
\bibitem{Baringer:2011nh} 
  P.~Baringer, K.~Kong, M.~McCaskey and D.~Noonan,
  ``Revisiting Combinatorial Ambiguities at Hadron Colliders with $M_{T2}$,''
  JHEP {\bf 1110}, 101 (2011)
  [arXiv:1109.1563 [hep-ph]].

\bibitem{Choi:2011ys} 
  K.~Choi, D.~Guadagnoli and C.~B.~Park,
  ``Reducing combinatorial uncertainties: A new technique based on MT2 variables,''
  JHEP {\bf 1111}, 117 (2011)
  [arXiv:1109.2201 [hep-ph]].
  
\bibitem{Shim:2014aua} 
  J.~H.~Shim and H.~S.~Lee,
  ``Improving Combinatorial Ambiguities of ttbar Events Using Neural Networks,''
  Phys.\ Rev.\ D {\bf 89}, 114023 (2014)
  [arXiv:1402.3907 [hep-ph]].

\bibitem{Baer:1990rq} 
  H.~Baer, D.~D.~Karatas and X.~Tata,
  ``Gluino and Squark Production in Association With Gauginos at Hadron Supercolliders,''
  Phys.\ Rev.\ D {\bf 42}, 2259 (1990).

\bibitem{Agrawal:2013uka} 
  P.~Agrawal, C.~Kilic, C.~White and J.~H.~Yu,
  ``Improved mass measurement using the boundary of many-body phase space,''
  Phys.\ Rev.\ D {\bf 89}, 015021 (2014)
  [arXiv:1308.6560 [hep-ph]].

\bibitem{Matchev:2009iw} 
  K.~T.~Matchev, F.~Moortgat, L.~Pape and M.~Park,
  ``Precise reconstruction of sparticle masses without ambiguities,''
  JHEP {\bf 0908}, 104 (2009)
  [arXiv:0906.2417 [hep-ph]].

\bibitem{Cho:2012er} 
  W.~S.~Cho, D.~Kim, K.~T.~Matchev and M.~Park,
  ``Probing Resonance Decays to Two Visible and Multiple Invisible Particles,''
  Phys.\ Rev.\ Lett.\  {\bf 112}, no. 21, 211801 (2014)
  [arXiv:1206.1546 [hep-ph]].
  
\bibitem{BK}
  E.~Byckling and K.~Kajantie, 
  ``Particle Kinematics", Wiley, John \& Sons, 1973.
  
\bibitem{Giudice:2011ib} 
  G.~F.~Giudice, B.~Gripaios and R.~Mahbubani,
  ``Counting dark matter particles in LHC events,''
  Phys.\ Rev.\ D {\bf 85}, 075019 (2012)
  [arXiv:1108.1800 [hep-ph]].

\bibitem{Lester:2006yw} 
  C.~G.~Lester,
  ``Constrained invariant mass distributions in cascade decays: The Shape of the 'm(qll)-threshold' and similar distributions,''
  Phys.\ Lett.\ B {\bf 655}, 39 (2007)
  [hep-ph/0603171].
                
\bibitem{Costanzo:2009mq} 
  D.~Costanzo and D.~R.~Tovey,
  ``Supersymmetric particle mass measurement with invariant mass correlations,''
  JHEP {\bf 0904}, 084 (2009)
  [arXiv:0902.2331 [hep-ph]].

\bibitem{CLTASI} 
C.~Lester, ``Mass and Spin Measurement Techniques (for the LHC)",
in  ``The Dark Secrets of the Terascale : Proceedings, TASI 2011, Boulder, Colorado, USA, Jun 6 - Jul 11, 2011,''
Eds.~T.~Tait and K.~Matchev.
  
\bibitem{Lester:2006cf} 
  C.~G.~Lester, M.~A.~Parker and M.~J.~White,
  ``Three body kinematic endpoints in SUSY models with non-universal Higgs masses,''
  JHEP {\bf 0710}, 051 (2007)
  [hep-ph/0609298].

\bibitem{Burns:2008va} 
  M.~Burns, K.~Kong, K.~T.~Matchev and M.~Park,
  ``Using Subsystem MT2 for Complete Mass Determinations in Decay Chains with Missing Energy at Hadron Colliders,''
  JHEP {\bf 0903}, 143 (2009)
  doi:10.1088/1126-6708/2009/03/143
  [arXiv:0810.5576 [hep-ph]].

\bibitem{Chatrchyan:2013boa} 
  S.~Chatrchyan {\it et al.} [CMS Collaboration],
  ``Measurement of masses in the $t \bar{t}$ system by kinematic endpoints in pp collisions at $\sqrt{s}$ = 7 TeV,''
  Eur.\ Phys.\ J.\ C {\bf 73}, 2494 (2013)
  doi:10.1140/epjc/s10052-013-2494-7
  [arXiv:1304.5783 [hep-ex]].
    
\end{thebibliography}
\end{document}